# Molecular Tools for Non-Planar Surface Chemistry

*Taleana Huff[1,†], Brandon Blue[1,†], Terry McCallum[1,†], Mathieu Morin[1,†], Damian G. Allis [1,†], Rafik Addou[1,*], Jeremy Barton[1,*], Adam Bottomley[1,*], Doreen Cheng[1,*], Nina M. Ćulum [1,*], Michael Drew[1,*], Tyler Enright[1,*], Alan T.K. Godfrey[1,*], Ryan Groome[1,*], Aru J. Hill[1,*], Alex Inayeh[1,*], Matthew R. Kennedy[1,*], Robert J. Kirby[1,*], Mykhaylo Krykunov[1,*], Sam Lilak[1,*], Hadiya Ma[1,*], Cameron J. Mackie[1,*], Oliver MacLean[1,*], Jonathan Myall[1,*], Ryan Plumadore[1,*], Adam Powell[1,*], Henry Rodriguez[1,*], Luis Sandoval[1,*], Marc Savoie[1,*], Benjamin Scheffel[1,*], Marco Taucer[1,*], Denis A.B. Therien[1,*], Dušan Vobornik[1,*]*

[†]*Equal co-first-author; see Author Contribution for more details.*

[*]*Equal co-author, ordered by family name, see Author Contributions for more details.*

[1]CBN Nano Technologies, Inc., Ottawa, ON, Canada

## GTOC

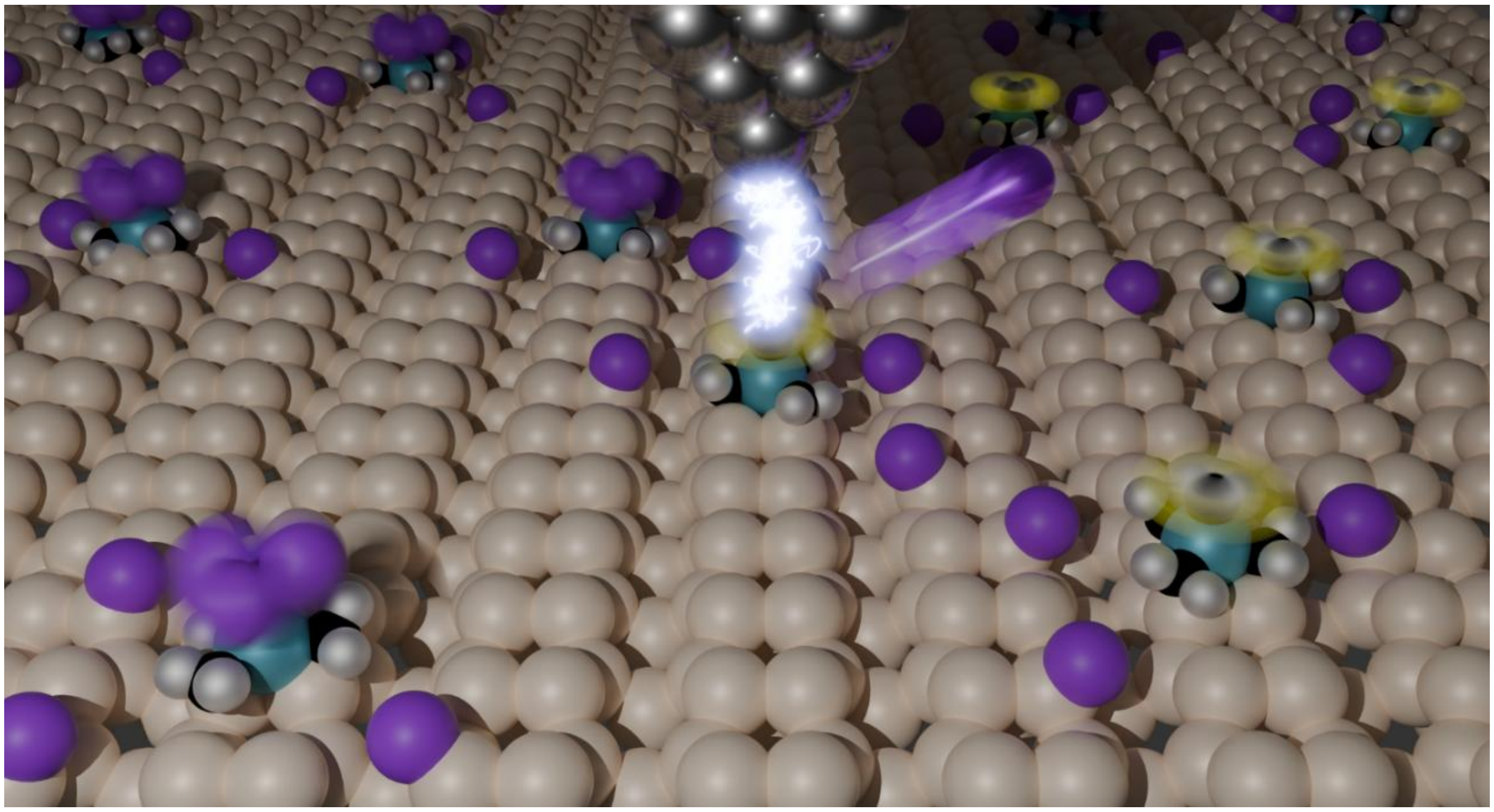

## Abstract

Scanning probe microscopy (SPM) investigations of on-surface chemistry on passivated silicon have only shown in-plane chemical reactions, and studies on bare silicon are limited in facilitating additional reactions post-molecular-attachment. Here, we enable subsequent reactions on Si(100) through selectively adsorbing 3D, silicon-specific "molecular tools". Following an activation step, the molecules present an out-of-plane radical that can function both to donate or accept molecular fragments, thereby enabling applications across multiple scales, *e.g.*, macroscale customizable silicon-carbon coatings or nanoscale tip-mediated mechanosynthesis. Creation of many such molecular tools is enabled by broad molecular design criteria that facilitate reproducibility, surface specificity, and experimental verifiability. These criteria are demonstrated using a model molecular tool tetrakis(iodomethyl)germane ($Ge(CH_2I)_4$; TIMe-Ge), with experimental validation by SPM and X-ray photoelectron spectroscopy (XPS), and theoretical support by density functional theory (DFT) investigations. With this framework, a broad and diverse range of new molecular engineering capabilities are enabled on silicon.

## Silicon Chemistry and 3D "Molecular Tools"

Silicon is the standard for modern devices due to its tunable electrical properties and decades of investment in tools to shape it. Thus, hybrid architectures with a silicon "base" present an attractive avenue for introducing new functionalities. One well-studied hybrid scheme is combining silicon with organic molecules which similarly possess mature and diverse tailorability.[1–5] Attaching molecules to silicon reproducibly is inherently challenging; difficulties are imposed both by silicon's strong covalent bonds and complicated surface energetics, as well as the molecules' energetically accessible binding geometries. [2–4,6] Some investigations accept multiple energetically accessible landing geometries on silicon,[7–11] whereas others have sought to exert adsorption control. These studies include working on passivated surfaces where deposited molecules can rearrange due to weak surface bonding,[12–14] selective adsorption reactions like cycloaddition of cyclooctynes,[15–17] alkenes,[18] or carbonyls[19,20] to silicon dimers, or guiding reactions on hydrogen-passivated silicon with "molecular molds" formed by STM-tip-patterned surface radicals.[21–25]

Of interest would be combining elements of these prior explorations to enable ordered and tailorable carbon-silicon coatings. That is, selective and reproducible adsorption of an organic molecule on reactive silicon in ultra-high-vacuum (UHV) conditions, but where the molecule possesses a reactive, spatially-accessible (non-planar) radical, enabling chemoselective *in situ* function as an acceptor or donor of molecular fragments. While organic molecular radicals have been created on silicon before, these studies have employed the passivated Si(111)$\sqrt{3} \times \sqrt{3}$R30°-B surface, with the radical in-plane to the sample for on-surface assembly investigations.[13,14,26] On bare silicon, groups have investigated tailored carbon-silicon coatings with "click-chemistry" layering, but some steps require challenging-to-integrate wet chemistry in an adjoining chamber.[17]

Herein, we enable UHV preparation of spatially-accessible radicals on 3D molecular scaffolds through a broad interdisciplinary effort in silicon-specific molecular adsorbate design. These molecular radicals are anticipated to be useful by enabling: (1) tailored carbon-silicon coatings where new functional groups added to the organic radical could impart interesting properties for photonic,[27,28] semiconductor,[29] and catalysis applications,[30–33] and (2) a route to atomically precise manufacturing in three dimensions, where a reactive scanning probe tip interacts with the reactive radical to "pick-and-place" a backing molecular fragment.[34]

To facilitate effective adsorption control and subsequent radical production, we present a framework for the design of functional 3D molecules ("molecular tools") which can be reproducibly deposited, experimentally validated, and subsequently modified. These criteria include:

1. High molecular symmetry;
2. Covalent bond formation;
3. "Loose legs, rigid body;"
4. Lattice-matching;
5. Confidence in surface-bound orientations; and
6. An accessible, stable radical on the surface-bound molecule.

We have applied these criteria on silicon with a newly synthesized model molecular tool, tetrakis(iodomethyl)germane ($Ge(CH_2I)_4$, or TIMe-Ge): a tetrahedral molecule with four iodomethyl "legs" ($CH_2I$) bound to a germanium atom core (Criteria 1). It was pre-screened by theoretical assessment to possess an optimal geometry and expected reactivity for consistent chemisorption with a minimum number of configurations on the Si(100)-(2×1) surface.[35,36] A threefold C−I dissociative addition chemisorption pathway is experimentally shown to present a single unreacted $CH_2I$ group out-of-plane to the surface on a 3D carbon "scaffold." [37–42] This $CH_2I$ is subsequently deiodinated by both photoactivation and STM tip interactions, and the produced radical is experimentally verified.

## Chemical Synthesis of TIMe-Ge

TIMe-Ge was synthesized using a homologation strategy featuring the addition of Turbo-Grignard (*i*PrMgCl-LiCl)[43,44] to a solution of $GeCl_4$ and $CH_2ICl$, followed by halogen exchange using Finkelstein reaction conditions[45] to produce the desired molecular tool (**Scheme 1**). The resultant product was characterized using nuclear magnetic resonance (NMR) spectroscopy, high-resolution mass spectrometry (HR-MS), and infrared spectroscopy (IR) prior to deposition on silicon. Complete details and reaction conditions are provided in *SI - TIMe-Ge Synthesis and Characterization*.

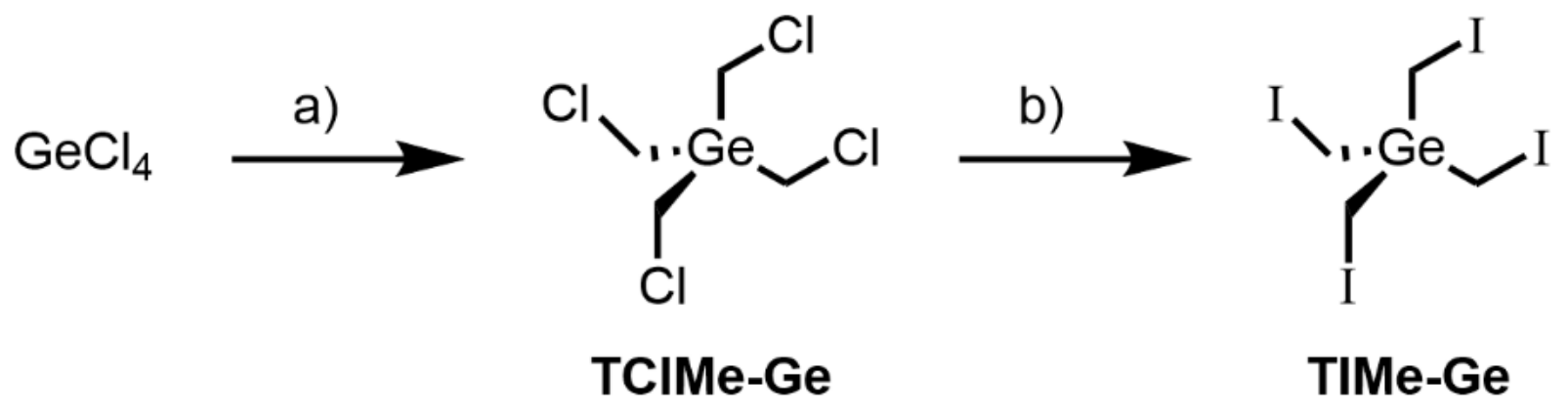

**Scheme 1**: **Synthesis of TIMe-Ge.** Reagents and conditions: **(a)** $CH_2ICl$/*i*PrMgCl-LiCl, THF, −78°C; **(b)** NaI, Acetone, Δ.

## Matching the Molecule to the Surface

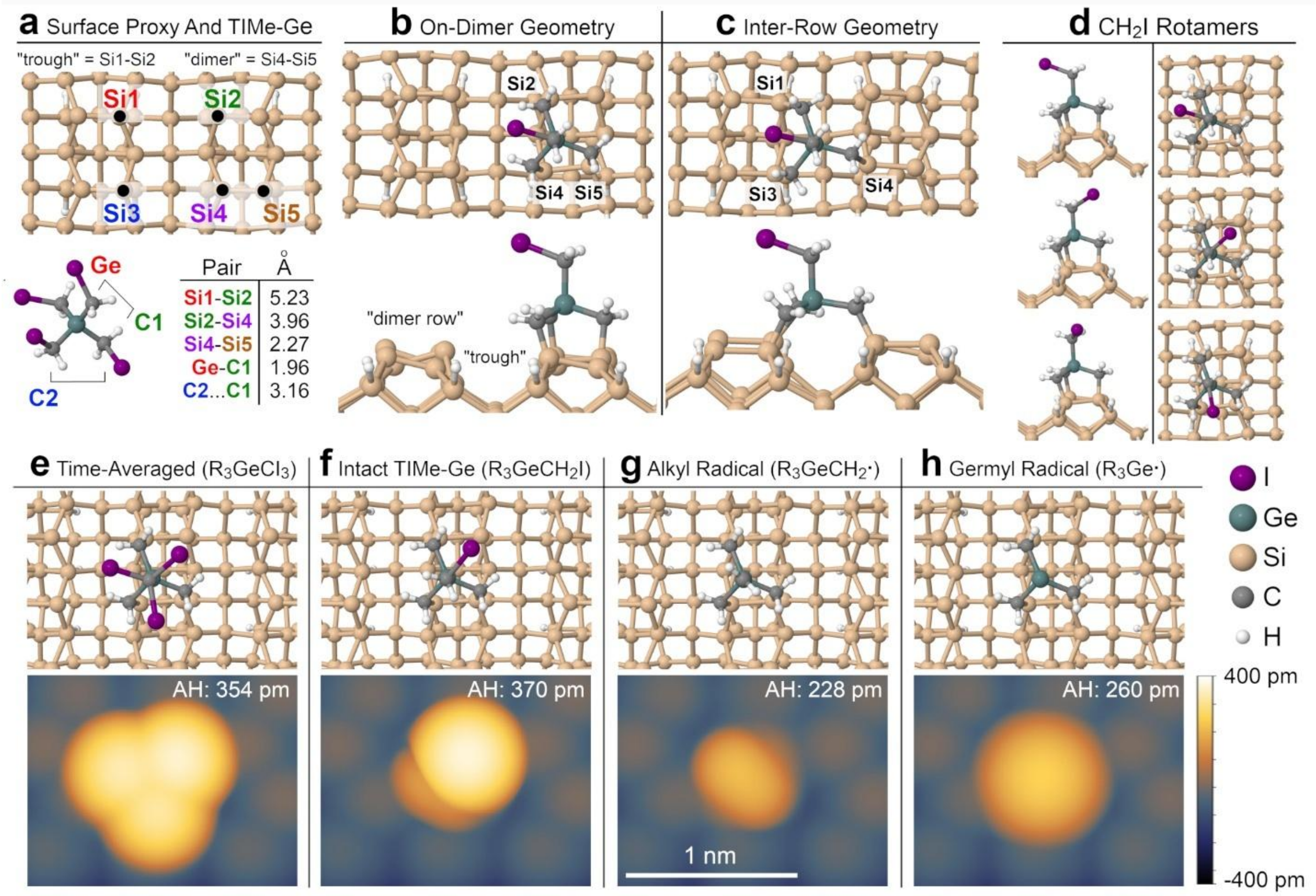

**Figure 1. DFT predictions of chemisorbed TIMe-Ge surface configurations.** (**a**) The Si(100)-(2×1) surface and gas-phase TIMe-Ge ($Ge(CH_2I)_4$) molecule with relevant computed dimensions for lattice matching. (**b**,**c**) DFT-optimized geometries for the on-dimer and inter-row three-legged surface bindings, respectively. (**d**) Representative three rotational isomers (rotamers) of the pendent $CH_2I$ group. (**e**-**h**) Simulated on-dimer STM images for (**e**) a $CI_3$ variant used as a time-averaged proxy for STM temperatures where the $CH_2I$ facilely rotates between rotamers, (**f**) an intact $CH_2I$ in a single "frozen" rotamer configuration, (**g**) the deiodinated, chemisorbed TIMe-Ge alkyl radical ($R_3GeCH_2^{\bullet}$; R = surface-bound $CH_2$), and (**h**) the deiodomethylated, chemisorbed TIMe-Ge germyl radical ($R_3Ge^{\bullet}$). "AH" refers to the STM "apparent height" of the tallest part of the molecule with respect to the local maxima of the Si dimer surface iso-current contour

(analogous to the experimental STM height measurements presented in this work). A legend of atom colors is provided.

A gas-phase model of the TIMe-Ge molecule and a top-down view of the Si(100)-(2×1) surface is provided in **Figure 1a**. From literature comparison, TIMe-Ge chemisorption on a Si(100) substrate was predicted to involve a series of sequential, spontaneous C−I dissociative addition reactions[37–41] forming three C−Si covalent bonds (Criteria 2). Rotation about the C−Ge bond for each $CH_2I$ leg[46] enables the molecule to access other nearby binding sites once the first C−Si bond has formed on a dimer, and the "rigid" Ge core prevents molecule deformation and restricts total surface binding geometries (Criteria 3). The molecule was predicted by DFT to adopt a stable tripodal geometry on the Si(100)-(2×1) surface in two potential surface bindings (Criteria 4): "on-dimer" (Si2-Si4-Si5, **Figure 1b**) or "inter-row" (Si1-Si3-Si4, **Figure 1c**). In either case, the fourth leg's Ge−C bond is oriented normal to the surface, available for interaction.

STM simulations were generated for anticipated intact ($R_3GeCH_2I$; R = surface-bound $CH_2$) and radical ($R_3GeCH_2^{\bullet}$ or $R_3Ge^{\bullet}$) presentations to guide experimental differentiability (**Figure 1e-h**). For intact presentations at elevated temperatures (≥77 K), the out-of-plane $CH_2I$ was expected to readily rotate between all three $CH_2I$ rotamers (**Figure 1d)** due to its DFT-calculated low rotational barrier (84 meV, $k_{77K} \geq 78$ MHz, *via* transition state energy differences, considering both an Eyring and Hindered-Rotor model; see *SI - Density Functional Theory Details*).[39,47] To reflect this STM appearance at ≥77 K, a $CI_3$ analog for its time-averaged appearance was generated, yielding an apparent "tri-lobe" (**Figure 1e**). For 4 K STM imaging, similar barrier arguments ($k_{4K} \leq 10^{-88}$ Hz) suggested the $CH_2I$ would be "frozen" into a single rotamer (**Figure 1f**).

STM simulations of a deiodinated alkyl radical ($R_3GeCH_2^{\bullet}$) and a deiodomethylated germyl radical ($R_3Ge^{\bullet}$) are provided in **Figure 1g,h**. Compared to intact TIMe-Ge, either radical would appear as a single lobe with a lower STM height (*i.e.*, a "mono-lobe," >100 pm shorter), allowing discrimination (Criteria 5). However, the heights of the radicals relative to each other (~30 pm difference) were anticipated to be difficult to consistently resolve. Only on-dimer simulations are presented in **Figure 1**, as it was the dominant configuration observed experimentally (*vide infra*). Partially-bound (**SI Figure 1**) and inter-row configurations (**SI Figure 2**) were also simulated.

Iodine was used as a "tagging" group for XPS characterization to enhance sensitivity to leg binding; the iodine 3d signal (I 3d) has a high relative sensitivity factor (RSF) of 6.206 (*e.g.,* ~20× that of the C 1s transition of 0.296), providing a strong signal for fitting.[48] C–I dissociative addition to the surface allows leg binding to be inferred by fitting the C–I and Si–I subpeaks in the I 3d doublet spectrum. Thus, leg binding and "intactness" could be evaluated for TIMe-Ge populations over a larger area (~$mm^2$, Criteria 5). Finally, the remaining out-of-plane iodomethyl leg of TIMe-Ge could be deiodinated to create a carbon-centred radical (Criteria 6) with two literature-supported paths: homolytic bond cleavage under UV[49] and localized tip-mediated removal.[50] Both are demonstrated here, with the aforementioned experimental indicators used to validate radical creation.

## Scanning Tunneling Microscopy (STM) Characterization

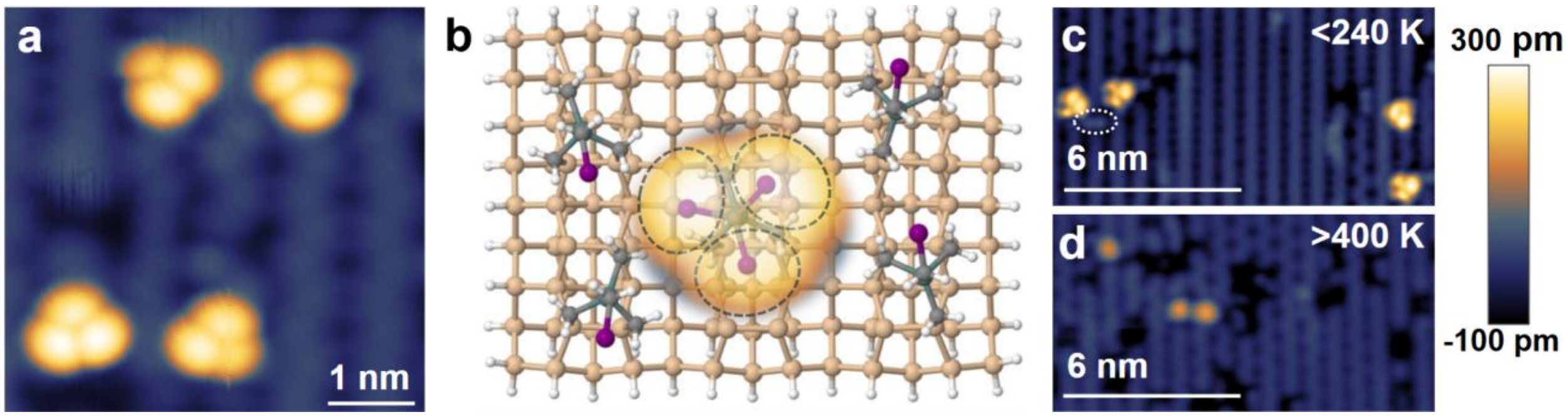

**Figure 2. TIMe-Ge molecules on Si(100): 77 K STM scanning. (a)** Experimental constant-current filled-states 77 K STM image of four TIMe-Ge molecules on Si(100)-(2×1) ($I$ = 50 pA, $V$ = –2.0 V). Molecule deposition was performed onto a room temperature Si(100) substrate. **(b)** The four symmetry-equivalent on-dimer configurations of TIMe-Ge. An experimental STM tri-lobe is overlaid on the central triply-iodinated proxy. **(c,d)** Representative 77 K STM images of TIMe-Ge molecules deposited on Si(100) substrates held at <240 K and >400 K, respectively, prior to imaging. The dashed white ellipse in (c) marks a pair of dissociated, surface-bound iodine atoms. A minor low-pass inverse fast Fourier transform filter has been applied to highlight TIMe-Ge positions on the Si lattice. z = 0 for STM heights is referenced to the highest point of the Si dimer-row surface.

TIMe-Ge was deposited with monodisperse coverage (~0.02 molecules/$nm^2$) on a flash-annealed Si(100)-(2×1) substrate (see *Methods*). Wafer temperature at the surface was calibrated against a diode, allowing temperatures to be estimated within ±70 K for LHe pre-cooling and ±50 K for $LN_2$ (see *SI - Experimental Methods Additional Details*). 77 K STM explorations of TIMe-Ge molecules deposited on room temperature (RT) substrates showed on-dimer tri-lobe features at filled-states biases (**Figure 2a**; compared to **Figure 1f**), supporting the predicted rotating $CH_2I$ group. With on-dimer TIMe-Ge requiring two adjacent silicon dimers for three-leg chemisorption, the molecules can adopt four symmetry-equivalent leg attachments (**Figure 2b**, **SI Figure 3**). The dark minima between lobes provide insight into the molecules' leg bonding configurations, with the $CH_2$–Ge legs acting as steric barriers to $CH_2I$ rotation.

For an RT deposition, the proportion of tri-lobes ($R_3GeCH_2I$) observed was 70% (±11%, with a 95% Wilson score[51] interval used throughout this work, $N$ = 62), with the remainder exhibiting mono-lobes ($R_3GeCH_2^{\bullet}$ or $R_3Ge^{\bullet}$). Some inter-row mono-lobe features were observed at an incidence of 9% (±6%, $N$ = 70), which could be inter-row alkyl or germyl radicals (**SI Figure 2**), but no inter-row tri-lobes were observed for any deposition temperature. For RT and higher temperature depositions, iodine atoms from C–I dissociative addition were rarely noted near their host molecules, instead migrating to defects as reported in other works.[52]

With a <240 K cooled surface (**Figure 2c**), tri-lobes comprised 91% of the surface-bound population (±4%, $N$ = 246), and iodine atoms were more often noted near their host molecules as individual or paired bright protrusions.[53] Few inter-row features were observed under cold substrate conditions (<2%, $N$ = 130). Conversely, at surface temperatures >400 K during deposition (**Figure 2d**), only 6% of molecules (±4%, $N$ = 140) were observed as tri-lobes, with on-dimer mono-lobe structures dominating. Under these conditions, inter-row features increased slightly to 15% (±6%, $N$ = 140, (**SI Figure 4**)), still with no tri-lobe character (<2%, $N$ = 140).

The effect of substrate temperature on depositions was further explored by XPS (*vide infra*), including colder temperatures that allow partially-bound two-leg-down configurations (**SI Figure 1**, **Row 2**). Results from 4 K STM scanning of TIMe-Ge agreed with rotational predictions; the $CH_2I$ is locked into a single rotamer but can be manipulated into others with tip interactions (**SI Figure 5**).

## Producing a Carbon-Centred Radical

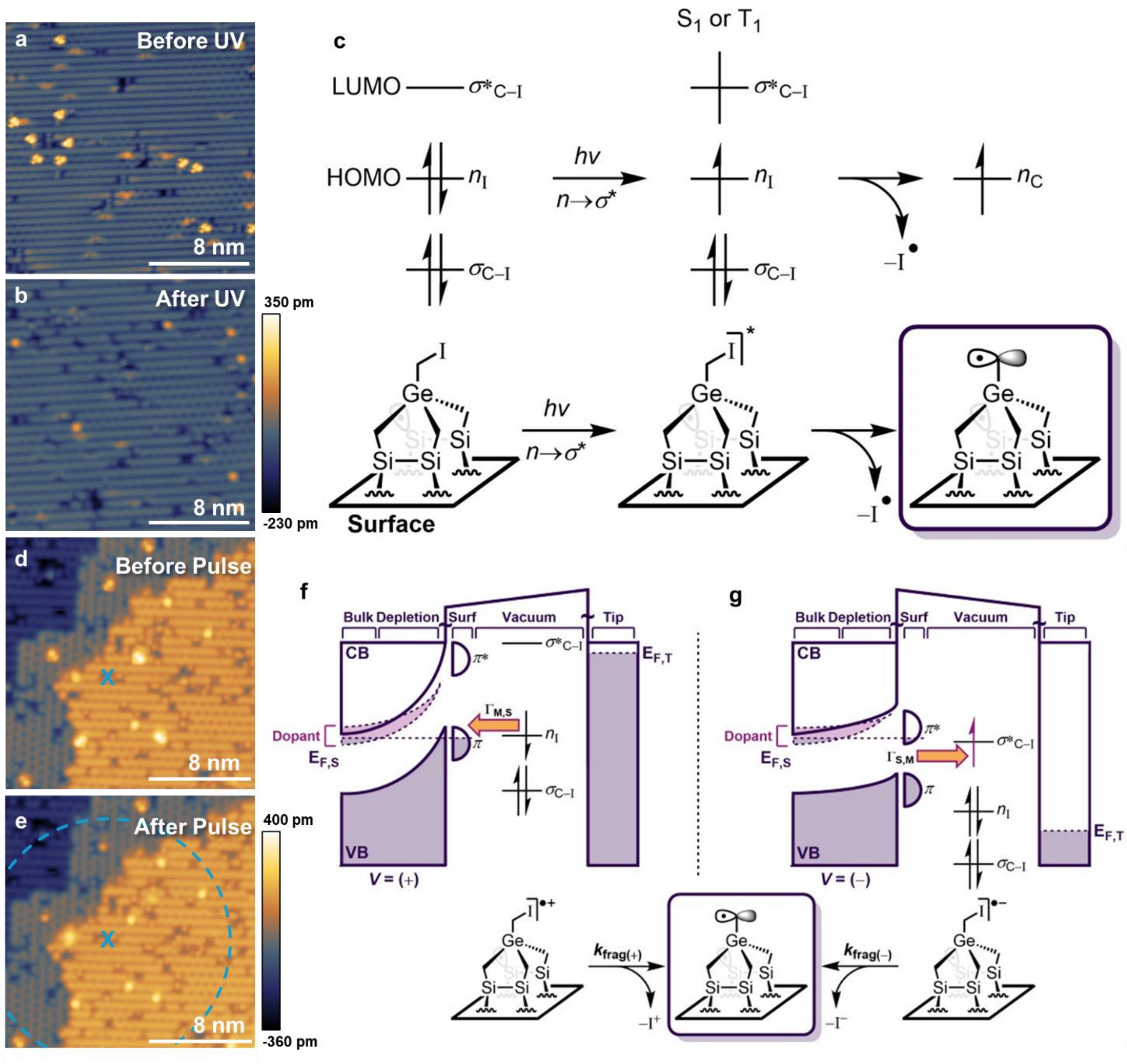

**Figure 3: Changes to TIMe-Ge *via* UV and bias pulse dehalogenation.** (**a,b**) Representative before and after STM ($V$ = –2.0 V, $I$ = 50 pA) images of a sample exposed to UV irradiation ($\lambda$ = 265 nm, irradiance = ~7 mW/cm$^2$, $t$ = 20 min). **(c)** Chemical reaction scheme outlining the dehalogenation mechanism for UV irradiation. (**d,e**) Before and after STM images ($V$ = –1.8 V, $I$

= 50 pA) of a region where a probe-retracted empty-states bias pulse ($z_{retract}$ = 5 nm, $\Delta V$ = +11.3 V, $t$ = 50 ms, ΔV relative to the STM imaging setpoint) was applied at the location marked with an X, and an approximate affected area noted with dashed lines. (**f,g**) Proposed chemical reaction schemes and band diagrams outlining the dehalogenation processes for probe-retracted (**f**) empty and (**g**) filled-states bias pulses. Proposed charge transfer pathways are denoted by $\Gamma_{M,S}$ for molecule-to-surface or $\Gamma_{S,M}$ for surface-to-molecule. TIMe-Ge molecular orbitals include the $\sigma$ (C–I; bonding), the $\sigma^*$ (C–I; anti-bonding), and $n$ (iodine atom; non-bonding) energy levels. $E_{F,T}$ and $E_{F,S}$ denote the tip and sample Fermi levels, respectively. π and π* are the delocalized Si(100)-(2×1) surface states. The dopant band for highly n-doped Si(100) is drawn as the dashed light purple region overlapping with the conduction band edge. z = 0 for all STM heights is referenced to the highest point of the Si dimer-row surface.

To produce a radical on the surface-bound molecule (Criteria 6), dehalogenation by UV irradiation and tip-mediated bias pulses were tested. **Figure 3a** displays a representative STM image before UV with tri-lobes as the dominant feature. After UV irradiation (see *Methods*), these tri-lobes entirely transformed into mono-lobes (103/103, 98±2%, **Figure 3b**). A proposed mechanism for UV deiodination is provided in **Figure 3c**, featuring the expected[49,54,55] $n\rightarrow\sigma^*$ transition of surface-bound TIMe-Ge. The reactive, excited-state intermediate undergoes C–I bond homolysis at near-unity quantum yields, producing mono-lobe alkyl radicals (as in **Figure 1g**). Deiodomethylation (*i.e.*, loss of $CH_2I$) is unlikely under these photochemical conditions.

Deiodination was also induced by retracting the STM tip and applying bias pulses. **Figure 3d,e** illustrates such deiodination for empty-states bias pulsing. Tri-lobes in a local area around the pulse location became mono-lobes of reduced height (see also **SI Figure 6**). Tip-induced reactions in literature have been attributed to electric-field effects,[56,57] charge transfer/tunneling mechanisms, or both.[58] The areal deiodination initially led to consideration of a field-driven bond-breaking model, but the required field strength for similar analogs (>2 V/Å)[56,59] far exceeded the estimated field here (0.12 ± 0.01 V/Å, **SI Figure 7**). Additionally, the bond being broken (C–I) is not aligned with the surface-normal field (~70° off-vertical), further reducing likelihood.

Thus, our proposed mechanism for bond breaking must incorporate both an electric field (to explain the areal effect) and charge transfer. However, tip removal from the surface limits charge transfer to between the molecule and sample only, as illustrated in **Figure 3f,g** where qualitative band diagrams for Si(100)-(2×1) are provided for empty- and filled-states bias pulses, respectively. Silicon surface states are denoted as $\pi$ and $\pi^*$,[60,61] which "pin" the band-bending upward in both polarities.[62]

As a 3D molecule, the tip electric field is posited to shift the molecular levels relative to the surface levels to facilitate charge transfer to/from TIMe-Ge orbitals. For an empty-state bias pulse (**Figure 3f**), the highest occupied molecular orbital (HOMO; iodine atom, $n_I$) could be aligned with states in the surface $\pi$ band, facilitating electron transfer from the molecule to the surface ($\Gamma_{M,S} > 0$) to produce a transient radical cation. Mesolytic cleavage at the C–I bond ($k_{frag(+)}$) would yield a carbon-centered radical as with UV, but with cationic iodine release. Sufficiently high field strength also can yield a Ge-centered radical *via* a deiodomethylation pathway upon radical cation fragmentation, or by a stepwise a-germyl cation rearrangement involving net oxidation of surface-bound TIMe-Ge by $2e^-$ (**SI Figure 8, 15**).

Similar conversions of tri-lobes to mono-lobes were observed for filled-state bias pulses (**Figure 3g**). Single electron tunneling events from the partially filled $\pi^*$ to the $\sigma^*_{C-I}$ ($0<\Gamma_{S,M}$) could result in transient radical anions, undergoing mesolytic C–I bond cleavage($k_{frag(-)}$) to produce a carbon-centered radical with release of an iodide anion. Unlike empty-state biases, no areal deiodination effect was observed, suggesting that either only molecules underneath the apex experienced sufficient field, or tunneling rates from surface$\rightarrow\pi^*\rightarrow\sigma^*_{C-I}$ were slow/inelastic, reducing yield. Additionally, a concerted dissociative electron transfer mechanism circumventing the formation of a formal radical anion intermediate may be possible toward the same products (carbon-centered radical, iodide anions).[63–65]

These proposed mechanisms qualitatively account for our observations based on molecular orbital alignment with a pinned, semi-conducting substrate. Additional experimental observations and other proposed mechanisms are explored in the SI (see *SI - Alternate Molecular Fragmentation Pathways,* **SI Figures 9-10 and 14-16**).

## X-ray Photoemission Spectroscopy (XPS) Characterization

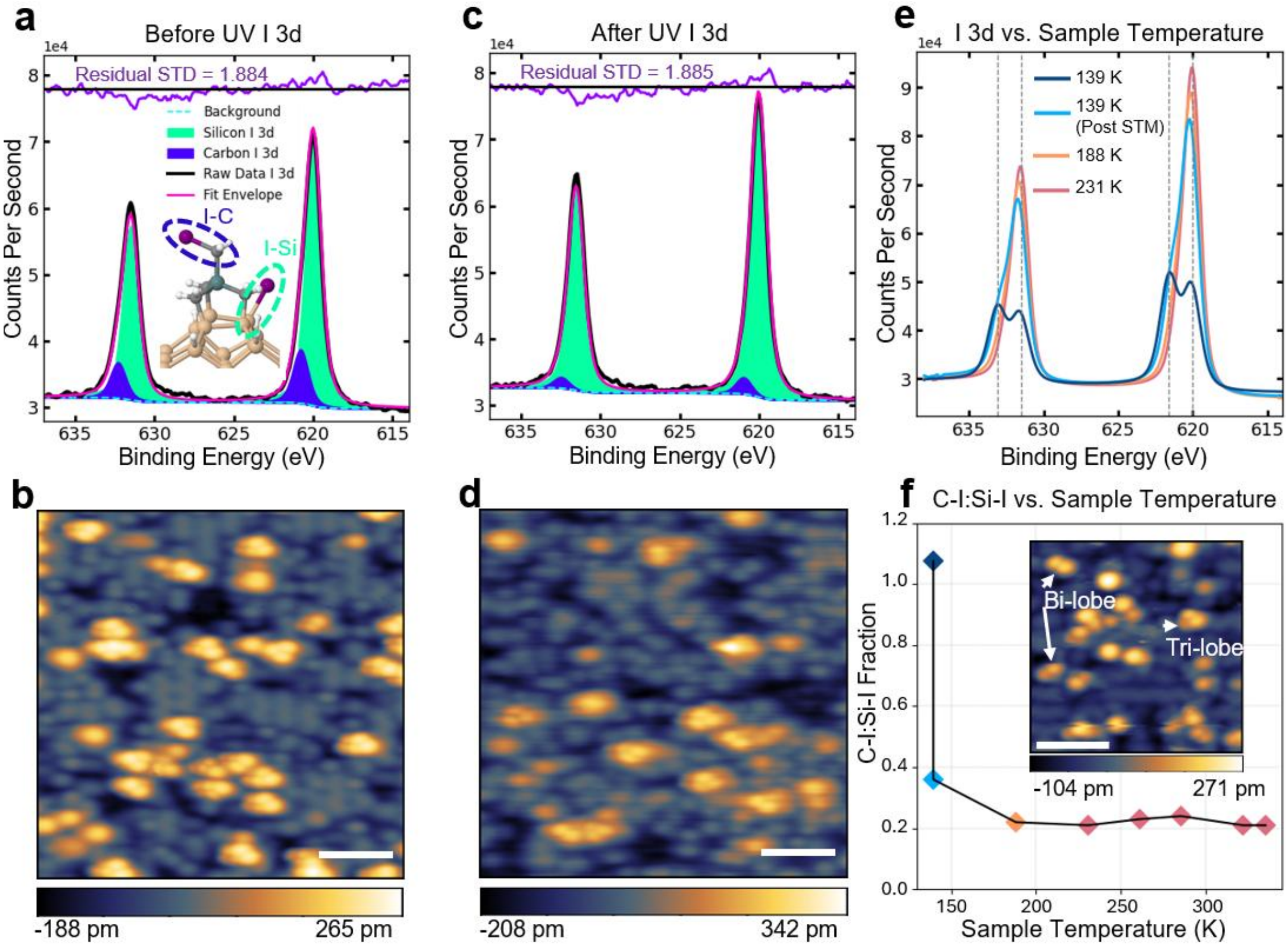

**Figure 4. XPS and STM characterization of TIMe-Ge on Si(100).** (**a**) I 3d XPS spectrum raw data (black) with C–I (dark blue) and Si–I (green) subpeak fits for a sub-monolayer TIMe-Ge deposition (substrate held at RT during deposition). The ball-and-stick model highlights the contributing bonds to the I 3d spectrum. (**b**) 77 K STM image acquired after (a). (**c**) I 3d XPS spectrum after UV irradiation of the same sample as (a,b) ($\lambda$ = 265 nm, irradiance = ~25 mW/cm$^2$, $t$ = 5 min). (**d**) 77 K STM image of the sample after UV. (**e**) I 3d XPS spectra acquired as a sub-monolayer deposition of TIMe-Ge molecules on Si(100) (substrate held at 139 K during deposition). Snapshots were taken at 139 K (before and after the STM in (f), dark and light blue, respectively) and other indicated temperatures. Vertical dashed grey lines mark the C–I and Si–I

subpeak centres for the doublet. **(f)** The C–I:Si–I ratio as a function of temperature for the same XPS spectra as shown in (e), color-coded to aid comparison. Inset: STM of the cold-transferred sample (dark blue) revealing mono-lobes, bi-lobes, and tri-lobes. All temperatures are the expected Si wafer temperature as determined by a calibration sample (see *SI – Experimental Methods Additional Details*). STM scale bars are 2 nm. XPS data display the Shirley background fits (bottom, dashed light-blue), fit envelopes (pink), and standardized residuals of the fits (top, purple). $V = -2.0$ V, $I = 50$ pA for all STM images. Other XPS spectra (C 1s, O 1s, Si 2p, Ge 3d, Ge 2p, and a Survey) are provided in **SI Figure 17**.

To complement the STM data, sub-monolayer depositions of TIMe-Ge were characterized through XPS. We focus on the I 3d spectra due to iodine's high RSF and lack of overlap with other transitions. An I 3d spectrum of sub-monolayer TIMe-Ge molecules on Si(100) is presented in **Figure 4a**. The doublet (black) contains two fitted subpeaks: C–I (dark blue, I $3d_{5/2}$ = 620.8 eV, I $3d_{3/2}$ = 632.3 eV) and Si–I (green, I $3d_{5/2}$ = 620.0 eV, I $3d_{3/2}$ = 631.5 eV). A "perfect" deposition of intact TIMe-Ge molecules would have each molecule with a single C-I bond and three Si–I bonds (**Figure 4a** inset) for a C–I to Si–I peak area ratio of 1:3 ($0.3\overline{3}$, with decimal values used in the remainder of this work). All XPS-extracted values assume that surface-cleaved iodine remains chemisorbed and that there is no molecular decomposition.

From the subpeak fits in **Figure 4a** (see **SI Table 2** for numerical values), the C–I:Si–I fraction was 0.209, suggesting ~69% of TIMe-Ge molecules landed in an intact three-leg down configuration (*SI – XPS Additional Information*). To support this finding, we examined molecule populations on the same sample through 77 K STM imaging (**Figure 4b**). If tri-lobes indicate a single C–I bond and three Si–I bonds, and mono-lobes signify four Si–I bonds, the STM C–I:Si–I value was 0.175 (±0.02, *N*=268).

The same deposition was then UV-irradiated, which decreased the C–I subpeak area and increased the Si–I area (**Figure 4c**) for a C–I:Si–I value of 0.053 (*i.e.*, ~20% TIMe-Ge molecules are tri-lobe). 77 K STM assessment (**Figure 4d**) corroborated fewer tri-lobes, more mono-lobes, and a denser background coverage of surface iodine. The extracted C–I:Si–I value was 0.090 (±0.021, *N*=175). Discrepancies between the STM and XPS ratios could be due to XPS fit uncertainties, non-uniform molecule deposition density, surface contamination or vacancies decreasing available reactive sites, or some iodine atoms being ejected into vacuum.

XPS also was used to assess how deposition on a temperature-calibrated (*SI – Experimental Methods Additional Details)* cold substrate behaved upon warming. Following deposition at 139 K, XPS (dark blue, **Figure 4e**) showed a C–I:Si–I value of 1.08 (**Figure 4f**, **SI Table 3**): a ratio only possible with a near-equal number of C–I and Si–I bonds. To investigate, the cold sample was imaged in STM, where some molecules presented as bi-lobes (**Figure 4f** inset). These bi-lobes often displayed local discontinuities during imaging, with some converting to tri-lobes from tip perturbation. This finding supported the existence of a "partially-bound" configuration of TIMe-Ge, with two out-of-plane $CH_2I$ legs (**SI Figure 1**, **Row 2**), consistent with the >1.0 XPS C–I:Si–I value. After STM analysis, the same sample was moved back to the 139 K manipulator arm for subsequent XPS scans (light blue, **Figure 4e**). The C–I:Si–I value decreased to 0.355 (attributed to thermal loading from the RT manipulator allowing some bi-lobes to fully bind). This sample was then counter-heated up to 335 K with XPS spectra acquired intermittently (**Figure 4f**). The C−I:Si−I value did not change significantly for sample temperatures above 188 K (leveling off at ~0.22), consistent with the proposed thermally-driven bonding dynamics.

## Conclusions

In this work, we presented six molecular design criteria deemed essential to enabling guided molecule reactivity. The criteria's utility was showcased on Si(100)-(2×1), guiding the synthesis and experimental validation of TIMe-Ge attachment. TIMe-Ge was shown to land predominantly in a single on-dimer, three-leg-down chemisorbed surface configuration when sample temperatures were held between 150-300 K during deposition. This three-leg-down configuration presented a $CH_2I$ group out-of-plane to the surface on a small 3D carbon scaffold, with unique presentation in STM and XPS. Normal-facing carbon radicals could then be prepared by deiodinating the $CH_2I$ group by either UV irradiation or tip-mediated bias pulses. These conclusions were supported by XPS studies, with the C−I:Si−I ratios in the I 3d spectrum suggesting surface-wide radical creation for UV irradiation.

This work thus serves two major purposes. Firstly, it demonstrates the utility of a multi-expertise approach in selecting molecular characteristics to combine desired surface interactions, characterization methods, and functionality. Secondly, we lay the groundwork for future applications of TIMe-Ge and similar "molecular tools" on Si(100); prepared radical intermediates are promising for on-surface synthesis and tailored carbon-coating applications. For example, one might functionalize a probe with a reproducible carbon entity (**SI Figure 16**) to then be used to make atomically-precise products in a "pick-and-place" SPM scheme, or one could accept new functional groups *en masse* for customizable coatings.[2,4,23] TIMe-Ge is expected to be a core motif for molecular engineering, especially with its versatile chemical framework making it suitable for diverse modification to its core, legs, and functional groups.

## Acknowledgments

The work presented in this manuscript was funded by the Canadian government under the Strategic Innovation Fund (SIF), and by Canadian Bank Note Company, Ltd. (CBN) under "CBN Nano Technologies, Inc." We gratefully acknowledge the efforts of Eduardo Barrera-Ramirez, Dr. Byoung Choi, Dr. William Cullen, Sheldon Haird, Dr. Alexei Ofitserov, Anthony Prior, Garrett Thomson, and Alex Vierich for their assistance in various hardware design, modification, and machining/assembly projects which contributed to this work. We thank Mark Jobes for programming assistance in developing tools for automated analysis. We thank Dr. Ryan Yamachika and Dr. Matthew Moses for fruitful discussions pertaining to the 77 K results. We thank Dr. François Magnan for assistance with infrared spectroscopy and Dr. Sharon Curtis for assistance with high-resolution mass spectrometry at the University of Ottawa. We thank Darian Blue for procuring consumables and organizing publication efforts. We thank Mike Meakin, John Turriff, and Steven DeSmet for project and program management efforts.

### Author Contributions

DA, RA, JB, MD, AH, TM, and M. Morin, contributed to the design criteria, synthesis, and/or *ex-situ* chemical characterization of TIMe-Ge. RA, NC, TE, TH, RK, SL, M. Morin, RP, HR and BS conceived, executed, analyzed, and/or consulted on the XPS experiments. DA, AB, BB, DC, NC, AH, AI, TH, RK, SL, AP, RP, BS, and DV conceived, executed, analyzed, and/or consulted on the 77 Kelvin STM experiments. TE, ATKG, HR, and DABT conceived, designed, and implemented the custom UV-LED deiodination module used for dehalogenation. RA, AB, DC, NC, TE, ATKG, RG, TH, RK, SL, HM, JM, OM, AP, RP, HR, BS, DABT and MT contributed to probe/sample/machine preparation used in this work. RG, TH, HM, JM, OM, HR, BS, MS, DABT and MT conceived, executed, analyzed and/or consulted on the 4 Kelvin STM/AFM experiments. DA, M. Kennedy, M. Krykunov, CM, and LS contributed to simulations and data interpretation. RA, DA, BB, JB, MD, TE, RG, TH, M. Kennedy, OM, MT, and DV provided supervision and management. RG, AI, M. Kennedy, SL, AP and DABT wrote software for data acquisition or processing. The core manuscript authoring and editing team consisted of DA, BB, TH, M. Morin, and TM, with additional input from AB, NC, TE, JM, RP, and DABT. All authors had an opportunity to review and propose edits to the manuscript before submission.

### Conflict of Interest Disclosure

The authors declare competing financial interests; patents have been filed on aspects of this work. All authors are or were affiliated with CBN Nano Technologies, Inc. a company seeking to commercialize atomically precise 3D manufacturing. The work presented in this manuscript was also funded in part by the Canadian government under the Strategic Innovation Fund (SIF, Agreement Number 813022).

## Methods

### Theoretical Modeling

Geometry optimizations were performed on molecular Si(100) proxies consisting of three dimer rows with four dimers per row (“3×4”) for on-dimer row geometries. A slightly expanded proxy with four rows and four dimers per row (“4×4”) was used for inter-row geometries. These calculations were done using the B3LYP hybrid density functional[66] and D3 version of the Grimme dispersion correction with Becke-Johnson damping[67] in Q-Chem 5.4.2[68] using program-option

SCF convergence = 8 ($\Delta E$ = $10^{-8}$) and threshold = 13 (two-electron integral cut-off = $10^{-13}$). Geometry optimizations were performed with surface C, H, and Si atoms treated with the 6-31G(d,p) basis set,[69] 6-311G(d,p) for I and Ge,[70] and 3-21G*[71] for subsurface (Si and H framework) atoms. Molecular orbital cube file generation was performed in Q-Chem 5.4.2 from the optimized geometries with the 6-311G(d,p) basis set used for all atoms, with these cube files provided as input to an STM simulation code based on the Tersoff-Hamann approximation.[72] To approximate the depassivated Si(100) surface and to simplify the generation of simulated STM images, a set of atom anchors and axis and plane constraints were imposed (***SI Table 1***). All simulated STM images generated with the Tersoff-Hamann approximation used the following current and voltage setpoints: ($I$ = 50 pA, $V$ = −1.5 V). Information about additional DFT considerations and rotational potential calculations is provided in *SI – Density Functional Theory Details*.

## Silicon Sample Preparation

The highly doped ($10^{18}$ arsenic atoms/$cm^3$) n–type Si(100) wafers used in this work were prepared by automated, pressure-limited (<$1\times10^{-9}$ Torr) ~1200 °C resistive flash anneals with remote-reading pyrometer feedback for 1 to 10 minutes in a UHV environment (baseline pressure of ~$5\times10^{-10}$ Torr or better). Clean crystalline samples were produced with 2×1 silicon surface reconstruction.

## TIMe-Ge Deposition

TIMe-Ge molecules were deposited onto silicon substrates through a Quad Cluster Source (QCS) effusion cell (Dr. Eberl MBE-Komponenten GmbH). The purified powdered molecules were loaded in quartz crucibles, packed firmly, and placed in the QCS heating receptacle. The QCS was then mounted to the UHV sample preparation chamber and pumped. The main chamber was baked to ~120 °C while providing cooling to the QCS with a closed-cycle PolyScience DuraChill chiller to attain UHV pressures (<$1\times10^{-10}$ Torr) without overheating the TIMe-Ge molecules, then opened to the sample preparation chamber. Due to the volatility of the molecules (which tended to sublimate above 5–15 °C in UHV), the QCS cells were always cooled *via* a 1:1 mixture by volume of ethylene glycol and deionized water chilled to −10 to 0 °C whenever the QCS was under vacuum. This cooling prevented water vapor condensation on the outer surface of the QCS while ensuring that the molecules did not unintentionally sublimate under baseline operating conditions. Molecules were degassed by counter-heating against the glycol chiller for several hours near their deposition temperature to remove water and other weakly-bound contaminants.

For depositions, crucibles were first heated to their aerosolization temperatures (~5 to 15 °C, depending on the QCS unit and its thermocouple placement), then the temperature was allowed to stabilize. The silicon samples, thermalized to the desired deposition sample temperature, were then placed in front of the crucible opening (~1 cm distance) with the line-of-sight shield closed. Upon opening the QCS shield, molecules were deposited ~1 to 10 minutes at a given flux exposure rate depending on the areal coverage desired.

## LT-UHV-STM

All STM results were gathered from one of three separate LT-UHV-STM systems with independent adjoining fast-entry load-lock, sample preparation, and molecular deposition chambers primarily composed of standard Scienta Omicron components. Each STM was controlled by Nanonis hardware and software (SPECS GmbH) with in-house developed automation utilities. Current was measured using DLPCA-200 current amplifiers at LN $10^9$ gain. The in-house designed and custom-machined STM heads were cooled *via* bath cryostats with either $LN_2$ or LHe for imaging at 77 K and 4 K, respectively. One system was a commercially purchased Omicron LT-STM/AFM system with an XPS system on a chamber adjoining the scan head.

### Sample Heating and Cooling

Different SPM systems used different heating and cooling schemes depending on available equipment. For all systems, silicon wafer surface temperature was calibrated against permanently mounted thermocouples *via* either tests with a special thermocouple sample (**SI Figure 18–22**) or the use of an externally-mounted PYROSPOT 10 pyrometer camera from DIAS Infrared.

To control silicon sample temperature during or after depositions on the combination STM/AFM/XPS Omicron system, we used a UHV-integrated translator arm with both an Omicron dual filament e-beam heater (PN03720) and a Vacuum Generator liquid nitrogen coil (ZLNHX). This setup allowed e-beam sample heating from the back, $LN_2$ cooling though the manipulator arm, or a combination of both simultaneously in the UHV environment (**SI Figure 19-22**). On all other STM instruments, direct-current heating was used, and cooling was performed by coming to thermal equilibrium in the SPM head (**SI Figure 18**). Complementary details can be found in *SI – Experimental Methods Additional Details*.

### Tungsten Probe Ex-Situ Preparation

Tungsten probes were prepared starting from a 0.25 mm diameter polycrystalline tungsten wire which was electrochemically etched (~20 nm radius of curvature or less) in a 5 M NaOH bath (**SI Figure 23a**). A field ion microscope using helium as the imaging gas was then used to assess the structure of the probe and remove oxide contamination (**SI Figure 23b**).[73] Nitrogen etching to reduce the radius of curvature was employed if needed to sharpen the probe.[73] Once clean and sharp, probes were either transported to the STM through a UHV vacuum suitcase (Scienta Omicron, <1E$^{-9}$ mB) or through air (<2 min ambient exposure).

## UV Dehalogenation

TIMe-Ge molecules were deiodinated *via* a custom UV-LED module (>38 mW, nominally 50 mW, $\lambda = 265$ nm, based on a ThorLabs M265L5) attached to a UV-transparent viewport (**SI Figure 24a,b**). Using a UV fused silica lens (ThorLabs, ASL5040-UV, numerical aperture of 0.65), the diffuse UV beam was loosely focused and guided to the desired sample region by two additional visible light lasers (**SI Figure 24c**) and an in-line compact microscope element. Approximately 7 mW/cm$^2$ of intensity was delivered to the sample surface; UV power density of the module was verified with an external optical power sensor, with typical values of ~19 mm beam diameter at 24 cm working distance. Irradiance was adjusted for systems with different working distances, as noted in figure captions. To mitigate the risk of eye exposure to UV light, the module was wrapped in an opaque safety fabric prior to use.

## XPS

XPS data were taken with a monochromated Al $K_\alpha$ x-ray source operated at 300 W and an Argus CU Spectrometer (Scienta Omicron). The pressure during data acquisition was <$5\times10^{-10}$ Torr. 100 eV passing energies were used for all presented data. At this passing energy, the FWHM of the Ag $3d_{5/2}$ peak measured on a sputtered silver foil sample was 0.95 eV. All data were collected such that the electron emission angle was normal to the sample. The analysis area for the spectrometer aperture used (aperture 5) was 1.54×4.09 mm$^2$. Samples were prepared (*i.e.*, annealed and deposited upon) in the analysis chamber using the methods described above. No charge neutralization was employed. For each I 3d measurement, O 1s, C 1s, Ge 2p, Ge 3d, and a survey scan were also collected (see **SI Figure 17*)***. Total acquisition times for all spectra scans in each run were ~20–30 min, with the I 3d spectra taking ~5 min. The spectra were acquired sequentially.

Peak fitting was performed using CasaXPS 2.3.25. All spectra were energy-corrected to the known Si peak energy of 99.4 eV. For all I 3d spectra, the background was fit with a Shirley function and the entire I 3d doublet with both spin-orbit components were used to fit the C–I and Si–I subpeaks. For constraints, the FWHM for all peaks were fixed, and between the spin components a splitting energy of 11.50 eV with an area constraint of 2:3 was used.[74–76] LA(83) line shapes were used to fit the peaks (*i.e.*, a convoluted Lorenzian-Gaussian shape with the 83 defining the gaussian width, and no asymmetry term(s)).[77]

### Molecule Height Extraction

Gwyddion[78] was used to extract and aggregate all theoretical and experimental STM height data. STM images were all three-point plane-leveled to tops of the Si dimers, with this plane defining z = 0 for the whole image. See *Supporting Information* for additional details.

# Supplementary Information: Molecular Tools for Non-Planar Surface Chemistry

*Taleana Huff*[1,†], *Brandon Blue*[1,†], *Terry McCallum*[1,†], *Mathieu Morin*[1,†], *Damian G. Allis* [1,†], *Rafik Addou*[1,*], *Jeremy Barton*[1,*], *Adam Bottomley*[1,*], *Doreen Cheng*[1,*], *Nina M. Ćulum* [1,*], *Michael Drew*[1,*], *Tyler Enright*[1,*], *Alan T.K. Godfrey*[1,*], *Ryan Groome*[1,*], *Aru J. Hill*[1,*], *Alex Inayeh*[1,*], *Matthew R. Kennedy*[1,*], *Robert J. Kirby*[1,*], *Mykhaylo Krykunov*[1,*], *Sam Lilak*[1,*], *Hadiya Ma*[1,*], *Cameron J. Mackie*[1,*], *Oliver MacLean*[1,*], *Jonathan Myall*[1,*], *Ryan Plumadore*[1,*], *Adam Powell*[1,*], *Henry Rodriguez*[1,*], *Luis Sandoval*[1,*], *Marc Savoie*[1,*], *Benjamin Scheffel*[1,*], *Marco Taucer*[1,*], *Denis A.B. Therien*[1,*], *Dušan Vobornik*[1,*]

[†]*Equal co-first-author; see Author Contribution for more details.*

[*]*Equal co-author, ordered by family name, see Author Contributions for more details.*

[1]CBN Nano Technologies, Inc., Ottawa, ON, Canada

## TIMe-Ge Synthesis and Characterization

### General Synthetic Chemistry Information

All reactions were performed in Pyrex glassware or vials equipped with a stir bar and capped with a septum. Commercial reagents were used without further purification. Yields refer to products that were isolated after purification. Reaction progress was tracked using thin layer chromatography (TLC) analysis and visualized with UV light (254 nm) and/or stains ($KMnO_4$, PA, $I_2/SiO_2$). Automated flash column chromatography was performed using a Yamazen Smart

Flash system equipped with an ELSD and UV detector (254 nm). $^{1}$H NMR (600 MHz) and $^{13}$C NMR (151 MHz) were recorded on a Bruker Avance III HD 600 spectrometer at the University of Ottawa; an artifact in $^{13}$C NMR spectra related to the hardware of the instrument was present at 105.14 ppm. NMR samples were dissolved in chloroform-d ($CDCl_3$). Chemical shifts are reported in ppm and referenced to the solvent residual ($^{1}$H NMR: δ = 7.26 ppm, $^{13}$C NMR: δ = 77.16 ppm). Peak multiplicities were described accordingly: s = singlet. High-resolution mass spectrometry (HR-MS) was performed by the John L. Holmes Mass Spectrometry Facility at the University of Ottawa on a Kratos Concept 2S (electron impact ionization, magnetic sector analyzer). Peaks pertaining to infrared spectroscopy are described accordingly: w = weak, m = medium, s = strong, vs = very strong.

## Synthesis and Characterization of Tetrakis(iodomethyl)germane (TIMe-Ge)

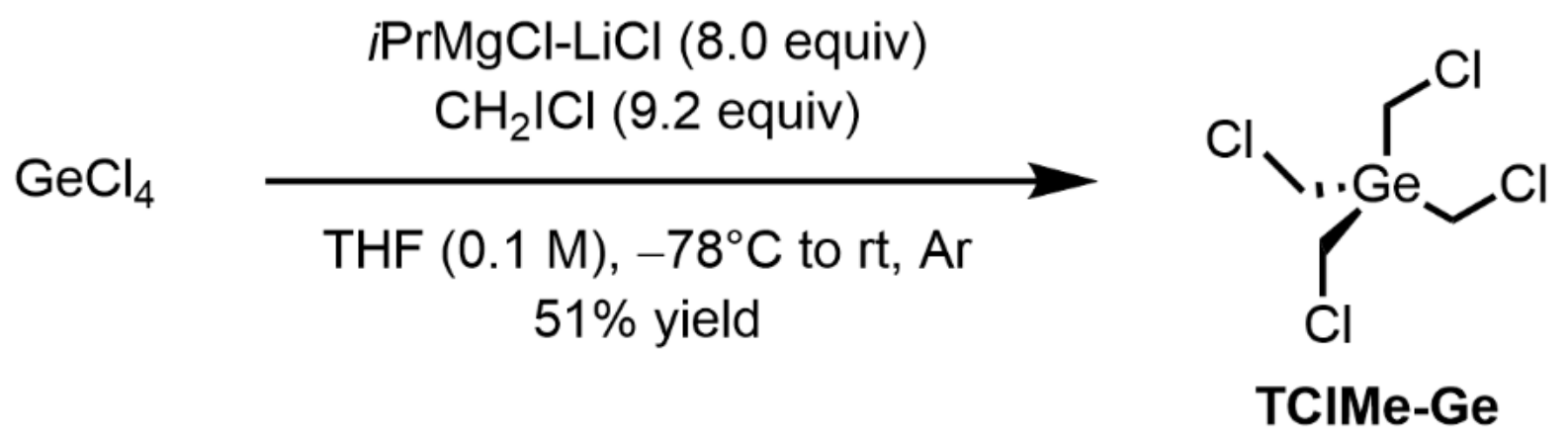

**Tetrakis(chloromethyl)germane (TClMe-Ge):**

Chloroiodomethane (5.46 mL (13.3 g), 75.0 mmol, 9.2 equiv) dissolved in tetrahydrofuran (THF; 30 mL) was added to a flame-dried 200 mL Schlenk flask that was equipped with a stir bar, topped with a septum, and cycled with vacuum and argon (5X), followed by $GeCl_4$ (945 µL (1.74 g), 8.1 mmol, 1.0 equiv). The solution was cooled to −78°C in a dry ice/acetone bath and stirred for 15 min. Turbo Grignard (*i*PrMgCl-LiCl 1.3 M in THF; 50.0 mL, 65.0 mmol, 8.0 equiv; 0.1 M THF final concentration) was added dropwise along the wall of the flask over 15 min at −78°C, where upon completion of the addition, the reaction mixture was allowed to stir at −78°C for 15 min. The cooling bath was removed, and the reaction was allowed to warm to room temperature. During this time (10–20 min), the reaction changed from a milky white mixture to a transparent solution. When the solution became transparent, the reaction was quenched with saturated aqueous $NH_4Cl$ (5 mL) and water (5 mL), where a white precipitate was formed. Excess water (50 mL) was added and eventually the white precipitate became soluble. The solution was transferred to a separatory funnel containing water and diethyl ether, where the aqueous phase was extracted with diethyl ether (3X). The organic phases were combined, dried over magnesium sulfate, and concentrated *in vacuo*. 2.44 g of crude oil was obtained, where 10 mL of pentane was added (solution did not precipitate at room temperature, even after sonication), and the solution was chilled at −20°C for over the weekend (o/w). A white precipitate was formed, and the pentane was removed using a syringe. To the crystals were added another 5 mL of pentane and the solution was chilled at −20°C for 2 h. The pentane was removed by syringe and residual solvent was removed *in vacuo*. Tetrakis(chloromethyl)germane (**TClMe-Ge**) was isolated as an amorphous white solid (1.12 g, 4.16 mmol, 51% yield).

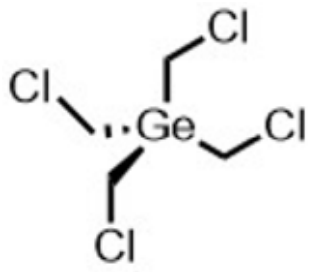

**$Ge(CH_2Cl)_4$**

**IR** (neat, $cm^{-1}$): 3012 (w), 2954 (w), 1384 (m), 1099 (m), 1021 (m), 1001 (m), 758 (s), 726 (vs), 703 (vs).

**HR-MS** (EI): *m/z* calc'd for $C_3H_6GeI_3$ $[M-CH_2Cl]^+$: 218.8756; found: 218.8751.

**$^1$H NMR** (600 MHz, $CDCl_3$): δ = 3.36 (s, 8H, 4 X $CH_2$, Ge−$CH_2$−Cl) ppm.

**$^{13}$C NMR** (151 MHz, $CDCl_3$): δ = 24.5 (4 X $CH_2$, Ge−$CH_2$−Cl) ppm.

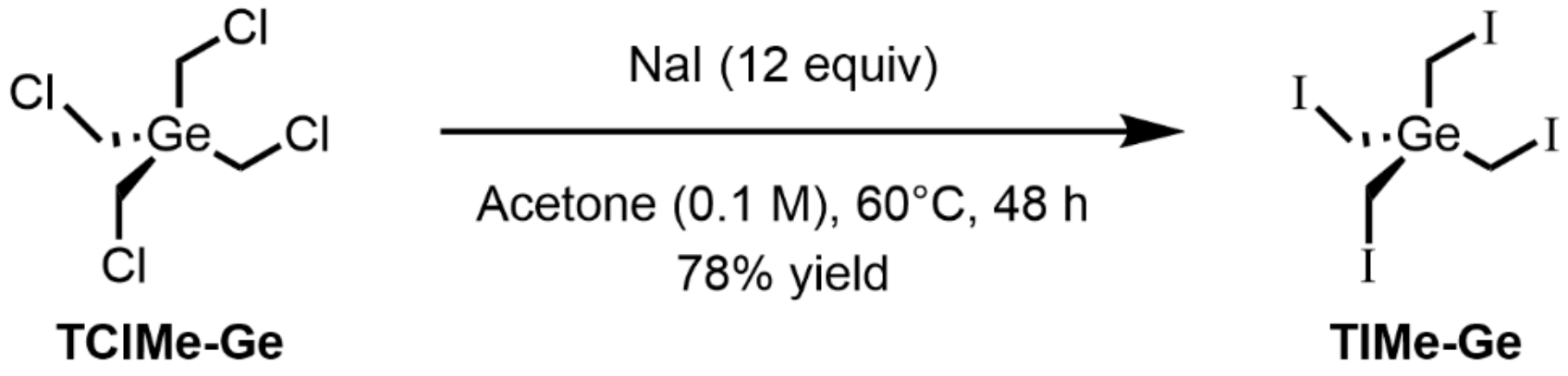

**Tetrakis(iodomethyl)germane (TIMe-Ge):**

Tetrakis(chloromethyl)germane (**TClMe-Ge**; 254 mg, 0.94 mmol, 1.0 equiv), acetone (9.4 mL, 0.1 M), and sodium iodide (1.69 g, 11.3 mmol, 12 equiv) were added to a vial equipped with a stir bar. The vial was capped, heated to 60°C, and stirred for 48 h. Upon completion and cooling, silica was added, and the mixture was concentrated, dry loaded onto the top of a silica plug, eluted with THF (250 mL), and concentrated *in vacuo*. The material retained yellow impurities (likely iodine) and was transferred to a separatory funnel containing organic (diethyl ether:tetrahydrofuran 10:1; 100 mL) and aqueous (saturated sodium thiosulfate; 50 mL) phases. The organic phase was extracted with saturated sodium thiosulfate (2X), where the combined aqueous phases were extracted with diethyl ether (50 mL). Brine (saturated aqueous sodium chloride) was not used as an aqueous phase to exclude deleterious chloride substitution of the desired product. The combined organic phases were dried over magnesium sulfate, filtered, and concentrated to a minimum of THF (~1 mL), *in vacuo*. The resulting light-yellow solution was added pentane (10 mL), triturated, and chilled to −20°C for 2 h. The yellow organic solvent was removed using a syringe, followed by addition of pentane (3 mL), where the solution was triturated and chilled to −20°C for 1 h. The colourless/transparent organic solvent was removed using a syringe and the residuals were removed *in vacuo*, leaving the title compound as a white, amorphous solid (467 mg, 0.73 mmol, 78% yield).

**Note**: Attempts to recrystallize the product in a variety of solvents and mixtures (hexane, pentane, ethyl acetate, acetone, diethyl ether, tetrahydrofuran, dichloromethane, and chloroform) did not produce useful conditions due to lack of solubility. Additionally, attempts to purify the compound using column chromatography (mixtures of hexane with ethyl acetate, diethyl ether, tetrahydrofuran, or dichloromethane) did not allow typical elution of the desired product (the compound precipitated in the column). In our hands, only THF was able to elute the compound from the column and this process did not remove all the impurities.

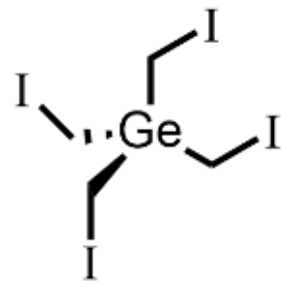

**Ge(CH2I)4**

**IR** (neat, cm$^{-1}$): 3000 (w), 1361 (w), 1066 (m), 1021 (m), 1001 (m), 713 (s), 659 (s).

**HR-MS** (EI): *m/z* calc'd for $C_4H_8GeI_3$ $[M-I]^+$: 506.7003; found: 506.7029.

**$^1$H NMR** (600 MHz, $CDCl_3$): $\delta = 2.48$ (s, 8H, 4 X $CH_2$, Ge−$CH_2$−I) ppm.

**$^{13}$C NMR** (151 MHz, $CDCl_3$): $\delta = -21.7$ (4 X $CH_2$, Ge−$CH_2$−I) ppm.

## NMR Spectra

Tetrakis(chloromethyl)germane (**TClMe-Ge**) $^{1}$**H NMR** (600 MHz, $CDCl_3$)

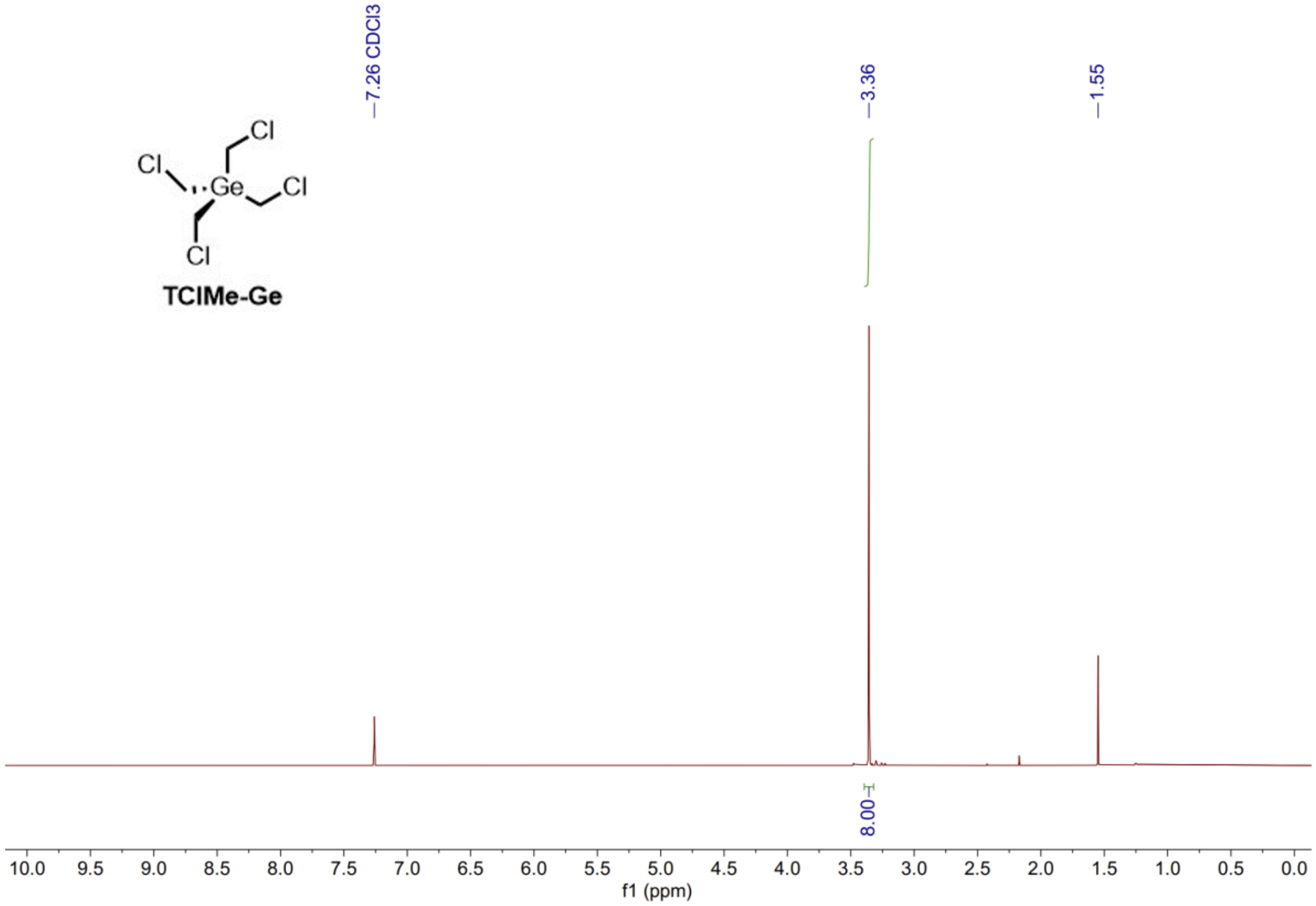

$^{13}$**C NMR** (151 MHz, $CDCl_3$)

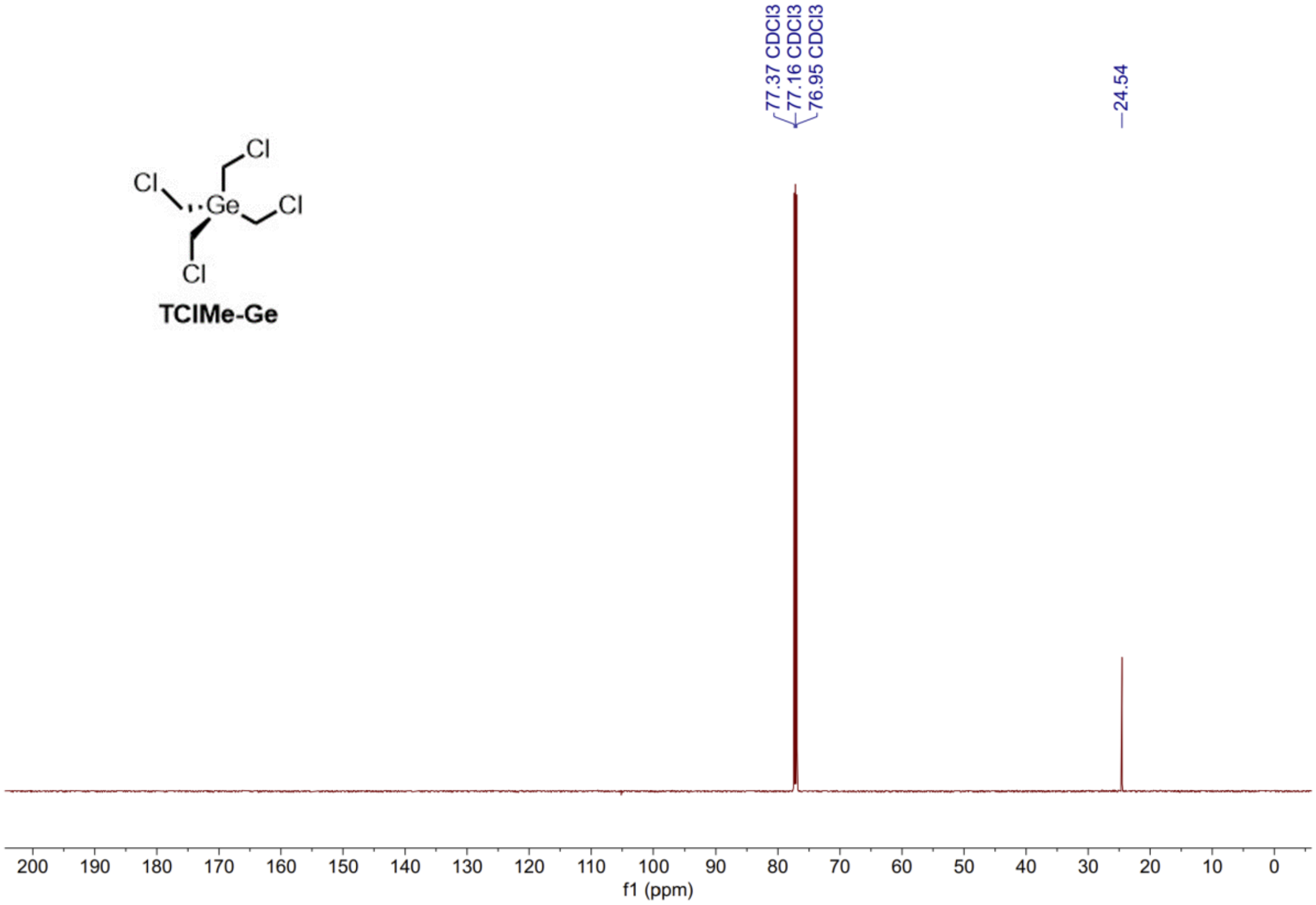

Tetrakis(iodomethyl)germane (**TIMe-Ge**) $^{1}$**H NMR** (600 MHz, $CDCl_3$)

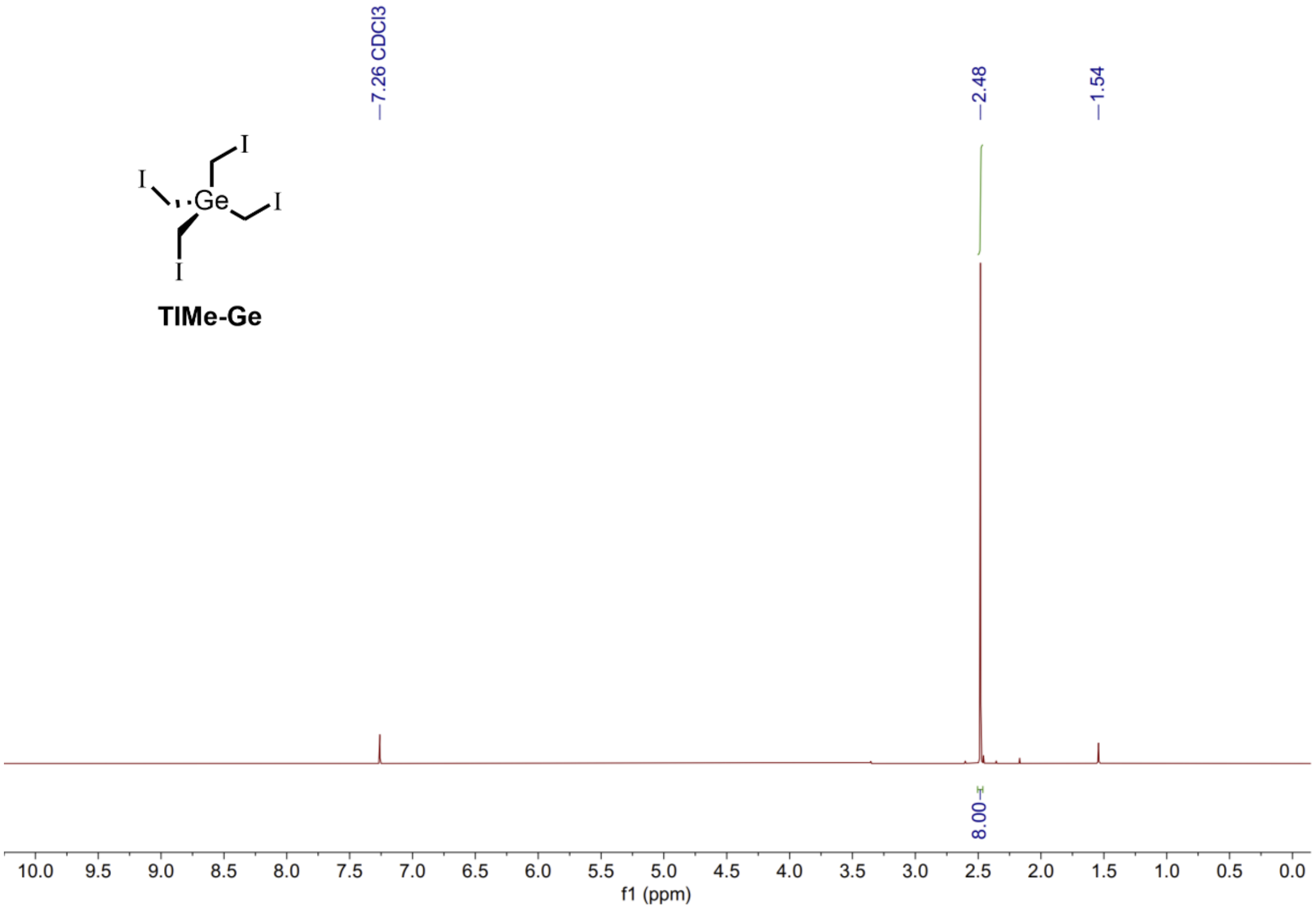

$^{13}$**C NMR** (151 MHz, $CDCl_3$)

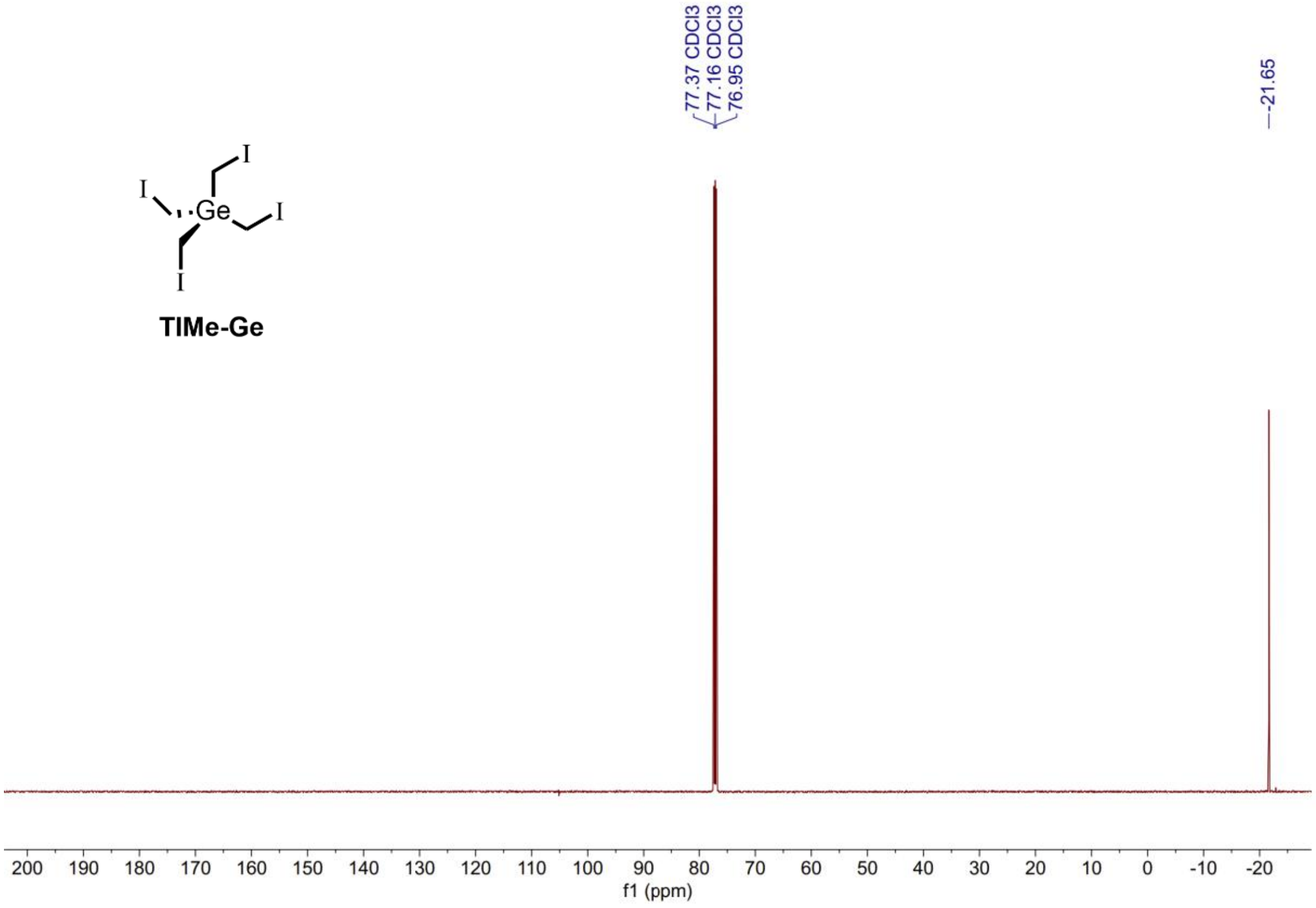

# Density Functional Theory Details

Rotational potentials for the pendent $CH_2I$ were produced from optimized geometries of a further reduced Si(100) proxy (one dimer row, two dimers ("2×2")) with the ωB97XD[1] hybrid density functional and 6-311G(d,p) basis set with Gaussian16[2] using program-option "tight" SCF convergence criteria ($DE < 10^{-8}$ a.u.) and ultrafine grid size (99 radial shells with 590 angular points per shell). The 6-311G(d,p) basis set for iodine was taken from the Basis Set Exchange.[3]

All geometries (in .xyz format) and representative images of all Si(100) proxies used in this study are provided. Ball-and-stick representations were generated in Jmol 14.32.3[4] and simulated STM images were processed in Gwyddion 2.60.[5]

| **Proxy** | **XY Plane** | **XZ Plane** |
|---|---|---|
| "3×4" | 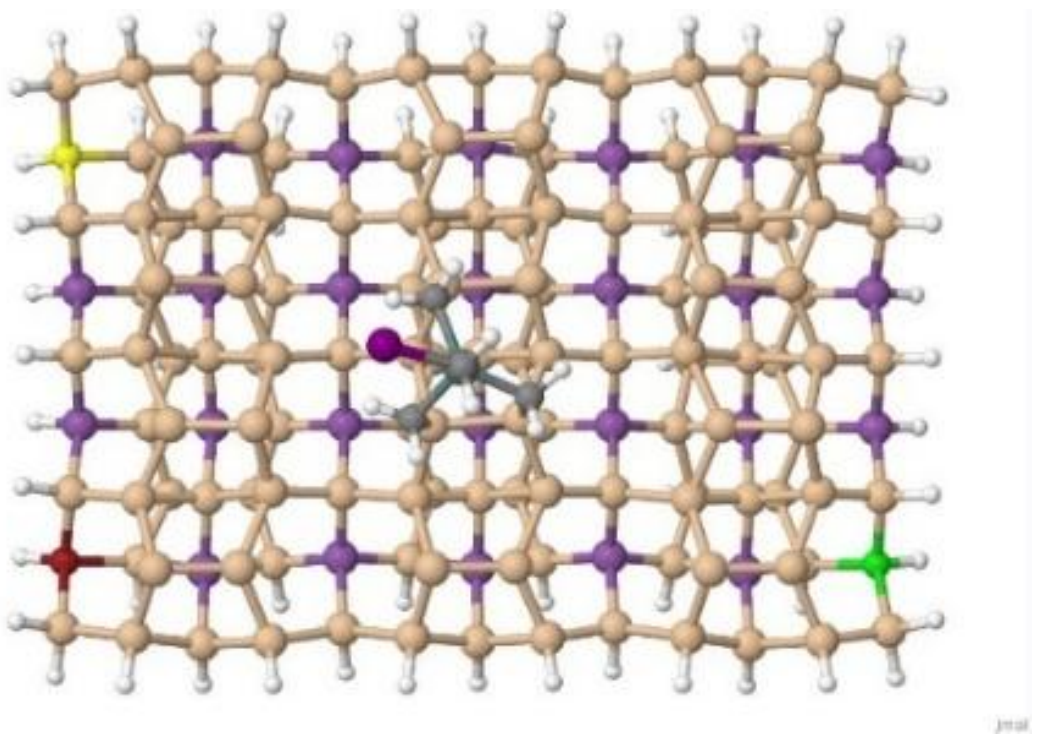 | 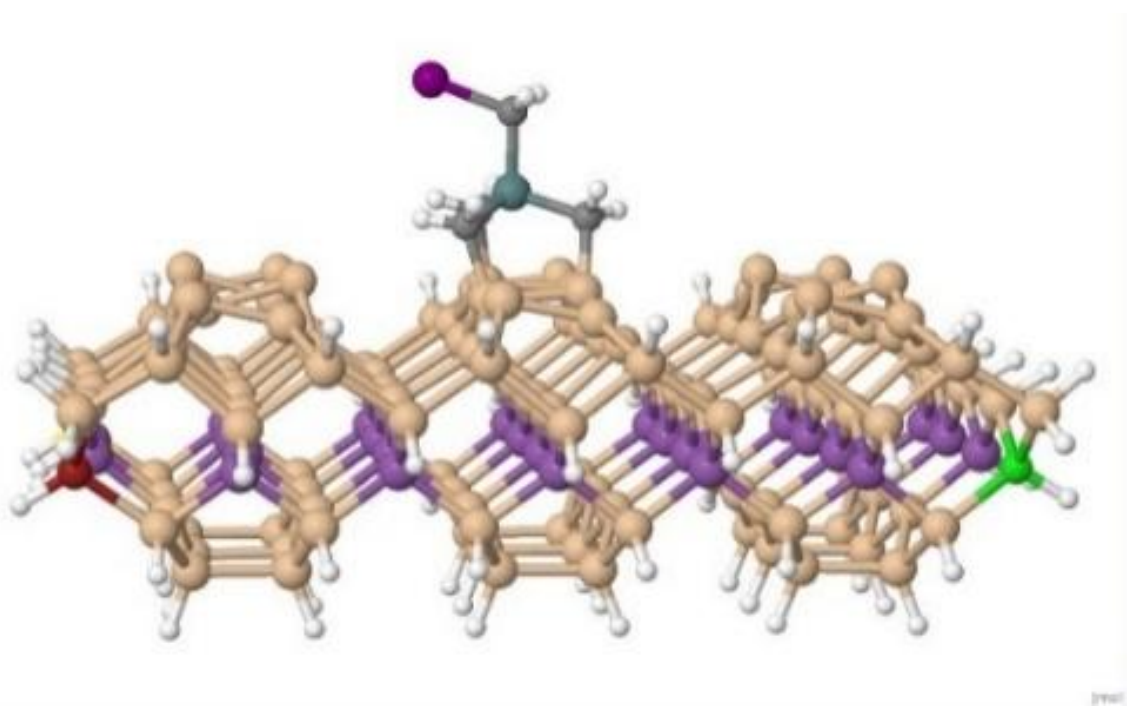 |

**SI Table 1: Atomic anchoring and constraint arrangement for the molecular proxies used in the DFT calculations.** The 3×4 proxy used to calculate the on-dimer (OD) TlMe-Ge geometries is shown with a typical OD TIMe-Ge. For consistency across all geometry optimizations and STM simulations, select atoms were constrained in their optimization. Red = anchored to the origin (X,Y,Z = 0,0,0), green = free motion only along the x-axis (X,Y,Z = N,0,0), yellow = free motion

only along the y-axis (X,Y,Z = 0,N,0). Purple = free motion only within the XY plane (X,Y,Z = N,N,0).

## DFT N-Legged Configurations

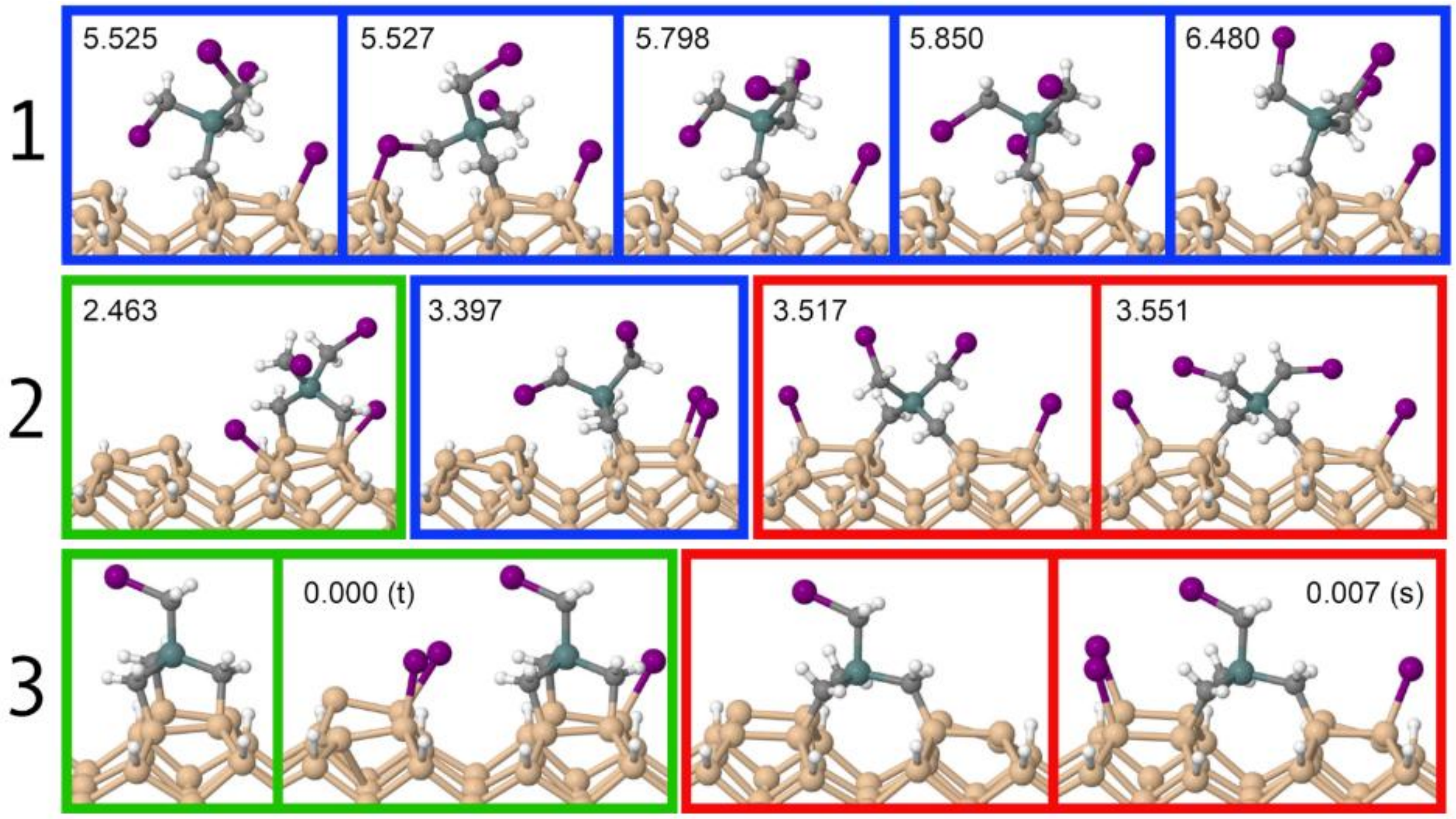

**SI Figure 1: N-Legged (N = 1, 2, 3) configurations and stabilization energies added by additional Si-C bond formation.** A selection of one-, two-, and three-legged surface-bound configurations of TlMe-Ge. For the one- and two-legged configurations, relative energies are given in the top corners (in eV) as referenced to the most stable three-legged configuration (OD, ground state triplet, Row 3, left) using the geometry optimization approach presented in the main text in *Methods*. (s) = ground-state singlet optimization, (t) = ground state unrestricted Kohn-Sham triplet optimization.

In **SI Figure 1**, TIMe-Ge configurations with different numbers of surface bound legs are examined in DFT. Structures with blue borders can adopt either OD or inter-row (IR) three-legged products. The green-bordered two-legged structure in “2” is one example configuration that can only produce an OD product (with the first two legs inhibiting IR binding). The red-bordered two-legged structures in “2” are geometries that can only adopt IR products. The blue-bordered structure in “2” can adopt either configuration type provided at least one iodine migration step occurs to enable an OD geometry. The sampling here is only a subset of configurations used for STM simulations to consider and exclude tri-lobal geometries that might have one- or two-legged origins and to aid in experimental image interpretation.

## DFT Inter-Row Three-Leg Configurations

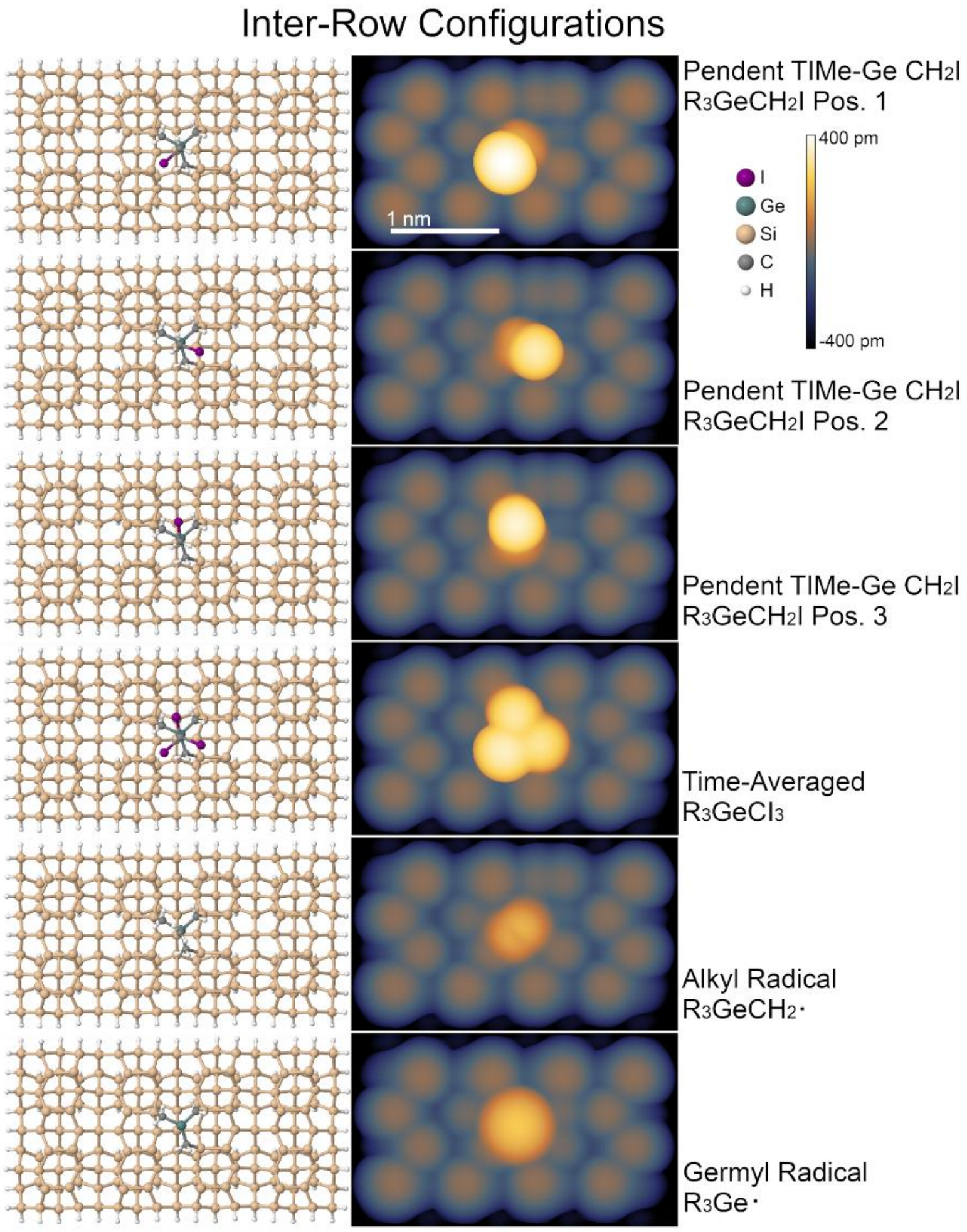

**SI Figure 2: DFT predictions of chemisorbed TIMe-Ge inter-row surface configurations.** A proxy containing four dimer rows and four dimers per row (“4×4”) was used to have sufficient coverage of dimers around the TIMe-Ge molecules. All structures were generated by the geometry optimization and STM simulation approach described for **Figure 1**. The time-averaging of the deiodinated ($CH_2$) product was not considered but is predicted to have a very low rotational barrier (0.024 eV at the ωB97DX/6-311G(d,p) level of theory). Buckled silicon dimer averaging was not considered because it was not expected to impact the appearance of the TIMe-Ge features.

## Attempt Rates of the CH2I Rotation at 4 K and 77 K

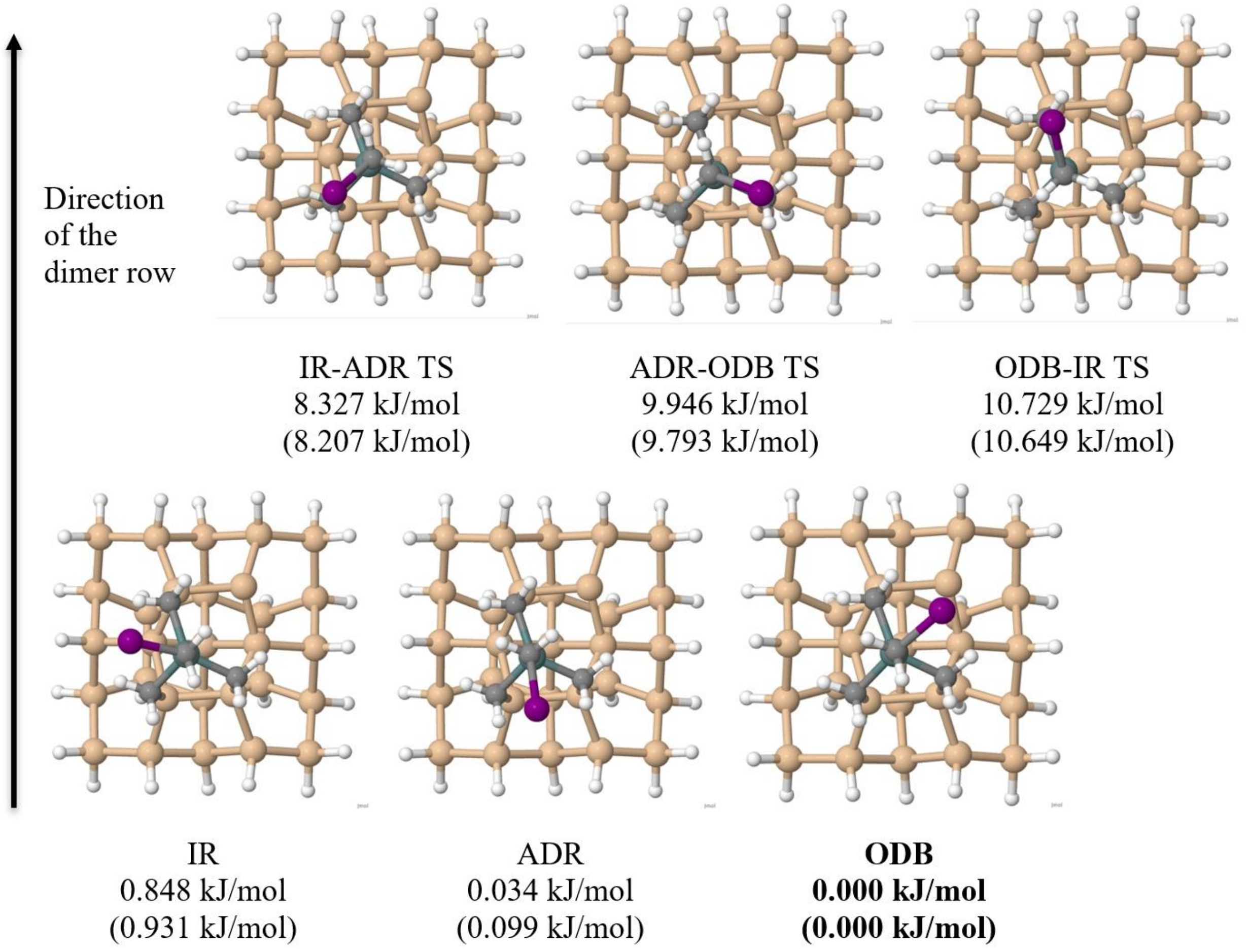

**SI Figure 3**. The IR, along-dimer row (ADR), and over-dangling bond (ODB) $CH_2I$ rotational minima and their associated transition states for the OD configuration. Relative electronic energies are provided at the ωB97XD/6-311G(d,p) and, in parentheses, ωB97XD/def2-TZVPD//ωB97XD/6-311G(d,p) levels of theory.

The IR, ADR, and ODB $CH_2I$ positions (labeling with respect to the position of the iodine atom on the small silicon surface proxies and not the molecular placement, see **SI Figure 3**) in the OD configuration are made asymmetric by placing three C−Si legs on three of four silicon atoms within a dimer pair. The relative minima and transition state (TS) energies for the $CH_2I$ rotations are all made unique by this asymmetry (as summarized in **SI Figure 3**) and relative to the lowest-energy configuration (ODB).

Geometry optimizations and normal mode analyses were performed at the ωB97XD/6-311G(d,p) level of theory as unrestricted Kohn-Sham doublets for all structures. For comparison, the relative electronic energies of these six configurations are provided in parentheses at the ωB97XD/def2-TZVPD//ωB97XD/6-311G(d,p) level of theory[6–8] in **SI Figure 3** to show the very reasonable predicted energies at the ωB97XD/6-311G(d,p) level.

To err on the side of the lowest barrier for estimating $CH_2I$ rotation rates at 4 K and 77 K, we consider the rotation of the pendent $CH_2I$ as a symmetric rotor despite there being three distinct minima and transition states for the full rotation. Estimates use the lowest energy TS: IR-to-ADR (IR-ADR TS, 8.327 kJ/mol based on the ODB global minimum and 7.478 kJ/mol based on the relative difference between the IR and IR-ADR TS energies). Both the Eyring TS Theory and Classical Hindered Rotor models were used in this analysis as another possible mitigation against the small surface proxy and theory level being insufficient to model the behavior of the entire molecule bound to the surface.

The attempt rate, $k_{Eyring}$, is calculated from the Eyring equation using the Gibbs free energy of activation ($\Delta G^{\ddagger}$, in J) for the rotational barrier as calculated for the IR-ADR transition state from normal mode analyses at the ωB97XD/6-311G(d,p) level of theory:

$$k_{\text{Eyring}} = \frac{k_B T}{h} \exp\left(-\frac{\Delta G^{\ddagger}}{k_B T}\right)$$

$\Delta G^{\ddagger}$ = 8069.4 J at 4 K and 7281.8 J at 77 K

$T$ = temperature (4 K, 77 K)

$k_B = 1.380649 \times 10^{-23}$ J·K$^{-1}$ (Boltzmann constant)

$h = 6.62607015 \times 10^{-34}$ J·s (Planck constant)

The Classical Hindered Rotor attempt rate, $k_{HR}$, for the same configuration is calculated from the equation:

$$k_{HR} = \frac{k_B T}{h} Q_{\text{HR}} \exp\left(-\frac{E_a}{k_B T}\right)$$

$E_a$ = the relative IR-ADR TS rotational barrier height (7478 J) based on the IR and IR-ADR TS energies (and not relative to the ODB global minimum)

The classical partition function at temperature T, $Q_{HR}(T)$, was computed numerically for the one-dimensional torsional potential as calculated from the equation:

$$Q_{\text{HR}}(\text{T}) = \frac{1}{\sigma h}\sqrt{2\pi \text{I} k_B T}\int_0^{2\pi} e^{-\frac{V(\phi)}{k_B T}}\, d\phi$$

$\sigma$ = symmetry number for the rotamer (1)

$I$ = moment of inertia (kg·m²) calculated from $CH_2I$ (along the C−Ge bond) in the ADR configuration

$V(\phi)$ = torsional potential for the $CH_2I$ group, treated symmetrically with the smallest barrier to rotation, given for the nondegenerate ($\sigma = 1$) rotor as:

$$V(\phi) = \frac{V_0}{2}[1 - \cos(\phi)]$$

The summary of attempt rates from both models is provided below:

| **Temperature (K)** | **ΔG‡ (J)** | **$k_{Eyring}$ ($s^{-1}$)** | **$k_{HR}$ ($s^{-1}$)** |
|---|---|---|---|
| 4 | 7953.4 | $5.42 \times 10^{-88}$ | $1.17 \times 10^{-93}$ |
| 77 | 8743.9 | $7.85 \times 10^{7}$ | $1.88 \times 10^{6}$ |
| 298 | 14202.3 | $7.50 \times 10^{12}$ | $2.01 \times 10^{10}$ |

The same attempt rate calculations are performed for the largest rotational barrier (ODB-to-ODB TS, $E_a$ = 10728.8 J) to show that the same behavior is expected as a function of temperature in this range. These results are provided below:

| **Temperature (K)** | **ΔG‡ (J)** | **$k_{Eyring}$ ($s^{-1}$)** | **$k_{HR}$ ($s^{-1}$)** |
|---|---|---|---|
| 4 | 10839.1 | $2.40 \times 10^{-131}$ | $1.65 \times 10^{-130}$ |
| 77 | 11660.9 | $1.98 \times 10^{4}$ | $4.06 \times 10^{5}$ |
| 298 | 16901.4 | $6.77 \times 10^{9}$ | $1.62 \times 10^{12}$ |

# Additional Experiments and Molecular Fragmentation Pathways

## Inter-Row STM Presentation

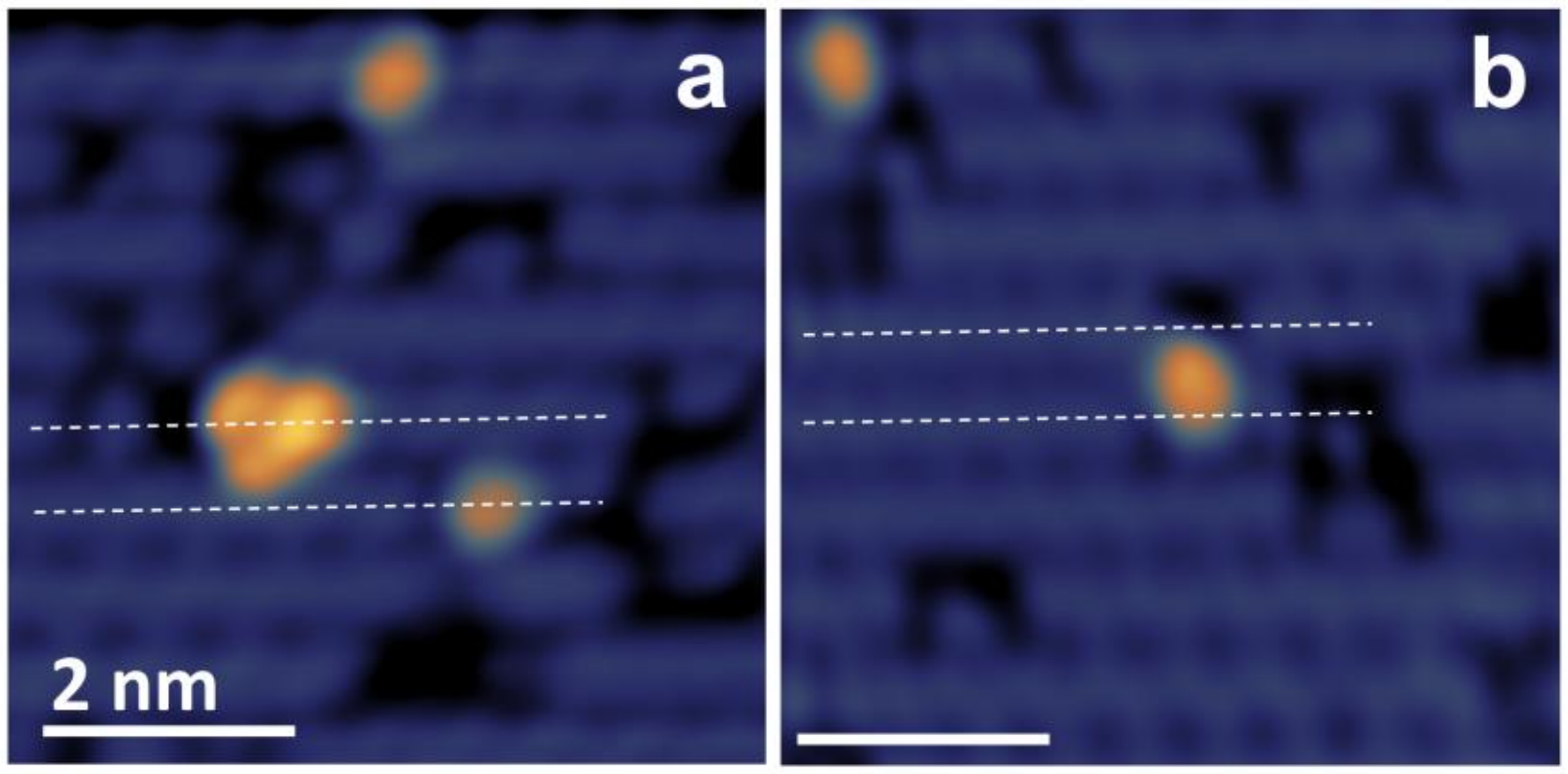

**SI Figure 4. A comparison of on-dimer and inter-row features after deposition at > 400 K.** (**a**) A region of the sample highlighting on-dimer features, including a tri-lobe and a mono-lobe. (**b**) A different region of the same sample with an inter-row mono-lobe. STM images were acquired at ($V$ = –2.0 V, $I$ = 50 pA). Dashed white horizontal lines highlight the centers of two adjacent dimer rows (running horizontal across the frame).

As discussed for main text **Figure 2d**, TIMe-Ge molecules deposited on hot substrates (> 400 K) had only 6% of molecules (±4%, $N$ = 140) observed as tri-lobes, with on-dimer mono-lobe structures dominating. Similarly, a surviving tri-lobe and two mono-lobe structures centered on dimer were observed in **SI Figure 4a** (dimer rows marked by white dashed lines). However, for hot depositions, some mono-lobes were observed inter-dimer at a rate of 15% (±6%, $N$ = 140) (**SI Figure 4b**). As temperature seemed correlated with a higher incidence of inter-row features, it is possible that higher surface temperature allowed TIMe-Ge to access inter-row three-leg-down surface configurations (**SI Figure 2**) that were then also deiodinated or demethylated from the elevated temperature. Further work would be required to verify.

### 4K STM Presentation of TIMe-Ge

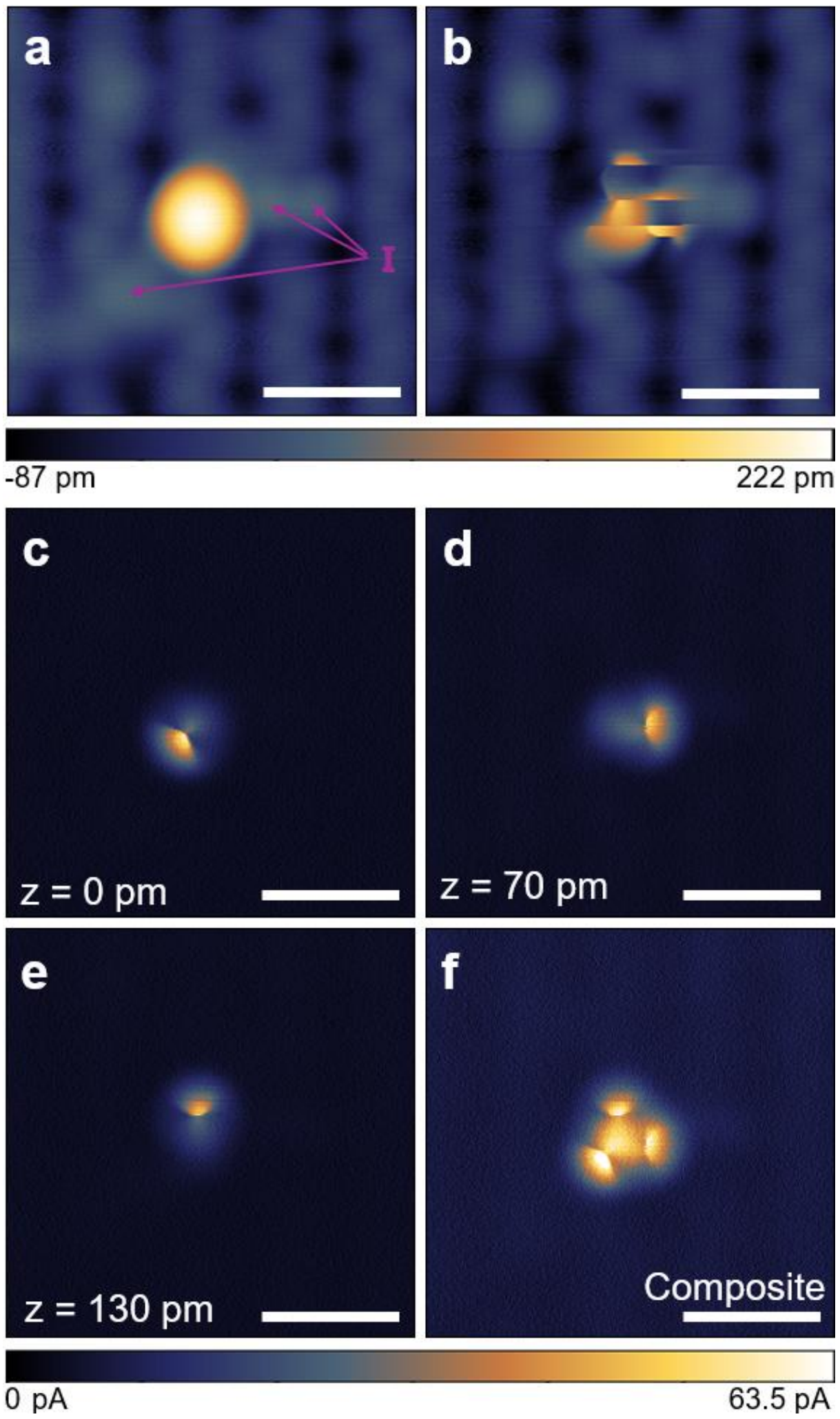

**SI Figure 5. TIMe-Ge molecules on Si(100): 4 K STM scanning. (a)** Constant-current STM image of a TIMe-Ge molecule at 4 K ($V$ = –2.0 V, $I$ = 30 pA). Purple arrows highlight surface iodine atoms from the dissociative leg attachment to the silicon. **(b)** "Streaky" constant-current STM image of the same molecule ($V$ = –2.0 V, $I$ = 30 pA). **(c-e)** Constant-height STM images of different $CH_2I$ rotamers of the same molecule. **(f)** Composite image of the three rotamers showing a tri-lobe appearance. Constant-height images were acquired at ($V$ = –2.0 V, $z_{ref}$ = 0 pm), with $z_{ref}$ defined by an STM setpoint of ($V$ = –2.0 V, $I$ = 20 pA) over the center of a TIMe-Ge molecule. Some molecule rotations had to be taken at larger tip-sample separations to prevented unwanted tip interaction. Additional offsets from $z_{ref}$ are presented in the lower left of the constant-height panels. All scale bars are 1 nm. Constant-current images have $z$ = 0 referenced to the background silicon-dimer-row tops, and constant-height images the 0 current background.

TIMe-Ge was also characterized by 4 K STM imaging. At 4 K, no tri-lobes were observed (**SI Figure 5a**; compared to main text **Figure 1f**), supporting the $CH_2I$ being "frozen" into a single rotamer. However, some uncommon tip compositions showed discontinuities over the molecule (**SI Figure 5b**), suggesting tip interactions could rotate the $CH_2I$. These interactions were utilized to induce a series of rotations amongst the three rotamer positions, with each characterized by non-perturbative constant-height STM imaging (**SI Figure 5c**-**e**). A composite image of the three constant height frames is provided in **SI Figure 5f**, reproducing the molecule's 77 K tri-lobe presentation.

## TIMe-Ge Height Change with UV Exposure and Tip-Removed Bias Pulsing

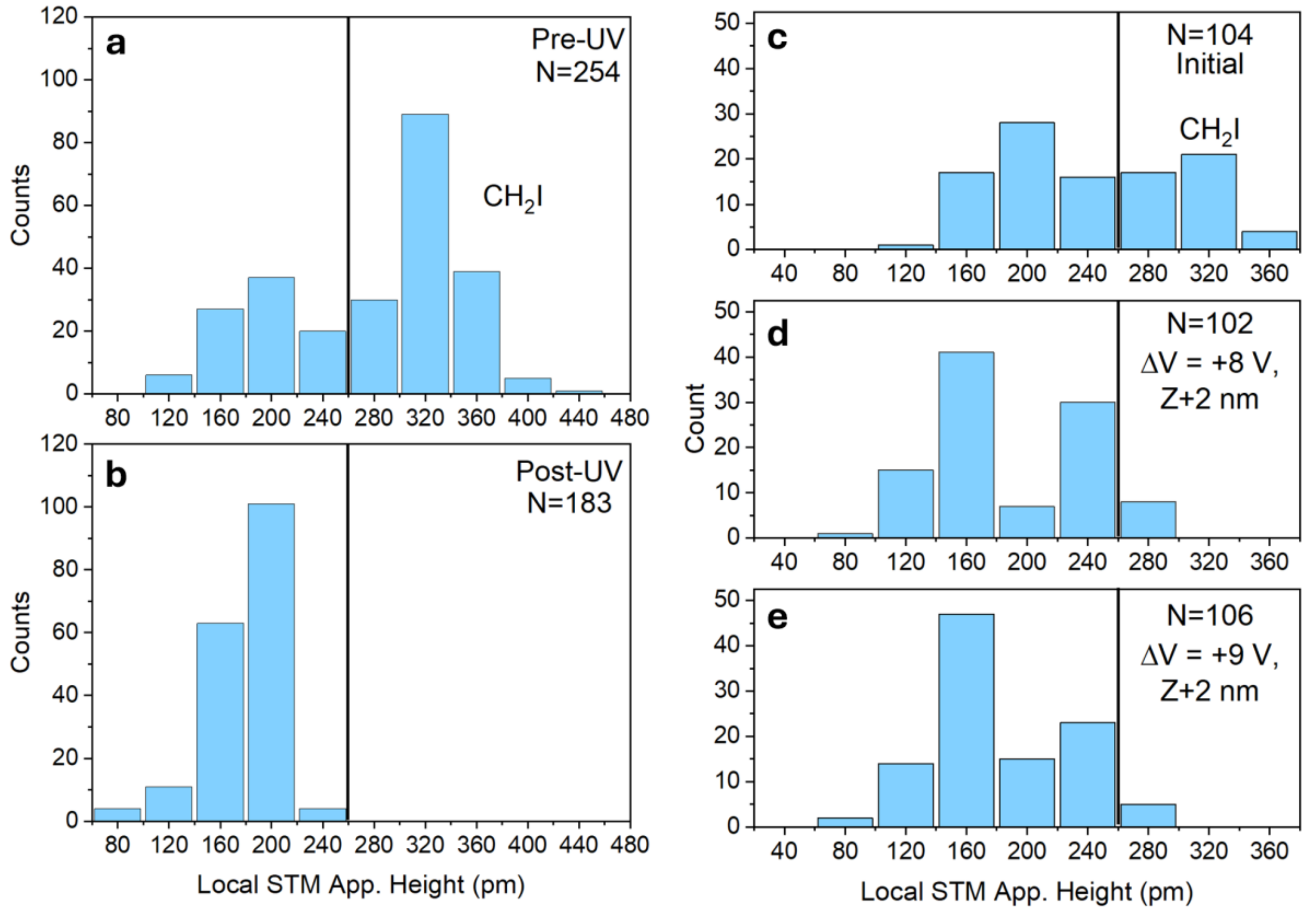

**SI Figure 6**. **STM apparent height *c*hanges with UV exposure and retracted-tip bias pulses.** (**a**,**b**) Comparison of measured STM heights of surface molecules before and after UV, respectively, and (**c**-**e**) positive bias pulses. To guide interpretation, an apparent height threshold of 260 pm is noted with a black vertical line, above which all molecules were observed to have tri-lobe character at 77 K in STM scanning. All STM images used to populate the histograms were taken with ($V$ = –2 V, $I$ = 50 pA). All histograms are an aggregation of many STM frames.

To explore the existence of both alkyl and germyl radicals, **SI Figure 6** presents the STM apparent heights of the molecular products after both UV and positive bias-pulse deiodination treatments. Gwyddion[5] was used to aggregate these data from dozens of STM images in threshold mode, with a set value of 80 pm and no other initial parameters. After automated grain selection, a manual annealing process of shrinking grains radially by 1–3 pixels then growing grains by the same amount was used to minimize the influence of extraneous grains. For each remaining grain, the maximum apparent height pixel with that grain was extracted for inclusion. Due to the relatively low number of counts, no peak fits were deemed appropriate in the generated histograms.

Histograms were plotted in OriginPro 2023 with manual bin widths of 40 pm (as approximated from applying the Freedman-Diaconis rule[9] to the available experimental data), starting at 20 pm to allow for the 260 pm threshold discussed below to be consistently indicated to the reader without confounding overlaps. We also note that minor tip changes occurred between the STM scans used to populate the histograms, which would add spread in the measured apparent heights. No mono-lobes observed in the collated datasets exhibited an apparent height greater than 260 pm across all tip states, motivating the cutoff marked by the black vertical line in all plots. Some tri-lobes were observed to have a maximum apparent height of 220–260 pm, leading to some mixing of the tallest mono-lobes and the shortest tri-lobes at that bin. Examining height changes induced by UV irradiation, all molecules above the cutoff in **SI Figure 6a** shift to lower heights in **SI Figure 6b**, with most post-UV products in the 160–200 pm bins.

A similar reduction of taller tri-lobe molecules was observed in the tip-removed positive bias pulse experiments (**SI Figure 6c-e**). Almost all molecules (within ~50 nm of the pulse site) above the cutoff reduced to molecules between 120–240 pm upon application of a +6 V absolute bias ($\Delta V$= +8 V) pulse (**SI Figure 6d)**. A second, stronger bias pulse of +7 V ($\Delta V$= +9 V) was then applied (**SI Figure 6e)**. Comparing **SI Figures 6d** and **6e**, the 240 pm bin redistributes some of its counts to the 200 and 160 pm bins. This redistribution could indicate alkyl radicals reducing further to germyl radicals, but is counter to the experimentally extracted height of a suspected germyl radical in **SI Figures 15** and **16**, which showed it as taller relative to the intact and alkyl radical forms of TIMe-Ge. We note the observations of the germyl being taller in **SI Figures 15** and **16** are counter to the DFT predictions of main text **Figure 1**, possibly due to the limitation of the DFT simulations or a minor STM tip change (as discussed for **SI Figure 16).** We do note that the spread in the data of **SI Figure 6e** is broader than the UV data of **SI Figure 6a,b** which could be attributed to the aggregation of STM data where tip changes occurred, or bias pulsing allowing access to additional products or fragmentation pathways (**SI Figure 8**).

## Tip Field Strength Calculation

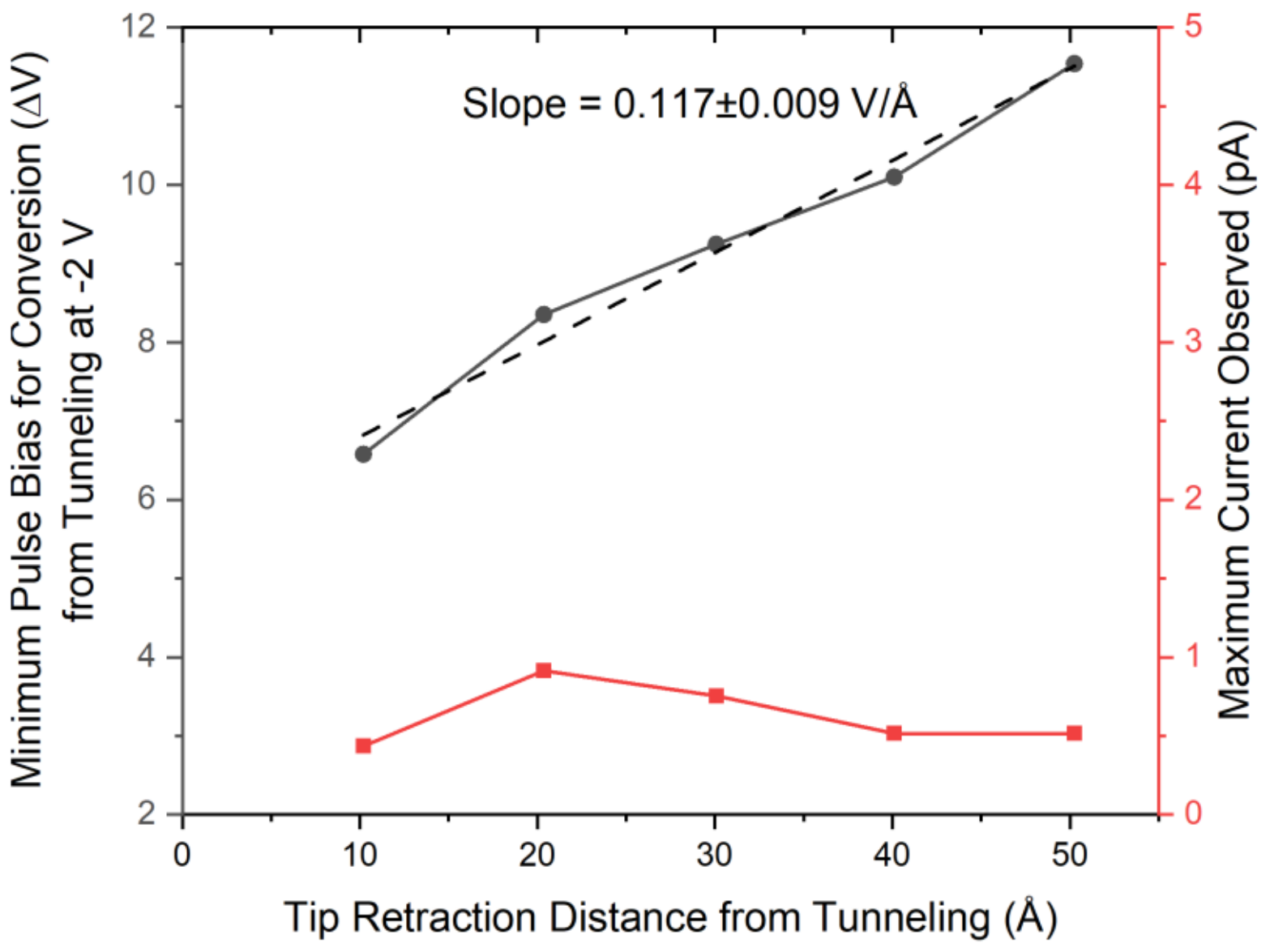

**SI Figure 7. Tip-removed electric field strength.** The minimum positive pulse bias required to alter surface TIMe-Ge molecules from tri-lobe to mono-lobe (black points) presentation for a given tip-sample height. Height is referenced to an STM setpoint defined by ($V$ = –2.0 V, $I$ = 50 pA). The simultaneously acquired peak observed current (including transient capacitive spikes) is in red, with its corresponding axis on the right. A linear fit to the V(z) data is plotted as the black dashed line.

STM imaging the same region between retracted-tip bias pulses was used to estimate the electric field strength and motivate that no tip-to-surface or tip-to-molecule charge injection occurred (**SI Figure 7**). For each height, the voltage was increased in 1 V increments until the post-pulse STM survey scan showed a conversion of molecules from tri-lobe to mono-lobe. The electric field strength was calculated from a linear fit to this extracted minimum threshold voltage for a given height: ~0.12 V/Å. We note here that the maximum current observed was largely capacitive in nature and could be reproduced even with the tip fully retracted. As such, the estimated tunneling current is well below the detection limit of the current pre-amplifier used in this work, suggesting a field-mediated mechanism and no charge transfer through the vacuum gap. However, the extracted field from the experimental data is below what would be expected to cleave the C–I bond, hence supporting the mechanism discussed in the main text of field-driven charge exchange with the Si surface.

## Alternate Molecular Fragmentation Pathways Under Bias Pulsing

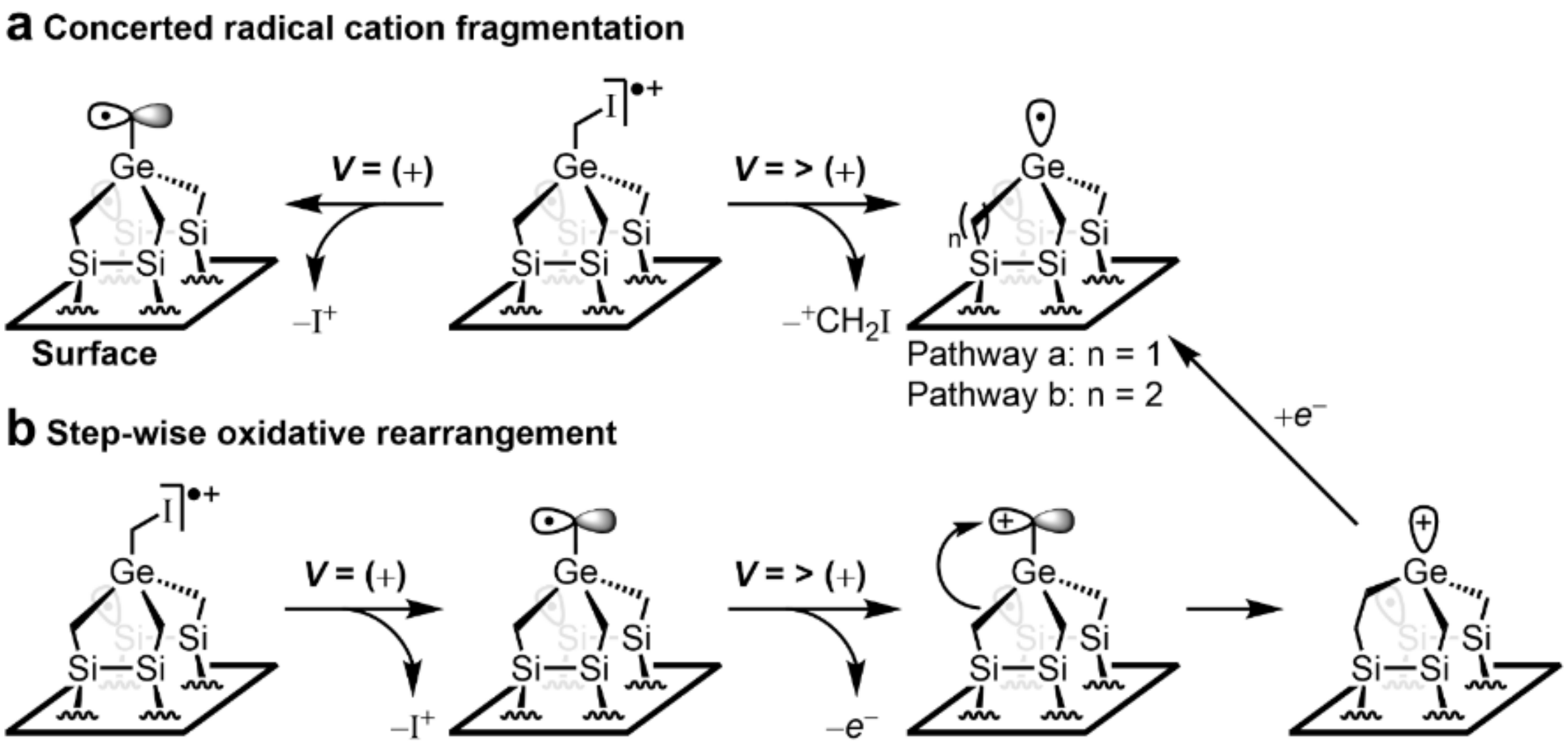

**SI Figure 8**. **Germyl radical formation under large positive potential**. Proposed mechanism for the formation of germyl radical intermediates via concerted (**a**) and stepwise (**b**) pathways under positive potential ($V = (+)$) and negligible tip-to-molecule tunneling conditions.

We suspected that germyl radicals could be produced either directly from intact TIMe-Ge or from the alkyl radical intermediate. The proposed mechanism for possible concerted and stepwise fragmentation pathways for TIMe-Ge under large positive bias voltages is illustrated in **SI Figure 8**. This fragmentation mechanism was prompted by preliminary experimental evidence that supported the formation of a germyl radical (**SI Figure 15)**, where a tip-removed (negligible current) positive bias sweep over an intact TIMe-Ge produced a taller and wider mono-lobe.

## Alternate Molecular Fragmentation Pathways Under Conditions with Measured Tunnel Current

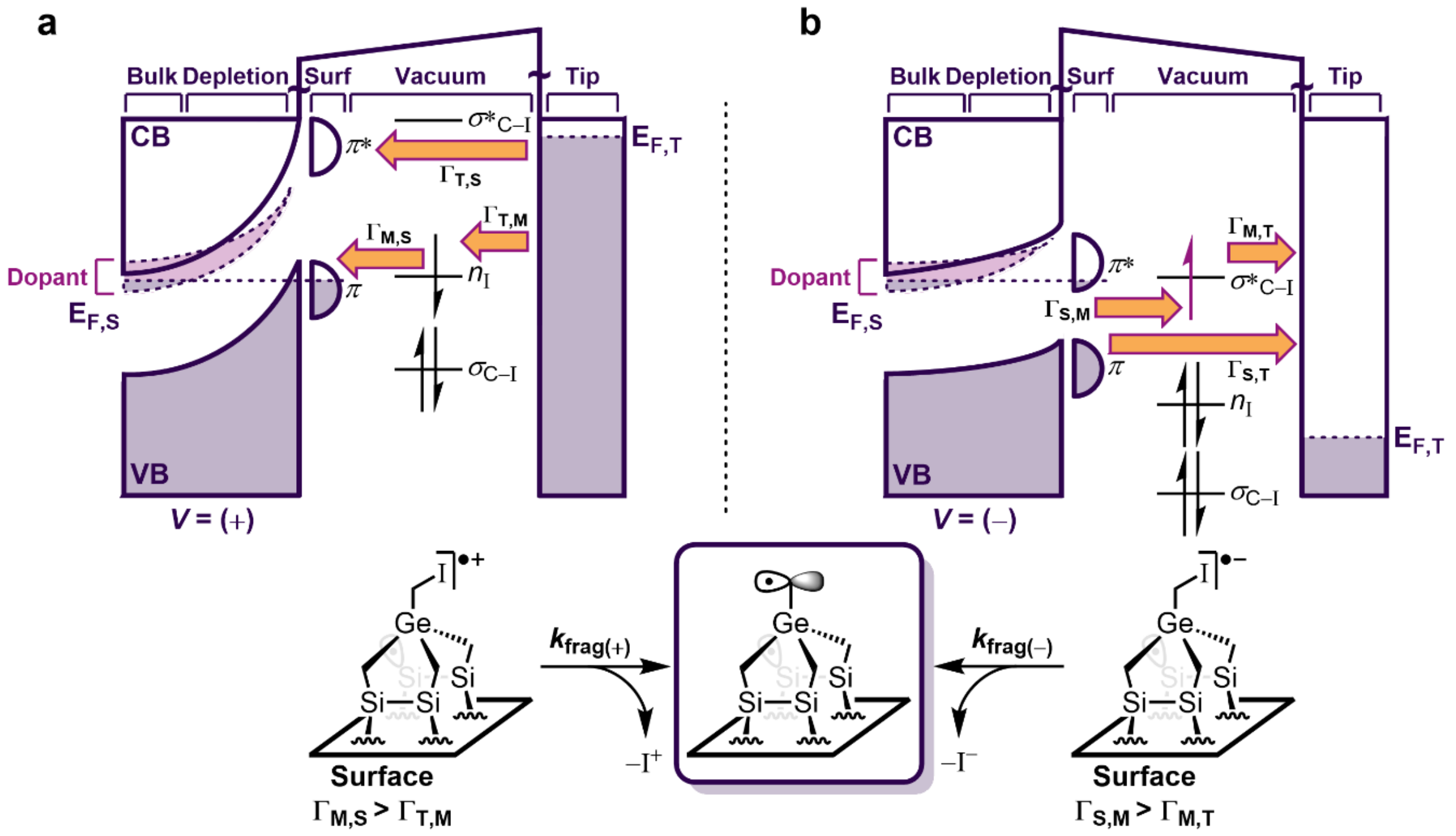

**SI Figure 9. Proposed in-tunneling alkyl generation scheme 1**. In-tunneling chemical reaction scheme outlining the dehalogenation mechanisms for **(a)** emptying the iodine non-bonding orbital ($n_I$) into the partially filled Si $\pi$ band at empty states STM bias (+V) and **(b)** populating the $\sigma^*$ (C–I; anti-bonding) orbital with an electron from the partially filled $\pi^*$ band at filled states STM bias (–V). Voltages are relative to the sample, with band diagrams highlighting conduction pathways between the surface-molecule-tip system: $\Gamma_{M,S}$ = molecule to surface, $\Gamma_{T,M}$ = tip to molecule, and $\Gamma_{T,S}$ = tip to surface. Molecular orbitals include the $\sigma$ (C–I; bonding), the $\sigma^*$ (C–I; anti-bonding), and the $n_I$ (iodine atom; non-bonding) energy levels. $E_{F,T}$ and $E_{F,S}$ denote the tip and sample Fermi levels, respectively. $\pi$ and $\pi^*$ are the delocalized Si(100)-(2×1) surface states. The dopant band for highly n-doped Si(100) is drawn as the dashed light purple region overlapping with the conduction band edge.

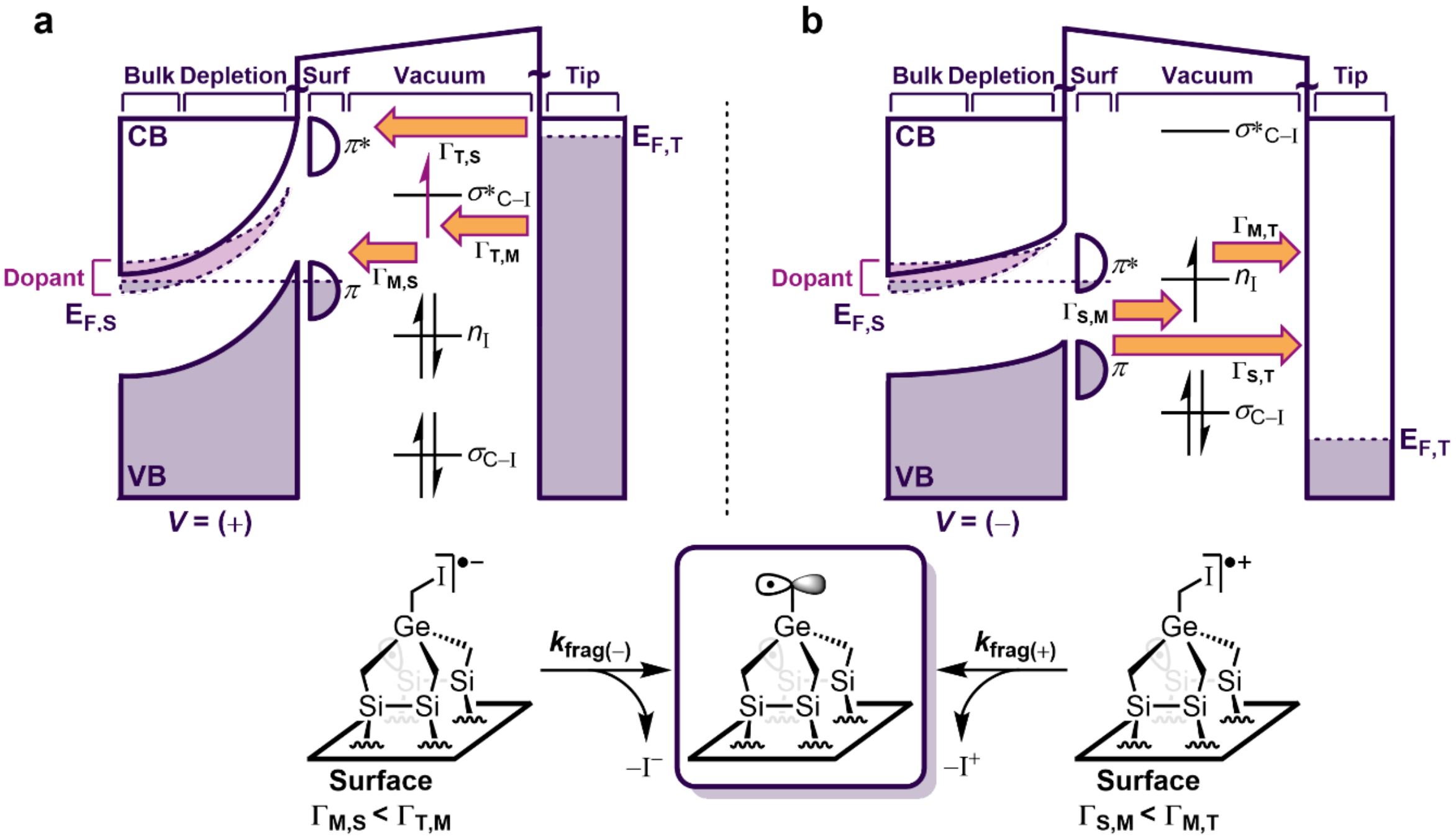

**SI Figure 10. Proposed in-tunneling alkyl generation scheme 2.** Chemical reaction scheme outlining the dehalogenation mechanisms for **(a)** filling the $\sigma^*$ (C–I; anti-bonding) orbital with an electron from the tip in empty states STM tunneling conditions (+V) and **(b)** emptying the iodine non-bonding orbital ($n_I$) into available empty states of the tip at filled states tunneling conditions (–V). Voltages are relative to the sample, with band diagrams highlighting conduction pathways between the surface-molecule-tip system: $\Gamma_{M,S}$ = molecule to surface, $\Gamma_{T,M}$ = tip to molecule, and $\Gamma_{T,S}$ = tip to surface. Molecular orbitals include the $\sigma$ (C–I; bonding), the $\sigma^*$ (C–I; anti-bonding), and $n_I$ (iodine atom; non-bonding) energy levels. $E_{F,T}$ and $E_{F,S}$ denote the tip and sample Fermi levels, respectively. $\pi$ and $\pi^*$ are the delocalized Si(100)-(2×1) surface states. The dopant band for highly n-doped Si(100) is drawn as the dashed light purple region overlapping with the conduction band edge.

Conversion of tri-lobes to mono-lobes was also experimentally observed at tip-sample distances where tunnel current was measured (**SI Figures 11**–**13**). Additionally, while no current was detected within our instrument limitations in tip-removed voltage sweeps or pulses over the molecule (main text **Figure 3** and **SI Figures 7**, **14**, and **15**), the possibility of transient electrons from tip-to-molecule or tip-to-sample cannot be entirely ruled out. Thus, **SI Figures 9** and **10** present two different hypothesized alkyl radical generation schemes with the inclusion of additional current pathways in the tip-molecule-sample system. Both schemes involve rate competition-limited charge transport between the iodine non-bonding orbital ($n_I$) and $\sigma^*$ (C–I; anti-bonding) orbital, with the relative rates of $\Gamma_{M,S}$ = molecule to surface, $\Gamma_{T,M}$ = tip to molecule, and $\Gamma_{T,S}$ = tip to surface governing charge exchange with different surface or tip elements.

The first surface-mediated fragmentation mechanism scheme is presented in **SI Figure 9**. Under positive bias tunneling conditions (empty states, **SI Figure 9a**), the deiodination reaction is hypothesized to proceed through a rate competition-limited radical cation fragmentation mechanism. Matching these to experimental observations under low positive bias tunnelling conditions (<2 V, $I$ = 50 pA), tip-induced band bending (TIBB) may be low enough to tune the tunnel barriers (qualitative rates through these are represented by the labeled Γs) such that the radical cations are readily neutralized by electrons tunneling from tip-to-molecule ($\Gamma_{T,M} > \Gamma_{M,S}$), preventing fragmentation. With larger positive bias (>2 V, $I$ = 50 pA), the competition between these rates may shift such that $\Gamma_{M,S} > \Gamma_{T,M}$, allowing an electron to be extracted from the $n_I$ level with resultant fragmentation at the C–I bond ($k_{frag(+)}$). Under this interpretation, the likeliness of fragmentation should scale with positive bias magnitude, which was experimentally observed. For example, STM scanning at +2.8 V showed that two out of four molecules in frame transformed to alkyl-radical appearing mono-lobes (**SI Figure 11**), suggesting frustrated rates where TIMe-Ge only occasionally fragments.

Under negative bias tunnelling conditions (filled states, **SI Figure 9b**), deiodination is hypothesized to proceed through a rate competition-limited anion fragmentation mechanism. Under bias conditions where $\Gamma_{S,M} > \Gamma_{M,T}$, an electron is transferred from the silicon's partially filled $\pi^*$ band to populate the $\sigma^*_{C-I}$ level, catalyzing the anionic fragmentation ($k_{frag(-)}$). Experimentally, we observed that mono-lobes most readily appeared in tunneling for conditions with both larger negative voltages and high currents ($V$ ≤ –3 V, $I$ > 1 nA, **SI Figure 13**), while remaining stable in STM scanning for smaller negative voltages. This observation supports a slow charge transfer rate between the surface and molecule in tunneling; bringing the tip closer would reduce the tunnel barrier (increase $\Gamma_{S,M}$) into the $\sigma^*_{C–I}$. We again note that detailed further experiments beyond the scope of this work would be required to validate this postulation. To explain the stability of TIMe-Ge molecules under modest negative bias (≈ –2 V, *I* = 50 pA), we hypothesize that the HOMO ($n_I$) of the surface-bound TIMe-Ge molecules participate in the tunnel current, but the $\sigma^*_{C–I}$ is still above sample occupied states.

While the above scheme is plausible to explain in-tunneling alkyl radical creation, the alignment of the molecular orbitals with the pinned semiconducting substrate is not yet fully detailed. Alternatively, the molecular orbitals could be lower in energy, as shown in **SI Figure 10.** The same molecular orbitals ($n_I$ and $\sigma^*_{C–I}$) are still involved in creation of an alkyl radical, but at the opposite polarity with the opposite fragmentation.

Under positive bias tunneling conditions (empty states, **SI Figure 10a**), the deiodination is now a rate competition-limited anionic fragmentation mechanism through the $\sigma^*_{C–I}$ level. Under low positive bias tunnelling conditions (<2 V, *I* = 50 pA), while the tip Fermi level is above the $\sigma^*_{C–I}$, the radical anion is readily neutralized by a low molecule to surface tunnel barrier ($\Gamma_{M,S} > \Gamma_{T,M}$), allowing the electron to tunnel out before anionic fragmentation can occur. With larger positive bias (>2 V, *I* = 50 pA), the competition between these rates may shift such that $\Gamma_{M,S} < \Gamma_{T,M}$, allowing fragmentation at the C–I bond ($k_{frag(-)}$).

Under negative bias tunnelling conditions (filled states, **SI Figure 10b**), deiodination is hypothesized to proceed through a rate competition-limited cation fragmentation mechanism. With sufficiently negative bias, an electron can be removed from the $n_I$ orbital ($\Gamma_{S,M} < \Gamma_{M,T}$) to begin C–I fragmentation ($k_{frag(+)}$). Lower negative STM scanning bias ( > –3.0 V) was stable as the $n_I$ orbital was always easily populated by surface electrons.

The presentation of multiple mechanisms in this work is due to qualitative conjecture of expected molecular orbital alignment with a pinned semi-conducting substrate. While the experimental data discussed in the main text supports the formation of alkyl radicals under zero current conditions, the magnitude of bias required for deiodination combined with modest tip withdrawals leaves some doubt as to the complete removal of tip-to-surface or tip-to-molecule charge transport pathways. A systematic series of tests identifying how height, current, and voltage parameters effect yield would be required to fully disentangle dominant processes and level alignment.[10] This investigation should be combined with band-bending calculations for a semi-conducting substrate,[11] and could be further fortified with pump-probe type experiments.[12] These proposed additional experiments are beyond the scope of this work.

## Positive Bias (Empty State) STM Scanning Dehalogenation

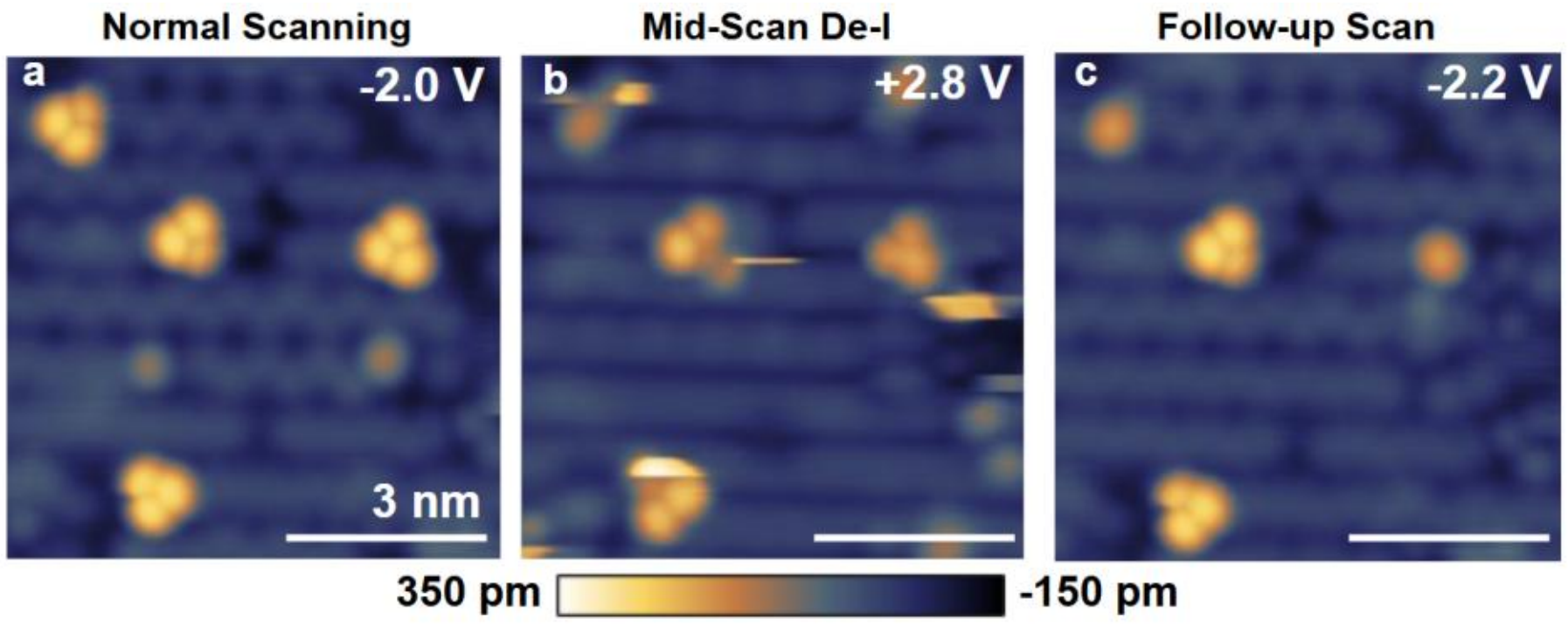

**SI Figure 11**. **In-tunneling empty state STM scanning deiodination. (a)** Filled state 77 K STM image of a TIMe-Ge deposition showing four stable tri-lobes in frame ($V$ = –2 V, $I$ = 50 pA). **(b)** Empty state STM image of the same molecules now displaying streaky discontinuities over several of the tri-lobes ($V$ = +2.8 V, $I$ = 50 pA). **(c)** Filled state STM image after (b) showing two of the tri-lobes have reduced to mono-lobe presentations. Surface iodines (mono-lobes from (a)) have also migrated out of frame from the tip field.

## Negative Bias (Filled State) STM Pulse Dehalogenation

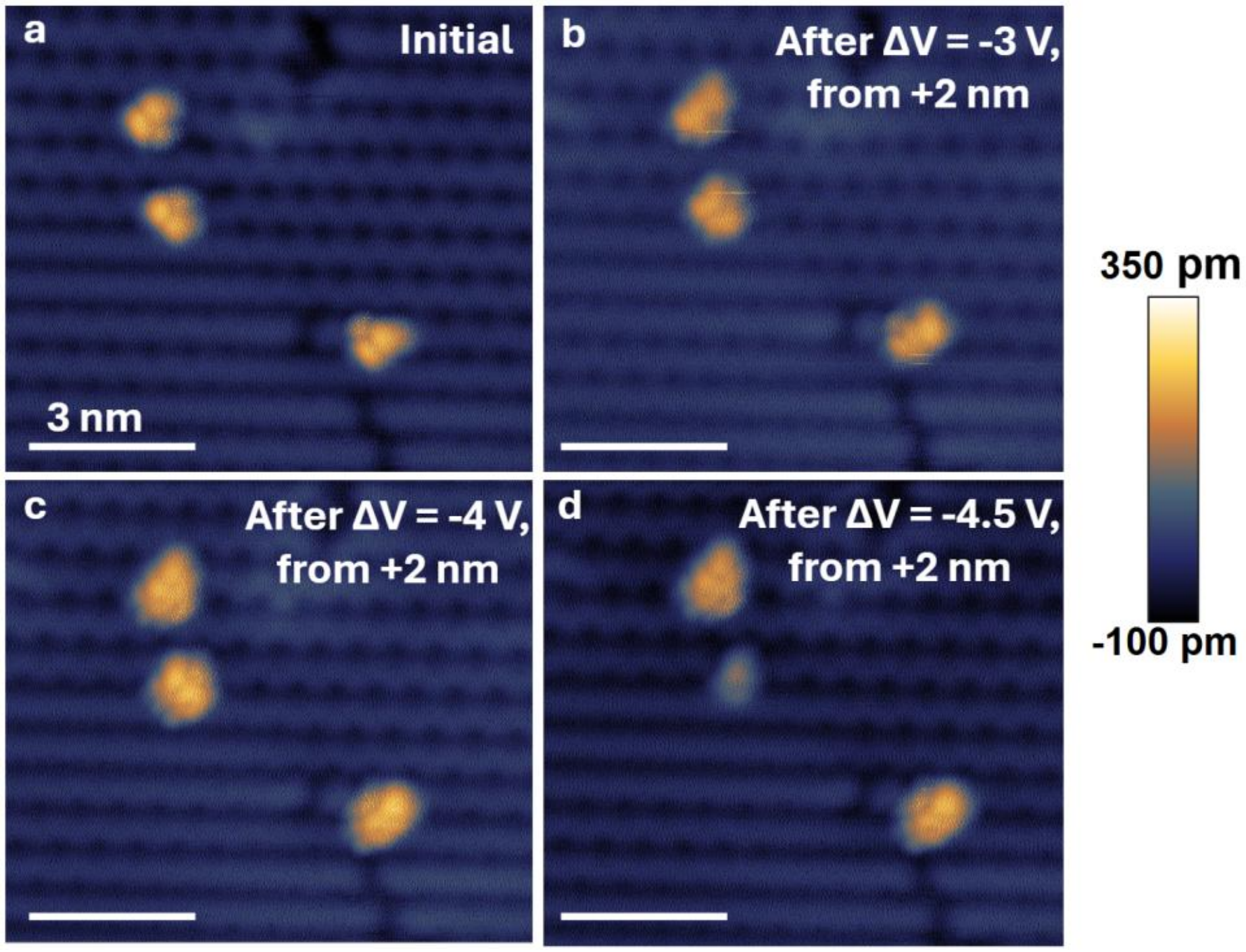

**SI Figure 12. Tip-removed negative bias pulses over TIMe-Ge.** A sequence of filled state STM images ($V$ = –2 V, $I$ = 50 pA) of tri-lobes subjected to sequential negative bias pulses from a zero tunneling current condition (z = +2 nm, retracted from the STM imaging setpoint). **(a)** Initial STM image prior to pulsing. **(b,c)** Images still displaying purely tri-lobe imaging character despite minor changes to the tip geometry leading to STM artifacting. **(d)** The first pulse in this region to yield a change in one of the molecules: both shorter and mono-lobed.

## Negative Bias (Filled State) STM Ramping Dehalogenation

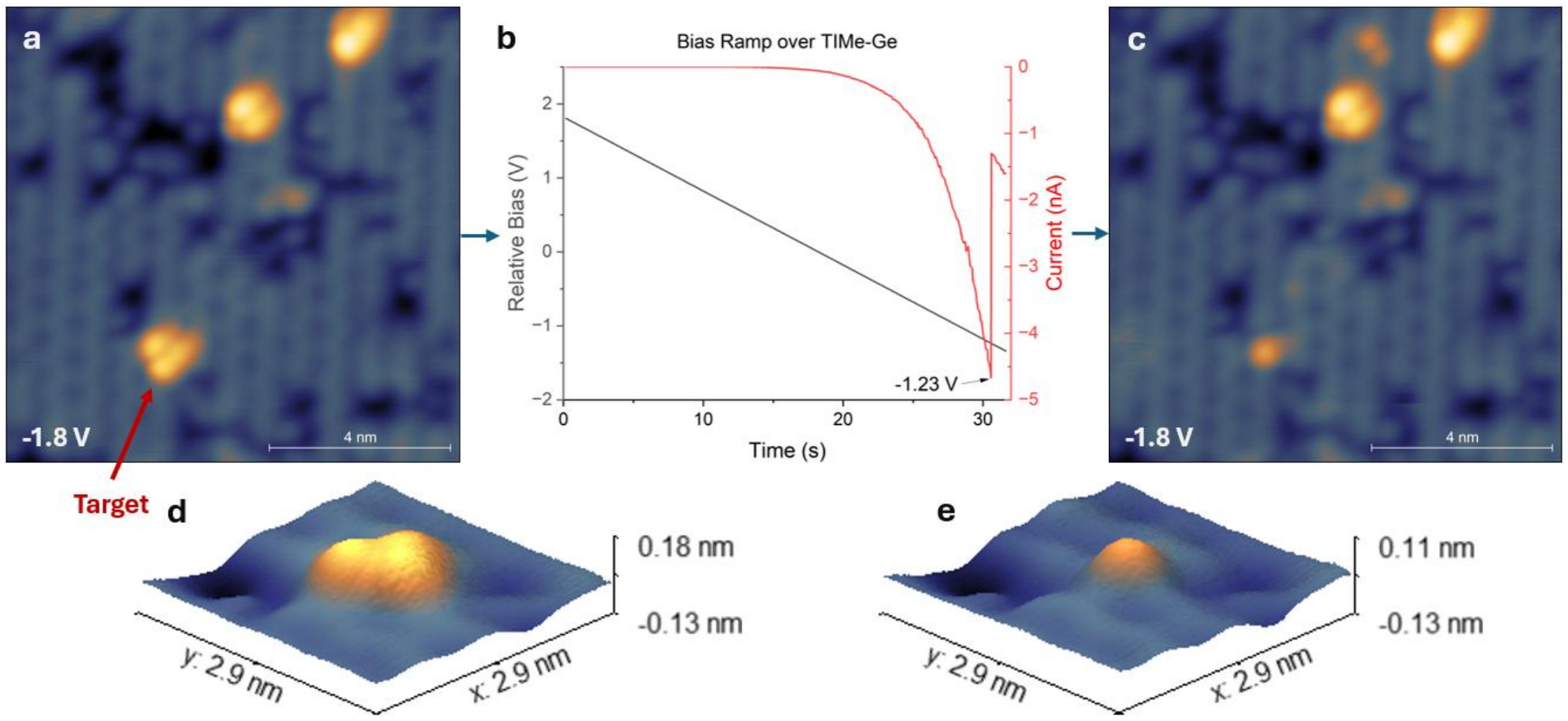

**SI Figure 13. Negative bias ramping over TIMe-Ge.** Before and after images of a tri-lobe subjected to an increasing voltage profile from a constant height determined by the initial tunneling conditions (−1.8 V, $I$ = 50 pA). **(a)** Initial STM image prior to bias ramp with target molecule indicated (−1.8 V, $I$ = 50 pA). **(b)** Bias ramp profile performed directly over top of the target molecule in constant height mode. The probe was positioned above the target and left stationary while the voltage was increased from +1.8 V until a discontinuity in the current profile was observed at −1.23 V ($I_{max}$ = −4.67 nA). **(c)** STM image taken after the bias ramp showing the now mono-lobed target and unaffected neighbors (−1.8 V, $I$ = 50 pA). **(d,e)** Orthographic images of the target before and after, respectively, showing a loss of apparent height.

Positive Bias (Empty States) and Negative Bias (Filled States) Ramping Dehalogenation

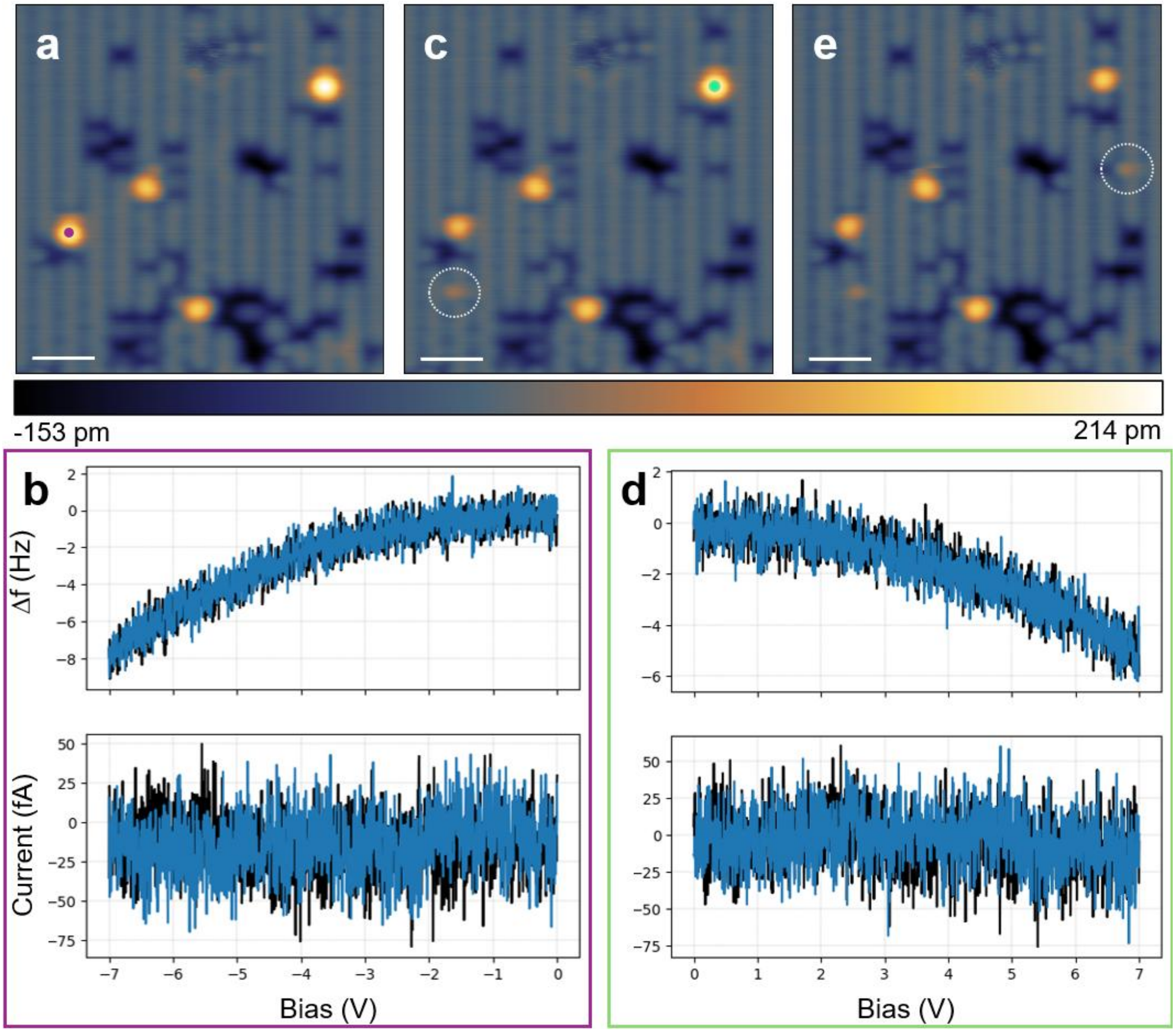

**SI Figure 14. 4 K bias ramping for deiodination. (a)** 4 K STM image of four molecules. The tall left (purple dot) and upper right molecules are intact TIMe-Ge, while the center two are thermally-activated alkyl radicals ($R_3GeCH_2^{\bullet}$). **(b)** Δf(V) and I(V) spectra for a negative bias sweep performed with the probe removed from the surface ($z$ = +1 nm, $V_{sweep}$ = 0 to −7 V) over the molecule marked by a purple dot in **(a)**. **(c)** STM image after the sweep from **(b)** showing a reduction in height for the target molecule. **(d)** Δf(V) and I(V) spectra for a positive bias sweep performed over the molecule marked by a green dot in **(c)** with the probe removed from the surface (*z* = +2 nm, $V_{sweep}$ = 0 to +7 V). **(e)** STM image after the sweep from **(c)** showing a reduction in height of the target molecule. All STM images were taken in dynamic STM scanning mode with a qPlus AFM probe (*V* = –2.0 V, *I* = 50 pA, *Osc. Amp.* = 50 pm). Probe removal heights were referenced the STM imaging setpoint over the center of a silicon dimer. White ellipses mark surface iodine atoms near the molecules where the bias sweeps were performed.

4 K STM studies of TIMe-Ge dehalogenation were conducted at both bias polarities in zero current regimes. Four in-frame molecules are shown in **SI Figure 14a**. The center left and upper right are intact TIMe-Ge molecules presenting with the DFT-predicted "shelf" indicative of intactness at 4 K filled state STM imaging (main text **Figure 1f**). The center two bright protrusions are dehalogenated TIMe-Ge molecules ($R_3GeCH_2^{\bullet}$), with dehalogenation occurring from sample exposure to elevated temperatures.

The tip was then removed from the surface to prevent current transport from tip-to-molecule or tip-to-surface and positioned over the molecule marked by the purple dot in **SI Figure 14a**. The spectra acquired from the qPlus probe during this negative bias voltage sweep showed no current (within the detection limits of the STM current amplifier), nor any anomalous signal in Δf (a step or jump), suggesting the tip is sufficiently removed to only alter the surface potential. The molecule population was then re-examined (**SI Figure 14b**), showing that the target molecule had shortened to the same height as the other in-frame alkyl radicals, and the iodine had landed on the surface nearby (white dashed ellipse). To test that the same alkyl radical product is created regardless of bias sweep polarity, the second intact molecule (green dot, **SI Figure 14c**) underwent a positive bias sweep with the tip similarly removed to prevent tip-to-surface current transport. The current and Δf signals during this sweep ($V_{sweep}$= 0 to +7 V) similarly showed no signals (**SI Figure 14d**), suggesting a passive field enhancement role in the deiodination process. Subsequent STM examination (**SI Figure 14e**) again displayed a shortened molecule with the same height as the other alkyl radicals in frame, with the iodine landing nearby (white dashed ellipse). Thus, we suspect that the alkyl product can be produced by bias sweeps of either polarity over the intended molecules. Additionally, while the sweeps should produce iodine atoms with different charge (main text **Figure 3** and **SI Figure 9** and **10**), the iodine atoms landing near their parent molecules suggests that the charged lifetimes are transient, charge exchange of the cation/anion occurs with the surface to neutralize it, or they are reactive with the partial charges or appropriate sign in a given buckled silicon dimer.

## Demethylation by Positive Bias (Empty States) Ramping and Mechanosynthesis

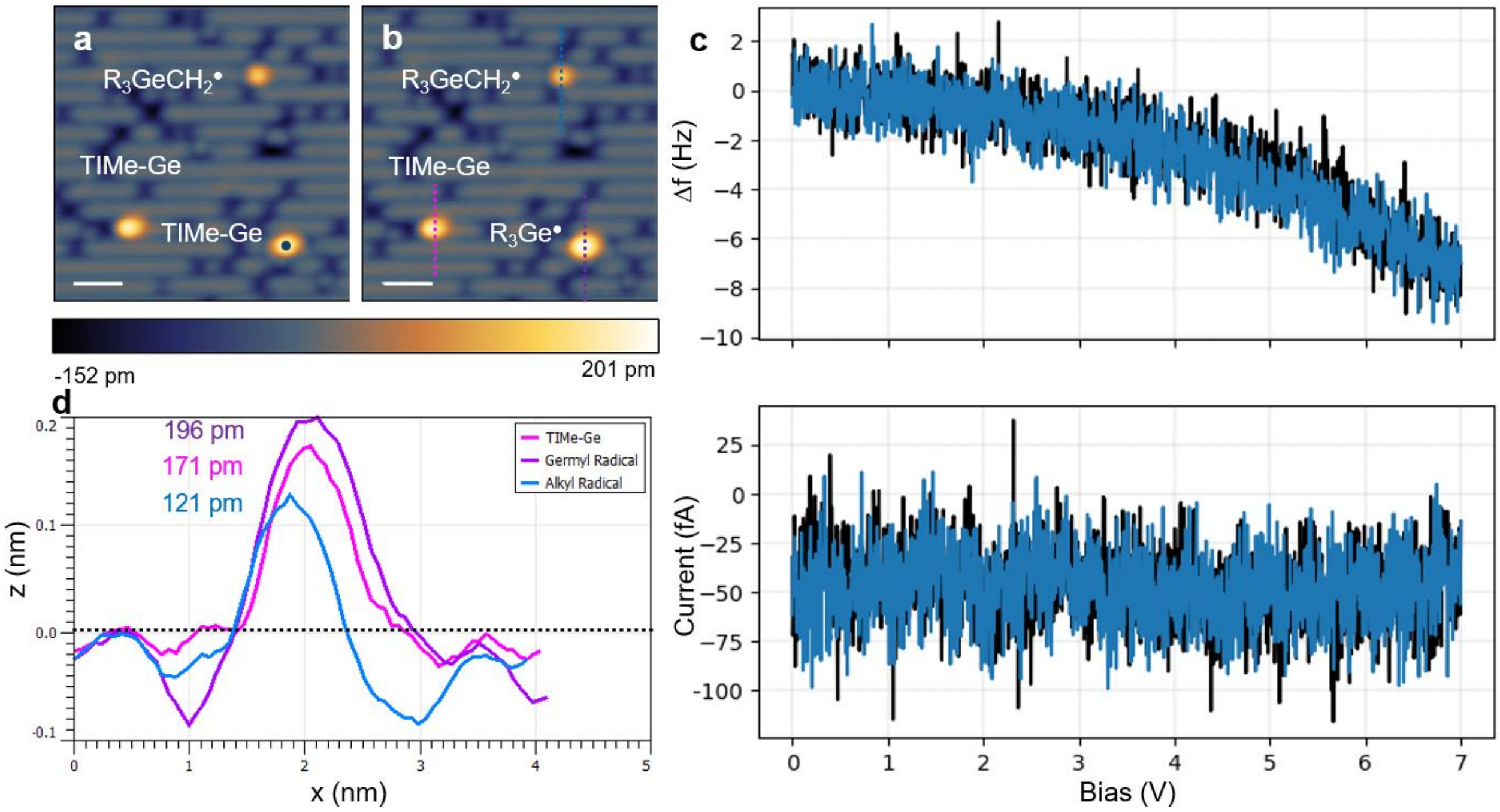

**SI Figure 15. 4 K bias ramping for demethylation. (a)** 4 K STM image of three molecules. Two are intact (TIMe-Ge) and one is an alkyl radical ($R_3GeCH_2^{\bullet}$). **(b)** 4 K STM image after a tip-removed positive bias sweep ($z$ = +2 nm, $V_{sweep}$ = 0 to +7 V) is performed over the molecule marked by a blue dot in **(a)**. **(c)** The frequency shift (Δf) and current signal collected during the positive bias sweep over the molecule. **(d)** Cross-section heights of the molecular products after the bias sweep from **(c)**, including a suspected germyl radical ($R_3Ge^{\bullet}$). The cross sections in **(d)** are marked and colour-coded to dashed lines in **(b)**. The black dotted line in **(d)** marks the height reference of the highest part of a silicon dimer. All STM images were taken in dynamic STM scanning mode with a qPlus AFM probe ($V$ = –2.0 V, $I$ = 50 pA, *Osc. Amp.* = 50 pm). Probe removal heights were referenced the STM imaging setpoint over the center of a silicon dimer.

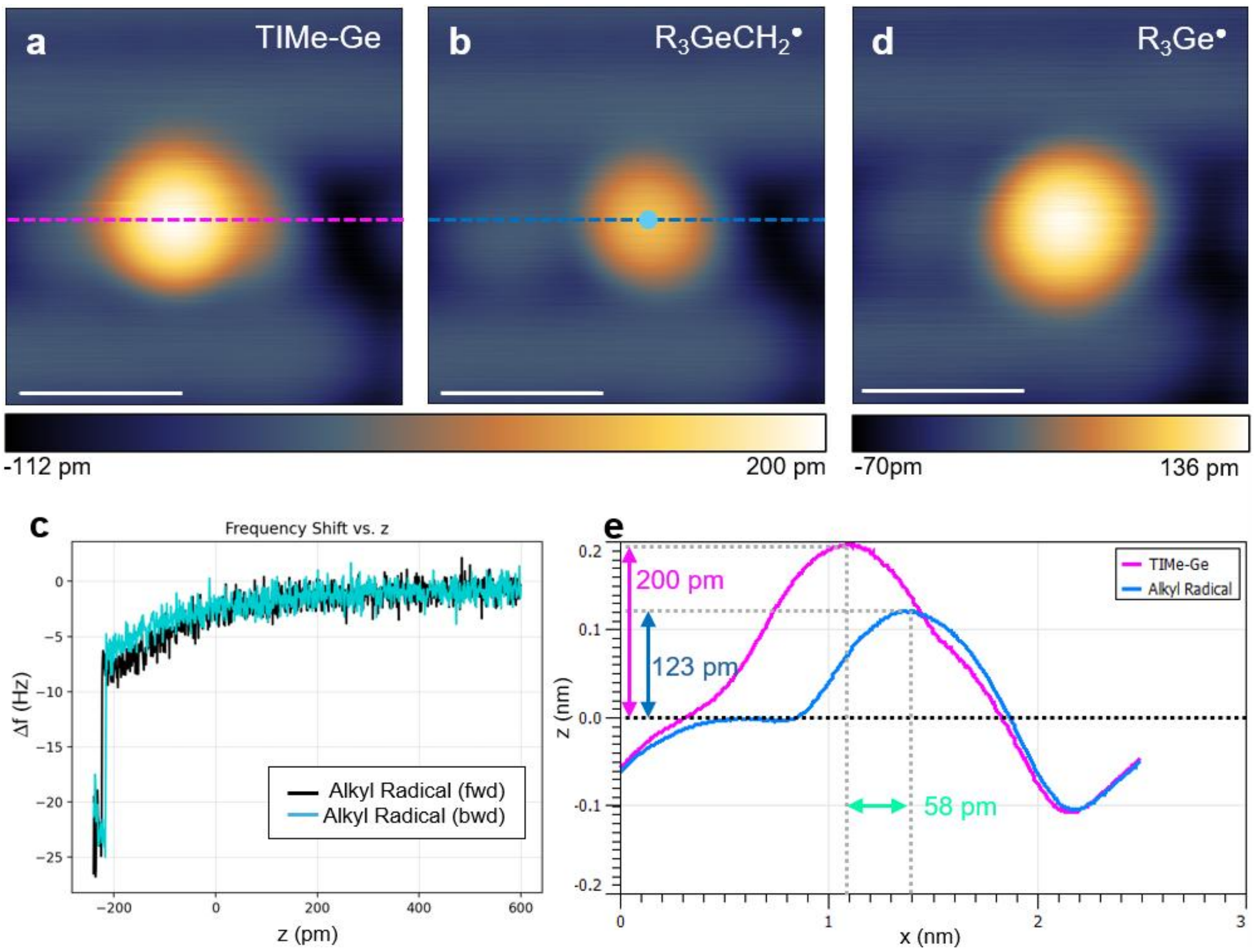

**SI Figure 16. 4 K mechanosynthetic demethylation. (a)** 4 K STM image of a TIMe-Ge molecule. **(b)** 4 K STM image after a tip-removed positive bias sweep is performed over the molecule ($z$ = +2 nm, $V_{sweep}$ = 0 to +7 V) to create an alkyl radical ($R_3GeCH_2^{\bullet}$) **(c)** Δf(z) spectra performed on the center of the alkyl molecule in **(b)**. **(d)** STM image of a suspected germyl radical ($R_3Ge^{\bullet}$). **(e)** STM cross sections of the intact and alkyl radical presentation of the molecule from **(a,b)**. Cross sections are marked with the colour-coded dashed lines in **(a,b)**. Height and horizontal shift differences are marked in **(e)**. The horizontal dashed black line marks the height reference for the top of the Si dimer. All STM images were taken in dynamic STM scanning mode with a qPlus AFM probe ($V$ = –2.0 V, $I$ = 50 pA, *Osc. Amp.* = 50 pm). Probe removal heights were referenced the STM imaging setpoint over the center of a silicon dimer.

While tip-removed positive bias sweeps most often yielded mono-lobes with alkyl radical character, occasionally a taller and wider third species was observed instead. Two intact TIMe-Ge molecules and a single alkyl radical are shown in **SI Figure 15a**. The tip was subsequently removed to a zero-tunnel current distance and a voltage sweep was performed (**SI Figure 15c**) over the intact molecule marked with a blue dot in **SI Figure 15a**. Subsequent examination of the area revealed an intact TIMe-Ge, the alkyl radical, and a new third species suspected to be the germyl radical (**SI Figure 15b**). The suspected germyl radical is larger in diameter than the intact or alkyl forms (**SI Figure 15b**,**d**), agreeing with the DFT prediction from main text **Figure 1h**. Unlike the DFT prediction, however, it is taller in apparent STM height than either the alkyl radical or intact forms of TIMe-Ge, possibly due to DFT limitations in simulating a germanium atom. A reaction scheme for this conversion from intact to germyl is presented in **SI Figure 8**. These suspected germyl radicals could also be produced by a subsequent zero current positive bias sweep over an alkyl radical.

To further corroborate its identity as a germyl radical and preliminarily test TIMe-Ge's ability to donate defined carbon fragments to a SPM tip, we created the same species mechanosynthetically (**SI Figure 16)**. An STM image of an intact TIMe-Ge molecule is provided in **SI Figure 16a**. The tip was then withdrawn to a zero tunnel current regime and a voltage sweep to +7 V was performed over the molecule to convert it to an alkyl radical (**SI Figure 16b**). Finally, the probe was switched from STM scanning mode to AFM scanning mode, and a frequency shift *vs* distance curve ($\Delta f(z)$, **SI Figure 16c**) was taken over the center of the molecule. At $z = -200$ pm, a sharp discontinuity was observed in the forward curve (black), with the retraction curve (blue) showing a different measured reactivity. The subsequent STM examination of the molecule (**SI Figure 16d**) revealed the same taller and wider species as discussed in **SI Figure 15**. Cross sections of the intact molecule and alkyl radical conversion of the molecule are plotted in **SI Figure 16e**, which showed not only a height reduction but a horizontal shift of the molecule center with the iodine removal. The suspected germyl radical is not included in these cross sections as the STM imaging character changed after the $\Delta f(z)$ spectra were acquired for **SI Figure 16c,** altering the conductivity of the probe and preventing objective comparison to the prior STM frames (see z scale bars). Nevertheless, the suspected germyl radical was still taller and larger in diameter than either the intact or alkyl radical forms, as measured following tip change with other available local references. The hysteric curves in **SI Figure 16c,** reduced probe conductivity, and alteration of the molecule strongly support functionalization of the apex with a $CH_2$ fragment and demethylation of the parent molecule, but further work is needed to confirm.

# XPS Additional Information

## Other XPS Spectra For Figure 4

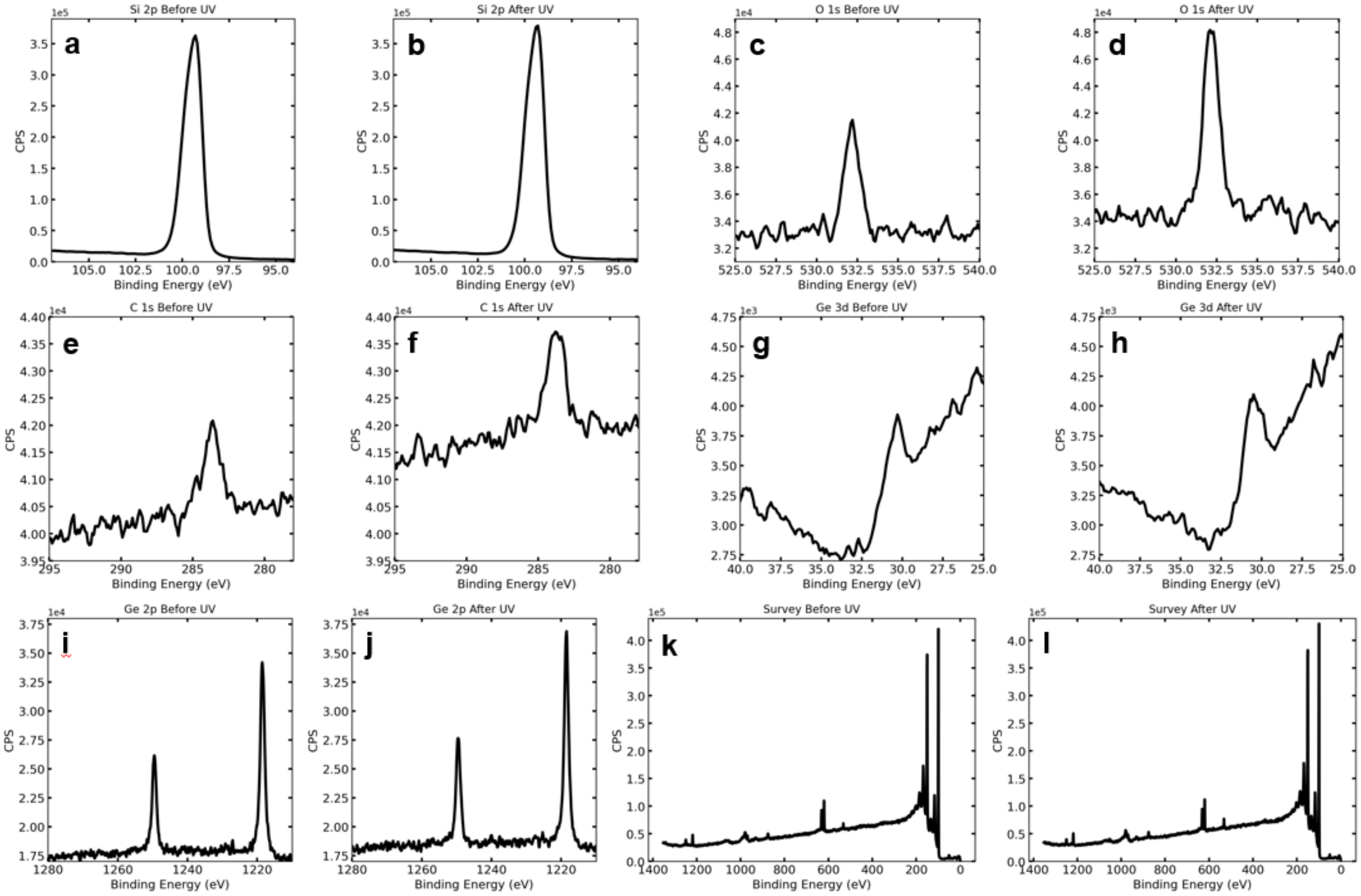

**SI Figure 17.** Additional XPS spectra before and after UV irradiation accompanying main text **Figure 4a,c. (a, b)** Si 2p, **(c, d)** O 1s, **(e, f)** C 1s, **(g, h)** Ge 3d, **(i, j)** Ge 2p, **(k, l)** Survey. y-axis and x-axis scales have been standardized for before and after.

## I 3d XPS Fit Areas

**SI Table 2. Main text Figure 4a,c STM counts of molecule populations and XPS area ratios.** For XPS ratios, C–I and Si–I area values are taken from the I $3d_{5/2}$ peak areas in the I 3d spectrum as the constraints (see *Methods – XPS* in the main text) enforce the same ratio for both spin-orbit peaks.

| **UV Status** | **XPS** | | | **STM** | | |
|---|---|---|---|---|---|---|
| | **C–I3d$_{5/2}$ Area** | **Si–I3d$_{5/2}$ Area** | **C–I:Si–I** | **Mono-lobes** | **Tri-lobes** | **C–I:Si-–I** |
| Before UV | 12354 | 59235 | 0.209 | 108 | 160 | 0.175 CI(95 %)=(0.155,0.195) |
| After UV | 3939 | 69981 | 0.056 | 117 | 58 | 0.090 CI(95 %)=(0.071,0.112) |

The ratios above are given for C–I:Si–I ratio, not C–I:(Total I). To convert between them, the following relation can be used.

$$tri - lobes = \frac{4r}{r+1}$$

Where ***tri-lobes*** is the C–I:(Total I) ratio expressed as a decimal, and ***r*** is the C–I:Si–I ratio, also expressed as a decimal. For example, a C–I:Si–I ratio = 0.209 would have a C–I:(Total I) ratio of 0.691 (*i.e.*, 69.1% of the molecules are three-leg-down with an intact normal-facing $CH_2I$ group).

**SI Table 3. Main text Figure 4e XPS ratios as a function of temperature.** For XPS ratios, C–I and Si–I area values are taken from the I $3d_{5/2}$ peak areas in the I 3d spectrum as the constraints (see *Methods – XPS* in the main text) enforce the same ratio for both spin-orbit peaks. The row marked with asterisks (*) was captured post-STM.

| Temperature (K) | XPS | | |
| --- | --- | --- | --- |
| | C–I $3d_{5/2}$ Area | Si–I $3d_{5/2}$ Area | C–I:Si–I |
| 140 | 40027 | 37230 | 1.075 |
| 140* | 32866* | 92115* | 0.356* |
| 190 | 22392 | 98984 | 0.226 |
| 230 | 21547 | 100538 | 0.214 |
| 260 | 22915 | 100375 | 0.228 |
| 285 | 23991 | 99060 | 0.242 |
| 322 | 21786 | 102984 | 0.212 |
| 335 | 21932 | 106077 | 0.207 |

# Experimental Methods Additional Details

## Calibrating the Silicon Wafer Surface Temperature under "Dynamic" Conditions

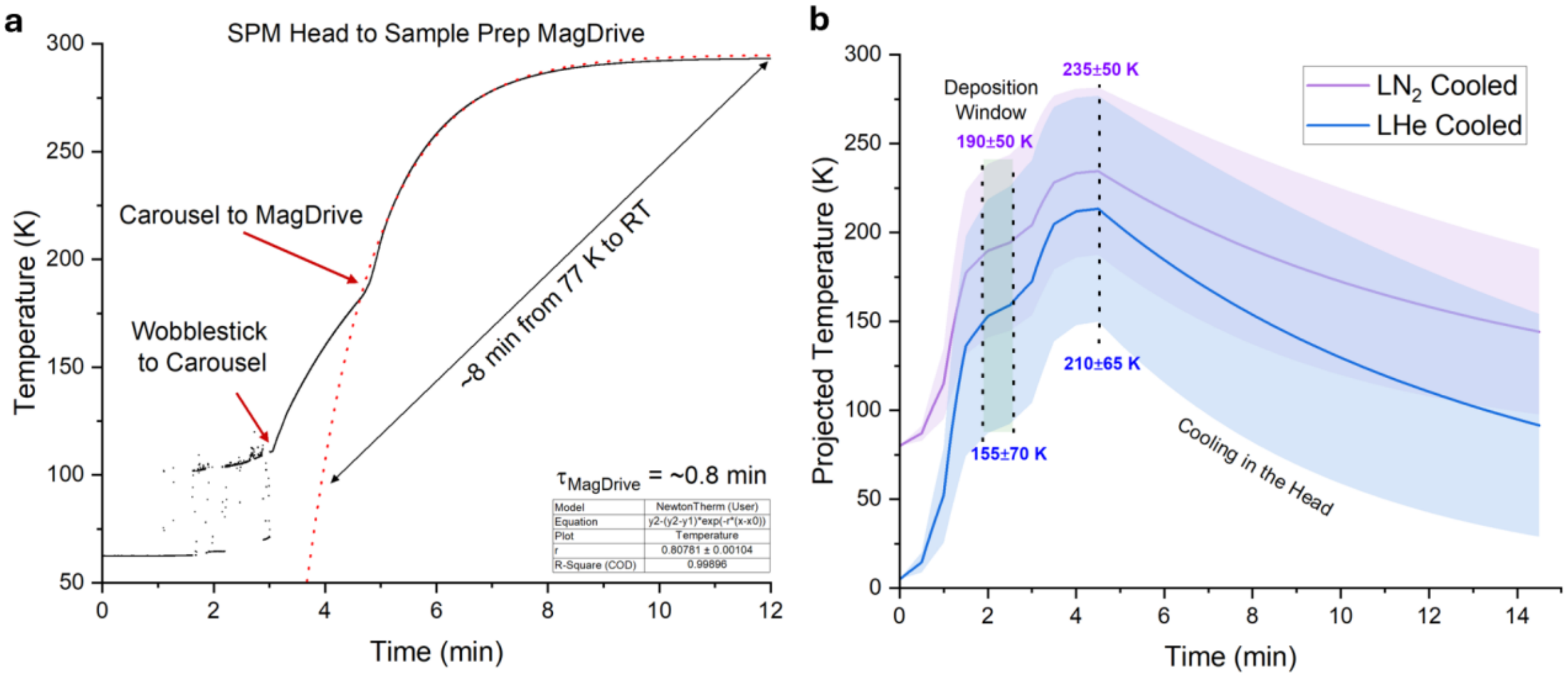

**SI Figure 18. Heating and cooling in SPM systems without a dedicated transfer arm for thermal control. (a)** Measured temperature of the thermocouple-equipped sample plate upon removal from the 77 K-cooled SPM head to the ScientaOmicron MagDrive *via* the sample storage carousel. The dashed line indicates a projected fit back to 77 K for illustration purposes. **(b)** Estimated thermal windows theoretically calculated from the measured heating curves for each individual step.

In SPM systems not equipped with a cooling stage for the silicon sample, assessments of the substrate temperature effect on the deposition quality and number of intact TIMe-Ge molecules were made by either preheating or precooling the substrate in the minutes before deposition. The STM head was generally used for cooling, while heating was done *via* direct current on the wafer, with a ~30 min waiting period in the sample storage carousel after flash-annealing for near-room temperature depositions. Using the custom sample temperature measuring devices discussed later in **SI Figure 19c**, effective thermal "half-lives" were empirically estimated for most stages in the transfer process, especially for precooled samples. This calculation relies on a simple numerically modeled variant of the so-called "Newton's law" of cooling:[13]

$$T(t) = T_{RT} + (T(0) - T_{RT}) * exp\left(-\frac{t}{\tau}\right)$$

With $\boldsymbol{T_{RT}}$ being approximately 300 K and the thermal half-life $\boldsymbol{\tau}$ being estimated from fit, assuming the mechanisms of heat transfer are similar under the given UHV conditions (*e.g.,* no convection and similar degrees of radiative exposure throughout). We estimate that the precooled samples were approximately 190 ± 50 K when cooled with $LN_2$ during deposition *versus* 155 ± 70 K when using LHe (**SI Figure 18**). The average projected peak temperature experienced by the surface-bound TIMe-Ge molecules before returning to the STM head was 235 K and 210 K for $LN_2$ and LHe cooling environments, respectively. This method of temperature estimation was used for sample temperatures in main text **Figure 2**.

## Calibrating the Silicon Wafer Surface Temperature -- The STM-XPS System

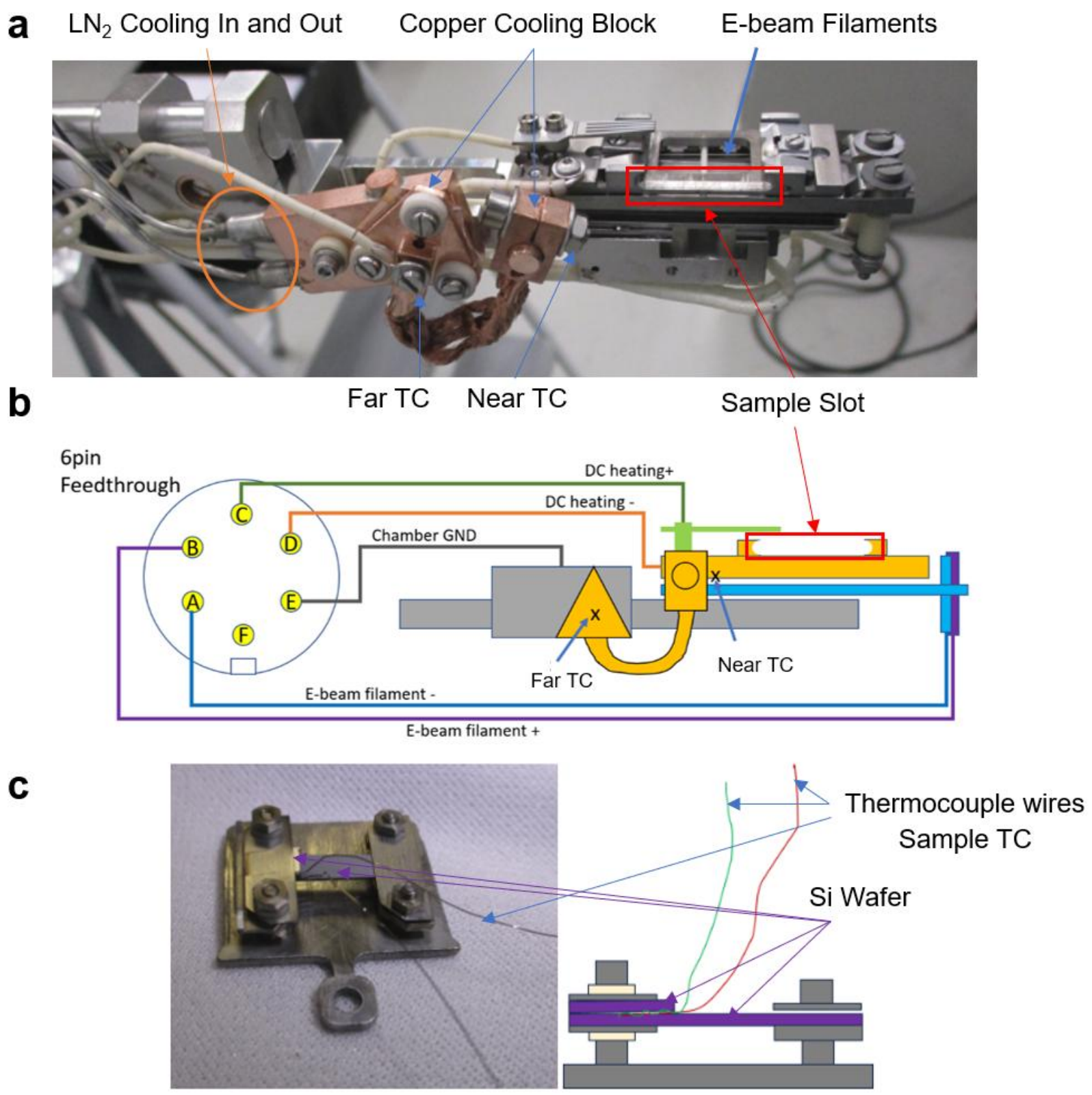

**SI Figure 19. Silicon wafer surface temperature calibration. (a)** The combination STM-XPS variable temperature manipulator arm. Cooling is conducted *via* $LN_2$ flow to a solid copper block attached to the sample slot, while heating is done either *via* resistive direct current or e-beam filament. The permanently mounted thermocouples (TCs) are labeled "Near TC" and "Far TC." **(b)** Schematic of the system electrical connections. The arm is mounted in the UHV system, with connections outside of UHV facilitated by the 6-pin vacuum feedthrough. **(c)** K-type **t**hermocouple leads sandwiched between two pieces of silicon on the left side of the sample plate. These leads are run to a UHV feedthrough for *ex-situ* temperature measurement (not pictured).

In the STM-XPS system, a dedicated variable temperature manipulator arm was available with two permanently mounted thermocouples for temperature measurement: Near TC and Far TC in **SI Figure 19a,b**. However, they do not reflect the temperature at the silicon surface suspended across the sample holder (**SI Figure 19c**). To determine the temperature at the silicon surface more accurately, a special sample holder was manufactured in which two K-type thermocouple leads were sandwiched under a second silicon wafer piece on one side of the sample holder (**SI Figure 19c, "**Sample TC"). The K-type leads were then connected to an electrical feedthrough on the UHV system to measure temperature *ex situ* at the silicon surface. The system was vented and subsequently baked to facilitate the installation of this special calibration sample. We then performed a series of tests to map the permanent thermocouple readings of the Near and Far TCs to the temperature measured at the sample TC in the temporary calibration sample. Thus, the temperature on a standard sample wafer with no thermocouple could be inferred using the permanently mounted ones using the calibration procedure.

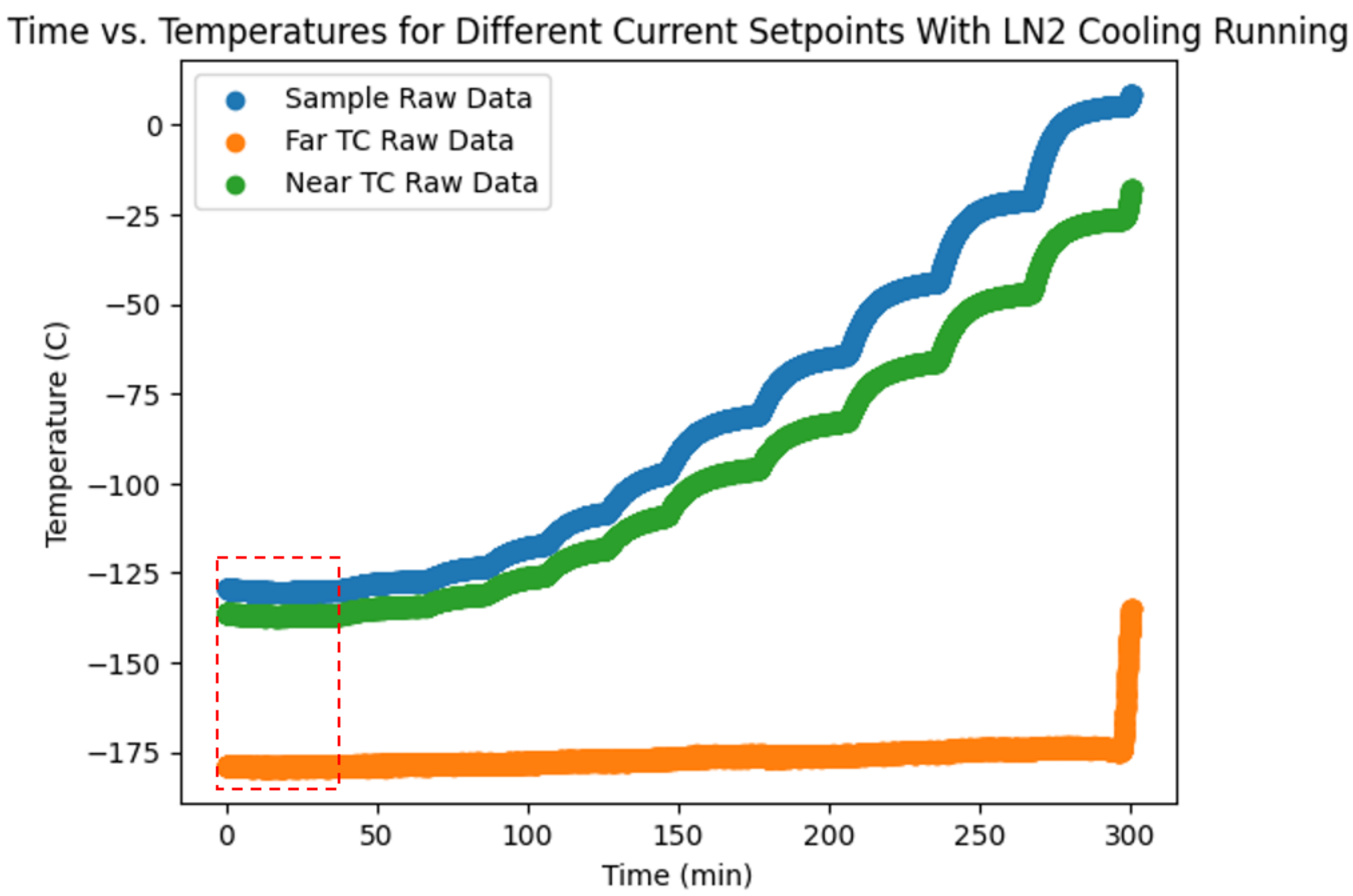

**SI Figure 20. Thermocouple offsets from counter heating a cold sample.** The Sample, Far, and Near TCs are denoted in blue, green, and orange, respectively. The "rises" (more obvious in the blue and green curves) are points where the e-beam filament current was increased in setpoint. The red-dashed box demarks where no e-beam heating is applied and all three thermocouples are in thermal equilibrium. Data were acquired using the setup described in **SI Figure 19**.

For the STM-XPS system, the arm was cooled to its minimum temperature and all three thermocouples were allowed to come to thermal equilibrium (red-dashed box in **SI Figure 20)**. Thermal equilibrium was defined as a change of <1 °C per 5 min. Due to the large thermal mass of the variable temperature manipulator arm, this process took between 15 to 45 min depending on the magnitude of the change. Differences were noted in measured offsets between all three thermocouples, with the Far TC significantly cooler than the Near or Sample TCs, likely due to being mounted directly on the copper block through which the $LN_2$ flows. E-beam radiative heating then was increased to different fixed current setpoints to counter heat the manipulator arm and raise it from its minimum temperature back up to RT (manifests in **SI Figure 20** as the rises seen periodically in time).

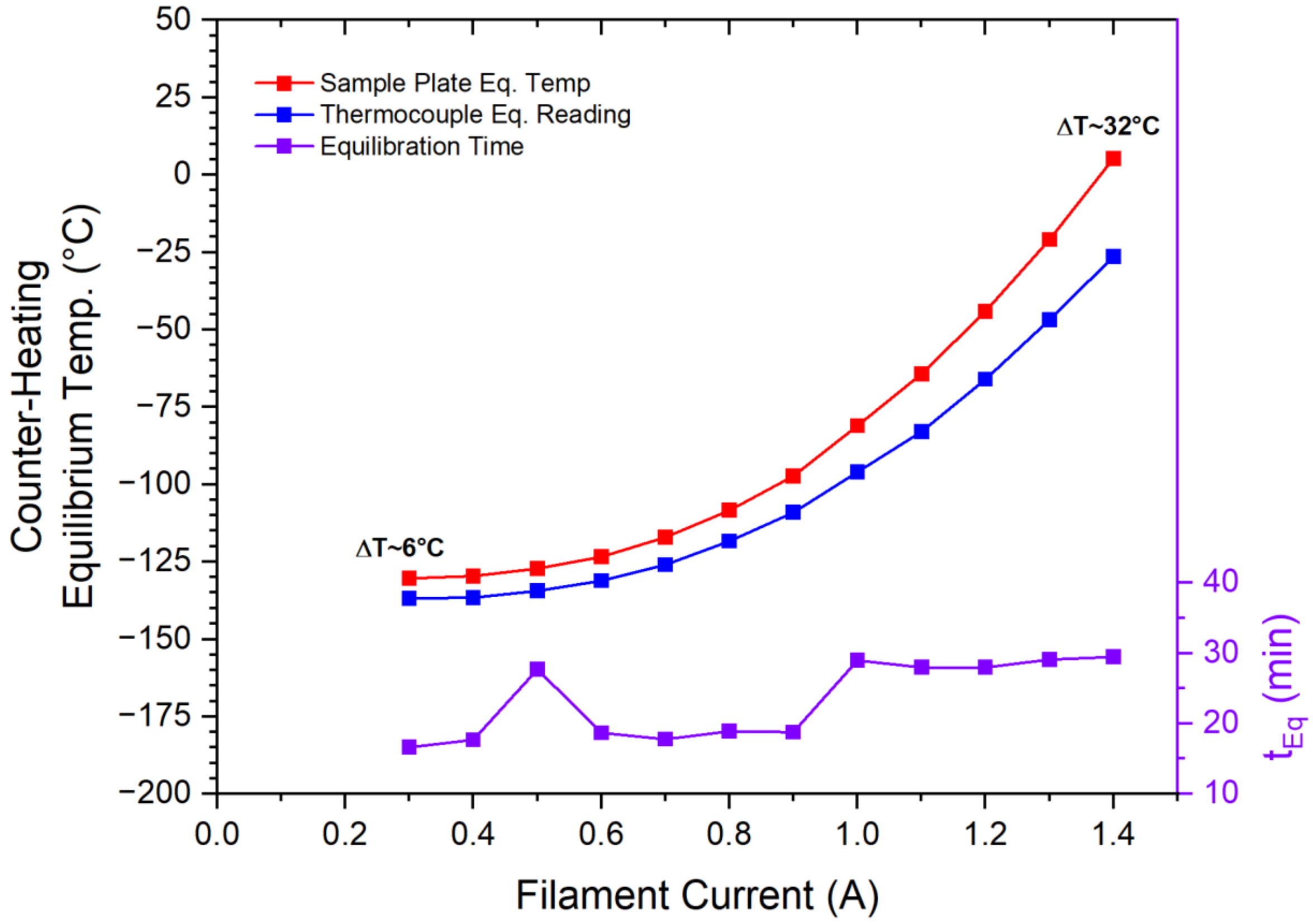

**SI Figure 21. Counter heating fits.** The time data from increases in **SI Figure 20** were extracted for a given e-beam current setpoint. From these time slices, the sample data were fit to the thermal transfer equation. Additionally, the experimentally measured Sample and Near TC temperatures were extracted at thermal equilibrium (<1 °C per 5 min). The time to thermal equilibrium and offset between the Sample and Near TCs were also calculated from the experimental data.

For each increase, we waited for thermal equilibrium was and performed a fit using a conductive thermal transfer equation of the form:

$$T(t) = (T_i - T_{inf}) * exp(b * t) + T_{inf}$$

Where ***T(t)*** is the sample temperature as a function of time, $\boldsymbol{T_i}$ is the initial temperature of the dataset, $\boldsymbol{T_{inf}}$ is the temperature at infinite time (full thermal equilibrium), **b** is the thermal decay factor, and **t** is time. In the fits, ***b*** and $\boldsymbol{T_{inf}}$ were allowed to vary as free parameters. These fits were performed to test if thermal equilibrium was truly reached; $\boldsymbol{T_{inf}}$ often agreed well with our experimentally measured thermal equilibrium (<1 °C per 5 min).

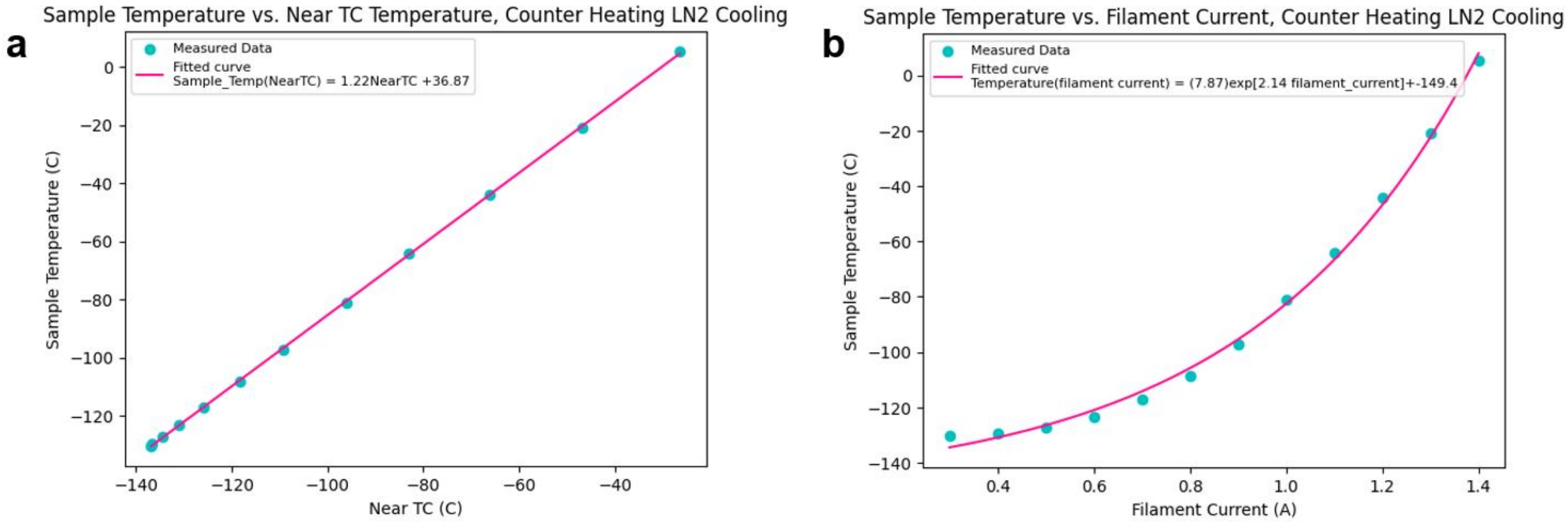

**SI Figure 22. Sample TC temperature related to Near TC temperature and e-beam current during counter heating. (a)** Mapping of the Sample TC temperature to Near TC temperature in thermal equilibrium (blue dots). The data were fitted with a linear expression (pink line). **(b)** Plot of the Sample TC temperature *vs.* e-beam filament current in thermal equilibrium (blue dots). The data were fitted with an exponential decay curve (pink line).

From **SI Figure 18** data, fits were performed on Sample TC temperature *vs.* Near TC temperature, as well as Sample TC temperature *vs.* e-beam filament current (**SI Figure 22**). The offset between the Sample and Near TCs displayed a highly linear relationship (**SI Figure 22a**), which was fit to produce the following relation:

$$T_{Samp}(T_{Near}) = 1.22 * T_{Near} + 36.87$$

Where $\boldsymbol{T_{Samp}}$ is the temperature of the silicon sample surface in °C and $\boldsymbol{T_{Near}}$ is the temperature measured at the permanently mounted Near TC in °C. Additionally, $\boldsymbol{T_{Samp}}$ was related to the e-beam counter heating filament current (**SI Figure 22b**) such that desired temperatures on the surface could be targeted:

$$T_{Samp}(I_{ebeam}) = 7.87 * exp(2.14 * \ I_{ebeam}) - 149.4$$

Where $\boldsymbol{T_{Samp}}$ is the temperature of the silicon sample surface in °C and $\boldsymbol{I_{ebeam}}$ is the current in amps sent to the e-beam filaments. We note that both equations are specific to this transfer arm, are only valid for cooling combined with e-beam counter heating, and require waiting for thermal equilibrium. This relation would have to be uniquely defined for every novel SPM system.

The relation defined above for the Near TC to the sample surface temperature was used to generate the x-axis for **Figure 4f** in the main text. 30 min was required to achieve thermal equilibrium between the XPS scans plotted in **Figure 4e** in the main text.

## Tungsten Probe Prep by a Field Ion Microscopy (FIM)

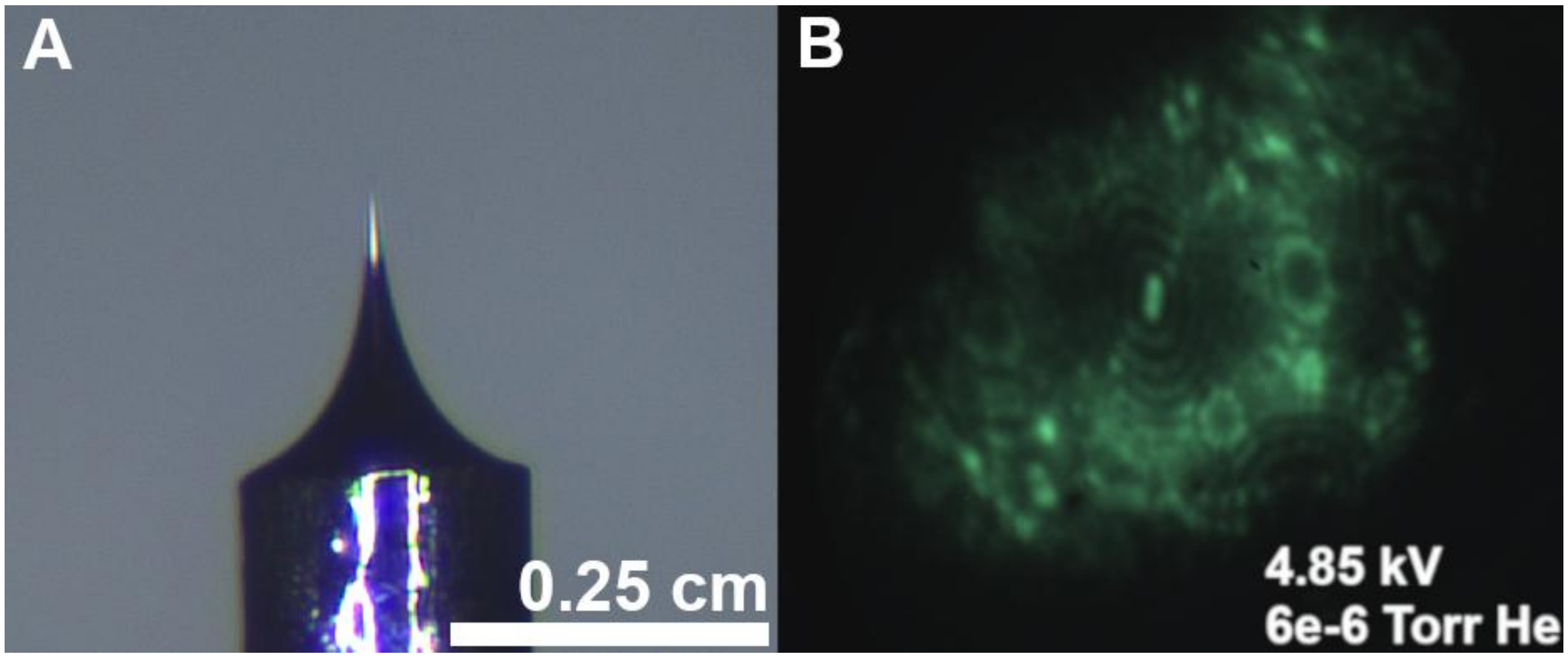

**SI Figure 23. *Ex-situ* tungsten probe characterization.** (**a**) Optical microscope image of a typical electrochemically-etched tungsten probe. (**b**) FIM image of the same probe ($V$ = 6 kV, He pressure = $4\times10^{-6}$ mB) showing atomic structure.

## UV Deiodination Module

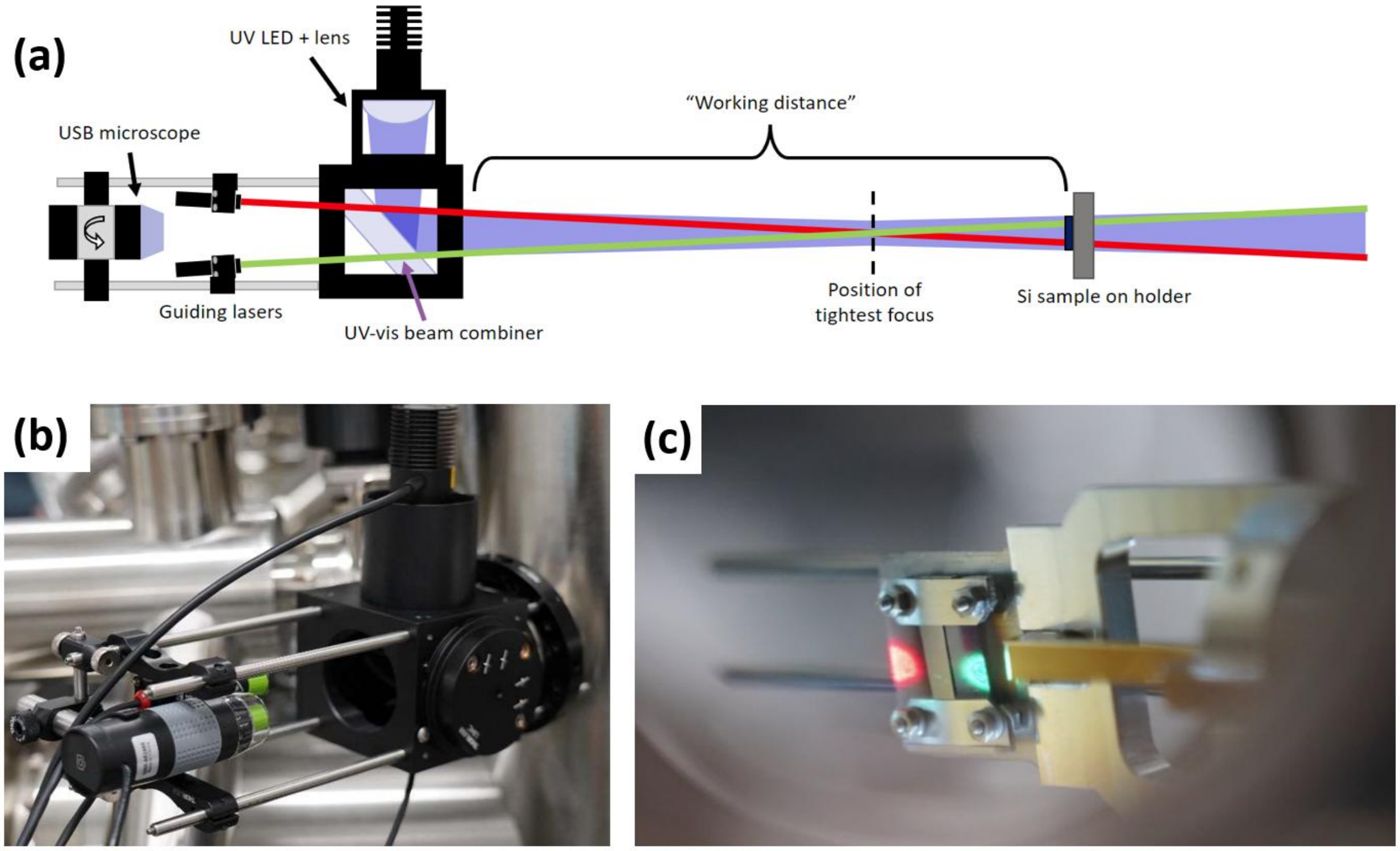

**SI Figure 24. Custom UV-LED dehalogenation module**. (**a**) Top-down diagram of the assembly. (**b**) Module flange-mounted to a UV-transparent 2.75 CF viewport. (**c**) Typical view through the beam combiner assembly including red and green guiding lasers for aiming the UV beam.